\documentclass{lmcs}
\pdfoutput=1

\usepackage{lastpage}
\lmcsdoi{17}{4}{24}
\lmcsheading{}{\pageref{LastPage}}{}{}%
{Feb.~08,~2019}{Dec.~23,~2021}{}

\usepackage[utf8]{inputenc}
\usepackage[T1]{fontenc}
\usepackage{amssymb}
\usepackage{wrapfig}
\usepackage{multicol}
\usepackage{multirow}
\usepackage{graphicx}
\usepackage{footnote}
\usepackage{stmaryrd,wrapfig}
\usepackage{wasysym}
\usepackage{booktabs}

\usepackage{tikz}
\usepackage{tikzit}
\tikzstyle{new style 0}=[fill=white, draw=black, shape=rectangle, tikzit shape=rectangle, tikzit fill=white, tikzit draw=black, minimum width=1.7cm, minimum height=1.7cm]
\tikzstyle{empty square}=[fill=none, draw=black, shape=rectangle, tikzit shape=rectangle, minimum width=5.7cm, minimum height=5.7cm]
\tikzstyle{new style 1}=[fill=none, draw=black, shape=rectangle, minimum width=10.8cm, minimum height=3.7cm]
\tikzstyle{czarny ma?y kwadrat}=[fill=black, draw=black, shape=rectangle]

\usetikzlibrary{arrows}
\usetikzlibrary{positioning,automata}
\newcommand{\cpo}{\ensuremath{\omega\mathsf{Cpo}}}

\newcommand{\defeq}{\triangleq}

\newcommand{\rightdcirc}{\makebox[1.1\width][l]{\ensuremath{%
\longrightarrow%
\makebox{$\mkern-24mu\color{white}{\bullet}\mkern+12mu$}%
\makebox{$\mkern-21mu\circ\mkern+10mu$}%
\ignorespacesafterend}}}%
\numberwithin{equation}{section}

\begin{document}

\title{A coalgebraic take on regular \texorpdfstring{\\}{} and \texorpdfstring{$\omega$}{omega}-regular behaviours}
\author[T. Brengos]{Tomasz Brengos}
\email{t.brengos@mini.pw.edu.pl}
\thanks{This work has been supported by National Centre for Research and Development Grant CYBERSECIDENT/456962/III/NCBR/2020}

\address{Faculty of Mathematics and Information Science\\
         Warsaw University of Technology\\
         ul. Koszykowa~75 \\
         00--662 Warszawa, Poland}


%%%%%%%%%%TITLE%%%%%%%%%%%%%%%%%%%%
\begin{abstract}
We present a general coalgebraic setting in which we define finite and
infinite behaviour with B\"uchi acceptance condition for systems whose type
is a monad. The first part of the paper is devoted to presenting a
construction of a monad suitable for modelling (in)finite behaviour. The second
part of the paper focuses on presenting the concepts of a (coalgebraic)
automaton and its ($\omega$-) behaviour. We end the paper with coalgebraic
Kleene-type theorems for ($\omega$-) regular input. The framework is
instantiated on non-deterministic (B\"uchi) automata, tree automata
and probabilistic automata.
\end{abstract}

\maketitle
%%%%%%%%%%%%%%%%%%%%%%%%%%%%%%%%%%%
%%%%%%%%%%%%%%%%%%%%%%%%%%%%%%%%%%%
\tableofcontents

\section{Introduction}
Automata theory is one of the core branches of theoretical computer science and
formal language theory. One of the most fundamental state-based structures
considered in the literature is a non-deterministic automaton and its relation
with languages. Non-deterministic automata with a finite state-space are known
to accept  \emph{regular languages}. These languages are characterized as subsets of words over a
fixed finite alphabet that can be obtained from simple languages
via a finite number of applications of
three types of operations:  union, concatenation and the Kleene star operation~\cite{Hopcroft:2000:IAT:557657,Kleene56}. This result is known under the name of
\begin{wrapfigure}[5]{r}{0.43\textwidth}
%\vspace{-0.3cm}
\resizebox{0.4\textwidth}{!}{
{ ${ { R ::= \varnothing \mid  a, a\in \Sigma_\varepsilon \mid  R+R \mid  R\cdot
R \mid   R^\ast}}$}
}
\caption{\small Regular expression grammar}
\end{wrapfigure}
 \emph{Kleene theorem for regular languages} and readily
generalizes to other types of finite input (see e.g.~\cite{pin:automata}).

On the other hand, non-deterministic automata have a natural infinite semantics
which is given in terms of infinite input satisfying the so-called B\"uchi
acceptance condition (or \emph{BAC} in short). The condition takes into account
the terminal states of the automaton and requires them to be visited infinitely
often. It is a common practise to use the term \emph{B\"uchi automata} in order
to refer to automata whenever their infinite semantics is taken into
consideration.

\begin{wrapfigure}[9]{r}{0.5\textwidth}
%\vspace{0.07cm}
\resizebox{0.45\textwidth}{!}{
\begin{tabular}{rcc}
\emph{input type} & \emph{Kleene theorem} &\emph{where}\\
  \toprule
  \multirow{2}{*}{$\omega$-words} & $\bigcup_{i=1}^n L_i\cdot  R_i^\omega$ &
$R_i,L_i=$ \\ &  &  regular lang.\\
  \multirow{2}{*}{$\omega$-trees} & $T_0\cdot  [T_1\ldots T_n]^\omega$ &$T_i=$
\\ & & regular tree lang.
\end{tabular}
}
\caption{\small Kleene theorems for $\omega$-regular input}
\end{wrapfigure}

Although the standard type of infinite input of a B\"uchi automaton is the set
of infinite words over a given alphabet, other types (e.g.\ trees) are also
commonly studied~\cite{pin:automata}. The class of languages of infinite words
accepted by B\"uchi automata can also be characterized akin to the
characterization of regular languages. This result is known under the name of
\emph{Kleene theorem for $\omega$-regular languages} and its variants hold for
many input types
(see e.g.\cite{NerodeAutomata,Buchi1990,Gradel:2002:ALI:938135,pin:automata}). Roughly
speaking, any language recognized by a B\"uchi automaton can be represented in
terms of regular languages and the infinite iteration operator
$(-)^\omega$. This begs the question whether these systems can be placed in a unifying
framework and reasoned about on a more abstract level so that the analogues of
Kleene theorems for ($\omega$-)regular input are derived. The recent
developments in the theory of coalgebra~\cite{DBLP:conf/cmcs/CianciaV12,silva2013:calco,urabe_et_al:LIPIcs:2016:6186,
rutten:universal} show that the coalgebraic framework may turn out to be
suitable to achieve this goal.

A coalgebra $X\to FX$ is an abstract (categorical) representation of a
 computation of a process~\cite{rutten:universal,gumm:elements}.
The coalgebraic setting has already proved itself useful in modelling finite
behaviour via least fixpoints  (e.g.~\cite{hasuo07:trace,silva2013:calco,brengos2015:lmcs})  and infinite behaviour
via greatest fixpoints of suitable mappings~\cite{DBLP:journals/entcs/Jacobs04a,DBLP:journals/entcs/Cirstea10,DBLP:journals/corr/UrabeH15}.
The infinite behaviour with BAC can be modelled by a combination of the two~\cite{urabe_et_al:LIPIcs:2016:6186,park1981:10.1007/BFb0017309}.

We plan to revisit the coalgebraic framework of (in)finite behaviour
from the perspective of systems whose type functor is a monad.
In the coalgebraic literature~\cite{brengos2014:cmcs,brengos2015:lmcs,
brengos2016:concur,brengos2015:jlamp,brengos2018:concur, BP19}
these systems are often referred to by the name of
\emph{systems with internal moves}. This name is motivated by
the research on a unifying theory of finite behaviour for
systems with internal steps~\cite{silva2013:calco,brengos2014:cmcs,brengos2015:lmcs,bonchi2015killing,
brengos2016:concur,brengos2015:jlamp}. They arise in a
natural manner in many branches of theoretical computer science, among which are
process calculi~\cite{milner:cc} (labelled transition systems
and their weak bisimulation) or automata theory (automata with
$\varepsilon$-moves), to name only two. Intuitively, these systems have a special
computation branch that is silent. This special branch, usually labelled by the
letter $\tau$ or $\varepsilon$, is allowed to take several steps and is, in some,
a neutral part of the process. As thoroughly discussed in~\cite{brengos2015:lmcs},
the nature of this type of transition suggests it is in fact (part of) the unit of a monad.
Hence, from our point of view the following terms
become synonymous:
\[
\text{coalgebras with internal moves }
=
\text{ coalgebras whose type is a monad.}
\]
This observation allows for an
elegant modelling of several coalgebraic behavioural
\begin{wrapfigure}[7]{r}{0.48\textwidth}
\vspace{-0.55cm}
\resizebox{0.48\textwidth}{!}{
\begin{tikzpicture}[shorten >=1pt,node distance=2cm,on grid]
  \node[state]   (q_0)                {$s_0$};
  \node[state]           (q_1) [right=of q_0] {$s_1$};
  \node[state] (q_2) [right=of q_1] {$s_2$};
  \path[->] (q_0) edge                node [above] {0} (q_1)
                  edge [loop above]   node         {$\varepsilon$} ()
                  edge [bend right]   node [below] {$0+1$} (q_2)
            (q_1) edge                node [above] {1} (q_2);
  \path[->] (q_2) edge [loop above]   node         {$1$} ();
\end{tikzpicture}
\hspace{0.5cm}
\begin{tikzpicture}[shorten >=1pt,node distance=2cm,on grid]
  \node[state]   (q_0)                {$s_0$};
  \node[state]           (q_1) [right=of q_0] {$s_1$};
  \node[state] (q_2) [right=of q_1] {$s_2$};
  \path[->] (q_0) edge                node [above] {0} (q_1)
                  edge [loop above]   node         {$\varepsilon$} ()
                  edge [bend right]   node [below] {$(0+1)1^\ast$} (q_2)
            (q_1) edge                node [above] {$1 1^\ast$} (q_2);
  \path[->] (q_1) edge [loop above]   node         {$\varepsilon$} ();
  \path[->] (q_2) edge [loop above]   node         {$1^\ast$} ();
\end{tikzpicture}
}
\caption{\small LTS with $\varepsilon$-moves and its saturation}
\end{wrapfigure}
equivalences which take silent steps into
account~\cite{brengos2015:jlamp,brengos2016:concur,brengos2015:lmcs}.
If the type $T$ of a coalgebra $\alpha:X\to TX$ is a monad then the map $\alpha$
becomes an endomorphism $\alpha:X\rightdcirc X$ in the Kleisli category for $T$: a natural
and simple setting to study composition and fixpoints.
For instance, if $T$ is taken to be the monad modelling labelled transition systems~\cite{brengos2015:lmcs}
then Milner's weak
bisimulation~\cite{milner:cc} of an LTS given by $\alpha$ is
 a strong bisimulation on its \emph{saturation}
$\alpha^\ast$, i.e.\ the smallest LTS over the same state space s.t.
$\alpha\leq \alpha^\ast$,  $\mathsf{id}\leq \alpha^\ast \text{ and }\alpha^\ast
\cdot \alpha^\ast \leq \alpha^\ast$ (where the composition and the order are
given in the Kleisli category for the LTS monad)~\cite{brengos2015:lmcs}.
Hence, intuitively, $\alpha^\ast$ is the reflexive and transitive closure of
$\alpha$ and is formally defined as the least fixpoint
$\mu x.(\mathsf{id}\vee x\cdot \alpha)$.
The fact that labelled transition systems'
weak bisimulation can be modelled via saturation of endomorphisms of a given Kleisli
category allows for a generalization of the setting to other systems
(e.g.\ probabilistic~\cite{brengos2015:lmcs,brengos2015:jlamp}). The only
requirement is that type functor is a monad
whose Kleisli category satisfies suitable conditions for the definition of $(-)^\ast$
to be meaningful.

 The reflexive and transitive closure $\alpha\mapsto \alpha^\ast$ is
understood as an accumulation of a \emph{finite} number of compositions of the
structure with itself. Hence, the concept of coalgebraic saturation is intrinsically
related to \emph{finite} behaviour of systems with a monadic type.
A~similar treatment of infinite
behaviour (and their combination used to model B\"uchi acceptance condition)
in the context of coalgebras whose type
is a monad has not been considered so far. The closest to this goal would be~\cite{DBLP:journals/corr/UrabeH15,urabe_et_al:LIPIcs:2016:6186}, where (in)finite
trace semantics is given in the setting
of $TF$-coalgebras for a monad $T$ and an endofunctor $F$. We take this treatment
one step further and embed $TF$ into a \emph{monad} $TF^\infty$ which is tailored to modelling
(in)finite behaviours and their combinations. The new setting allows us to
present clear definitions of coalgebraic (in)finite semantics  and reason
about them. In particular, it allows us to state Kleene theorems
for regular and $\omega$-regular behaviours which would be challenging without the monadic types.

\subsection{Motivations}   Our purpose is to build a single coalgebraic setting
that allows us to easily present definitions of (in)finite behaviours and reason about them aiming at their algebraic characterization.
By finding a suitable monad $T$ describing the type of systems taken into consideration we are able
to state generic Kleene theorems connecting syntax and semantics of languages: the former imposed
by the canonical algebraic nature of $T$ and the latter given by $T$-automata
and their behaviours.

By presenting a recipe to extend a functor to a suitable monad,
we automatically encompass systems with invisible steps. However, this should {not} be
viewed as our primary goal. Instead, from our point of view, it should be perceived as a
by-product.

\subsection{The aim of the paper} We plan to:
\begin{enumerate}[(A)]
\item revisit non-deterministic (tree) automata and their behaviour in the
coalgebraic context of systems whose type is a monad,
\item provide a type monad suitable for modelling (in)finite behaviour of
general systems,\label{item:1}
\item present a setting for defining (in)finite behaviour for abstract automata
whose type is a monad,\label{item:2}
\item state and prove coalgebraic Kleene theorems for ($\omega$-)regular
behaviour,\label{item:3}
\item\label{item:4} put probabilistic automata into the framework.
\end{enumerate}
The first point is achieved in Section~\ref{section:buchi_coalgebraically}
by describing non-deterministic
 (tree) automata and their finite and infinite behaviour in terms of different
coalgebraic (categorical) fixpoint constructions calculated in the Kleisli
category for a suitable monad. Section~\ref{section:buchi_coalgebraically}
serves as a motivation for the framework presented later in
Section~\ref{section:monads} and Section~\ref{section:automata}.

Originally~\cite{hasuo06,silva2013:calco}, coalgebras with internal moves were
considered as systems $X\to TF_\varepsilon X$ for a monad $T$ and an endofunctor
$F$, where $F_\varepsilon\defeq F+\mathcal{I}d$. Under some conditions
the functor $TF_\varepsilon$
can be embedded into the monad $TF^\ast$, where $F^\ast$ is the free monad
over $F$~\cite{brengos2015:lmcs}. The monad $TF^\ast$ is sufficient to model systems
with internal moves and their finite behaviour~\cite{bonchi2015killing,brengos2015:lmcs,brengos2015:jlamp}. However, it will
prove itself useless in the context of infinite behaviour. Hence, by revisiting
and tweaking the construction of $TF^\ast$ from~\cite{brengos2015:lmcs},
Section~\ref{section:monads} gives a general description of the monad
$TF^\infty$, the type functor $TF$ (or $TF_\varepsilon$) embeds into, which is used in the
remaining part of the paper to model the combination of finite and infinite
behaviour. The reason why we find the expressive power of $TF^\infty$  suitable is the following:
the object $F^\infty X$ is defined for any $X$ as the carrier of the
coproduct of the free
 algebra $F^\ast X$ over $X$ and the algebra $F^\omega$ obtained by inversing the final coalgebra
 map. Hence, by slighty abusing the notation, we can write
$
F^\infty = F^\ast \oplus F^\omega.
$

Item~\ref{item:2} in the above list is achieved by using two
fixpoint operators: the saturation operator $(-)^\ast$ and a new operator
$(-)^\omega$  defined in the Kleisli category for
a given monad. The combination of $(-)^\ast$ and
$(-)^\omega$ allows us to define infinite behaviour with BAC\@.

Kleene-type theorems of~\ref{item:3} are a direct
consequence of the definitions of finite and infinite behaviour with BAC using
$(-)^\ast$ and $(-)^\omega$.

Finally, in Section~\ref{section:probabilistic_systems} we put probabilistic
automata into the framework of (in)finite behaviour for systems whose type is a monad.

This paper is an extended version of~\cite{brengos2018:concur} with all missing
proofs and additional Section~\ref{section:probabilistic_systems} where
probabilistic automata are considered.
%%%%%%%%%%%%%%%%%%%%%%%%%%%%%%%%%%%%%%%%%%
%%%%%%%%%%%%%%%%%%%%%%%%%%%%%%%%%%%%%%%%%%%
%%%%%%%%%%%%%%%%%%%%%%%%%%%%%%%%%%%%%%%%%%%%
%%%%%%%%%%%%%%%%%%%%%%%%%%%%%%%%%%%%%%%%%%%%%
\section{Basic notions}\label{section:basics}
%
%%
%%%
%%%%%%%%%%%%%%%%%%%%%%%%%%%%%%%%%%%%%%%%%%%
%%%%%%%%%%%%%%%%%%%%%%%%%%%%%%%%%%%%%%%%%%%
We assume the reader is familiar with basic category theory concepts like a
category, a functor, an adjunction. For a thorough introduction to category
theory the reader is referred to~\cite{maclane:cwm}. See also e.g.~\cite{brengos2014:cmcs,brengos2015:lmcs,brengos2015:jlamp} for an extensive list
of notions needed here.
%\vspace{-0.4cm}
\subsection{Non-deterministic automata}%
\label{subsection:non-deterministic_automata}%
\label{subsec:buchi_tree}%
\label{subsection:non_determinist_automata}

The purpose of this subsection and the next one
is to recall basic definitions and properties of non-deterministic
automata and their tree counterparts: an automaton, its ($\omega$-)language and Kleene theorems
for regular and $\omega$-regular languages. Note that the aim of this paper
is to take these notions and statements and generalize them to the categorical setting.

Classically, a \emph{non-deterministic automaton}, or simply \emph{automaton}, is
a tuple $\mathcal{Q}=(Q, \Sigma, \delta,q_0, \mathfrak{F})$, where $Q$ is a
finite set of \emph{states}, $\Sigma$ is a finite set called \emph{alphabet},
$\delta:Q\times \Sigma \to \mathcal{P}(Q)$ a \emph{transition function} and
$\mathfrak{F}\subseteq Q$ set of \emph{accepting states}. We write
$q_1\stackrel{a}{\to}q_2$ if $q_2\in \delta(q_1,a)$. There are two standard
types of semantics of automata: finite and infinite. The finite semantics of
$\mathcal{Q}$ is defined as the
set of all finite words $a_1\ldots a_n\in \Sigma^*$ for which there is a
sequence of transitions $q_0\stackrel{a_1}{\to} q_1\stackrel{a_2}{\to}q_2\ldots
q_{n-1}\stackrel{a_n}{\to} q_n$ which ends in an accepting state $q_n\in
\mathfrak{F}$~\cite{Hopcroft:2000:IAT:557657}.
The infinite semantics, also known as the \emph{$\omega$-language} of
$\mathcal{Q}$, is the set of infinite words  $a_1a_2\ldots \in \Sigma^\omega$
for which there is a run $r=q_0\stackrel{a_1}{\to} q_1\stackrel{a_2}{\to}
q_2\stackrel{a_3}{\to} q_3 \ldots$ for which the set of indices $\{i \mid q_i
\in \mathfrak{F}\}$ is infinite, or in other words, the run $r$ visits the set
of final states $\mathfrak{F}$ infinitely often. Often in the literature, in
order to emphasize that the infinite semantics is taken into consideration the
automata are referred to as \emph{B\"uchi automata}~\cite{pin:automata}.
In our work we consider (B\"uchi) automata without the initial state
specified and define the ($\omega$-)language in an automaton for any
given state (see Section~\ref{section:buchi_coalgebraically} for details).

\subsubsection{Kleene theorems} Finite and infinite
semantics of non-deterministic automata can be characterized in terms of two Kleene theorems
(see e.g.~\cite{Hopcroft:2000:IAT:557657,pin:automata}). The first statement is the following.
A language $L\subseteq \Sigma^\ast$ is a language of
finite words of an automaton $\mathcal{Q}$ (a.k.a. \emph{regular language}) if and only
if it is a \emph{rational language}, i.e.\ it
can be obtained from languages of the form $\varnothing$ and $\{a\}$ for any $a\in \Sigma$
by a sequence of applications of finite union, concatenation and Kleene star operation
with the latter two given respectively by:
\begin{align*}
& R_1 \cdot R_2 \defeq  \{ w_1w_2\mid w_1\in R_1, w_2\in R_2\},\\
& R^\ast \defeq  \{ w_1\ldots w_n \mid w_i \in R \text{ and } n =0,1,\ldots \},
\end{align*}
for $R_1,R_2,R\subseteq \Sigma^\ast$.

The second Kleene theorem focuses on $\omega$-languages. A language $L_\omega\subseteq \Sigma^\omega$
is an $\omega$-language (a.k.a. \emph{$\omega$-regular language}) of an automaton $\mathcal{Q}$ if and only if it is
$\omega$-\emph{rational}, i.e.\ it can be written
as a finite union
\begin{align}
L_\omega = L_1\cdot R_1^\omega \cup \ldots \cup L_n \cdot R_n^\omega,\label{formula_omega_regular_language}
\end{align}
where $L_i,R_i$ are regular languages, the
language $R_i^\omega\subseteq \Sigma^\omega$ is given by\footnote{Our definition of $R_i^\omega$ and the one
presented in e.g.~\cite{pin:automata} differ slightly on $R_i$ with
$\varepsilon\in R_i$. Indeed, in \emph{loc.\ cit.},
$R_i^\omega = \{ w_1w_2\ldots \mid w_i\in R_i \text{ and } w_i \neq
\varepsilon\}$.
This small difference does not change the formulation of the Kleene theorem.
We choose our definition
of $(-)^\omega$ because it can be viewed as the greatest fixpoint
of a certain assignment. See the following sections for details.}:
\[R_i^\omega \defeq
\left \{ \begin{array}{cc}
\{w_1w_2w_3\ldots \mid w_i \in R_i \} & \varepsilon\notin R_i,\\
\Sigma^\omega & \text{ otherwise}
\end{array}
\right. \]
 and $L\cdot R_\omega \defeq \{w v \mid w\in L, v\in R_\omega \}$
for  $L\subseteq \Sigma^\ast$  and $R_\omega \subseteq \Sigma^\omega$.

\subsection{Tree automata}
There are several other variants of input for non-deterministic B\"uchi automata
known in the literature~\cite{pin:automata,Gradel:2002:ALI:938135}. Here, we
focus on \emph{non-deterministic (B\"uchi) tree automaton}, i.e.\ a tuple
$(Q,\Sigma,\delta,\mathfrak{F})$, where $\delta:Q\times \Sigma \to
\mathcal{P}(Q\times Q)$ and the rest is as in the case of standard
non-deterministic automata. The infinite semantics of this machine is given by a set of
infinite binary trees with labels in $\Sigma$ for which there is a run whose
every branch visits $\mathfrak{F}$ infinitely often~\cite{pin:automata,Gradel:2002:ALI:938135}. We recall these notions here below
(with minor modifications to suit our language) and refer the reader to e.g.~\cite{pin:automata} for more details.

\subsubsection{Trees}\label{subsection:trees}
Formally, a \emph{binary tree} or simply \emph{tree} with nodes in $A$ is a
function $t: P \to A$, where $P$ is a non-empty prefix closed subset of
$\{l,r\}^\ast$. The set $P\subseteq \{l,r\}^\ast$ is called the \emph{domain} of
$t$ and is denoted by $\mathsf{dom}(t)\defeq P$. Elements of $P$ are called
\emph{nodes}. For a node $w\in P$ any node of the form $wx$ for $x\in \{l,r\}$
is called a \emph{child} of $w$. A tree is called \emph{complete} if all nodes
have either two children or no children.  The \emph{height} of a tree $t$ is $\max
\{ |w| \mid w\in \mathsf{dom}(t)\}$. A tree $t$ is \emph{finite} if it is of a
finite height, it is \emph{infinite} if $\mathsf{dom}(t) = \{l,r\}^\ast$. The
\emph{frontier} of a tree $t$ is $\mathsf{fr}(t) \defeq \{ w\in
\mathsf{dom}(t)\mid \{wl,wr\} \cap \mathsf{dom}(t) = \varnothing \}$. Elements of
$\mathsf{fr}(t)$ are called \emph{leaves}. Nodes from $\mathsf{dom}(t)\setminus
\mathsf{fr}(t)$ are called \emph{inner nodes}. The \emph{outer frontier} of $t$
is defined by $\mathsf{fr}^+(t) \defeq \{wl, wr \mid w \in \mathsf{dom}(t)\} \setminus
\mathsf{dom}(t)$. I.e.\ it consists of all the words $wi\notin \mathsf{dom}(t)$
such that $w\in \mathsf{dom}(t)$ and $i\in \{l,r\}$. Finally, set
$\mathsf{dom}^+(t) \defeq \mathsf{dom}(t) \cup \mathsf{fr}^+(t)$.

Let $T_{\Sigma}X$ denote the set of all complete trees $t:P\to \Sigma+X$ with
inner nodes taking values
\begin{wrapfigure}[11]{r}{0.4\textwidth}
\resizebox{.4\textwidth}{!}{
\begin{tikzpicture}

\pgfmathsetmacro{\przesuniecieX}{0};
\pgfmathsetmacro{\x}{1};
\pgfmathsetmacro{\y}{0};
\pgfmathsetmacro{\shadow}{0.0};

\node (plus1) at (\x,\y) {$+$};
\node () at (\x-0.6,\y) {\tiny $t=$};
\node (times1) at (\x-0.3,\y-0.3) {$-$};
\node (var1) at (\x+0.3,\y-0.3) {$y_1$};
\node (var2) at (\x-0.6,\y-0.6) {$y_1$};
\node (var3) at (\x,\y-0.6) {$y_2$};

\draw[-] (plus1) -- (times1);
\draw[-] (plus1) -- (var1);
\draw[-] (times1) -- (var2);
\draw[-] (times1) -- (var3);

\draw[fill = white] (\x,\y) circle (0.12);
\node (plus1) at (\x,\y) {$+$};

\draw[fill = white] (\x-0.3,\y-0.3) circle (0.12);
\node (times1) at (\x-0.3,\y-0.3) {$-$};

\pgfmathsetmacro{\przesuniecieX}{\przesuniecieX+1.3};
\pgfmathsetmacro{\x}{\przesuniecieX};

\pgfmathsetmacro{\przesuniecieX}{0};
\pgfmathsetmacro{\x}{\przesuniecieX};
\pgfmathsetmacro{\y}{-1.1};
\pgfmathsetmacro{\shadow}{0.0};

\draw[fill = white] (\x,\y) circle (0.12);
\node (plus1) at (\x,\y) {$+$};
\node () at (\x-1,\y) {\tiny $g'\cdot f(x) =$};
\node (times1) at (\x-0.3,\y-0.3) {$-$};
\node (var1) at (\x+0.3,\y-0.3) {$t_1$};
\node (var2) at (\x-0.6,\y-0.6) {$t_1$};
\node (var3) at (\x,\y-0.6) {$t_2$};

\draw[-] (plus1) -- (times1);
\draw[-] (plus1) -- (var1);
\draw[-] (times1) -- (var2);
\draw[-] (times1) -- (var3);

\draw[fill = white] (\x,\y) circle (0.12);
\node (plus1) at (\x,\y) {$+$};
\draw[fill = white] (\x-0.3,\y-0.3) circle (0.12);
\node (times1) at (\x-0.3,\y-0.3) {$-$};
\pgfmathsetmacro{\przesuniecieX}{\przesuniecieX+2.1};
\pgfmathsetmacro{\x}{\przesuniecieX};
\pgfmathsetmacro{\y}{-1.1};
\pgfmathsetmacro{\shadow}{0.0};

\node (plus1) at (\x,\y) {$+$};
\node () at (\x-1,\y) {\tiny $g''\cdot f(x)=$};
\node (times1) at (\x-0.3,\y-0.3) {$-$};
\node (var1) at (\x+0.3,\y-0.3) {$t_2$};
\node (var2) at (\x-0.6,\y-0.6) {$t_2$};
\node (var3) at (\x,\y-0.6) {$t_1$};

\draw[-] (plus1) -- (times1);
\draw[-] (plus1) -- (var1);
\draw[-] (times1) -- (var2);
\draw[-] (times1) -- (var3);

\draw[fill = white] (\x,\y) circle (0.12);
\node (plus1) at (\x,\y) {$+$};
\draw[fill = white] (\x-0.3,\y-0.3) circle (0.12);
\node (times1) at (\x-0.3,\y-0.3) {$-$};
\pgfmathsetmacro{\przesuniecieX}{\przesuniecieX-1.2};
\pgfmathsetmacro{\x}{\przesuniecieX};
\pgfmathsetmacro{\y}{1.7};
\node(aaaaa) at (\x,\y+0.5) {$\Sigma = \{+,-\}$};
\node(aaaaa) at (\x,\y) {$f:\{x\}\to T_\Sigma \{y_1,y_2\}; x\mapsto t$};
\node(aaaaa) at (\x,\y-0.5) {$g':\{y_1,y_2\}\to T_\Sigma Z; y_1\mapsto
t_1,y_2\mapsto t_2$};
\node(aaaaa) at (\x,\y-1) {$g'':\{y_1,y_2\}\to T_\Sigma Z; y_1\mapsto
t_2,y_2\mapsto t_1$};

\end{tikzpicture}
}
\end{wrapfigure}
in $\Sigma$ and which have a finite number of leaves,
all from the set $X$. Note that trees from $T_{\Sigma}X$ of height $0$ can be
thought of as elements of $X$. Hence, we may write $X\subseteq T_\Sigma X$.
Moreover, trees of height $1$ can be viewed as elements from $\Sigma\times X \times X$. Thus,
$\Sigma\times X\times X\subseteq T_\Sigma X$.
Additionally, any $f:X\to Y$ induces a map $T_\Sigma f:T_\Sigma X\to T_\Sigma Y$
which assigns to $t\in T_\Sigma X$ the tree obtained from $t$ by replacing any
occurrence of a leaf $x\in X$ with $f(x)\in Y$. This turns $T_\Sigma(-)$ into a
$\mathsf{Set}$-endofunctor.  For two functions $f:X\to T_\Sigma Y$ and $g:Y\to T_\Sigma Z$ we may
naturally define $g\cdot f:X\to T_\Sigma Z$ for which $(g\cdot f)(x)$ is a tree obtained
from $f(x)$ with every occurence of a variable $y\in Y$  replaced with the tree
$g(y)\in T_\Sigma Z$. It is a simple exercise to prove that $\cdot$ is
associative. Moreover, if we denote the function $X\to {T}_\Sigma X; x\mapsto x$
by $\mathsf{id}$ then $\mathsf{id}\cdot f = f\cdot \mathsf{id}$. This follows
from the fact that $T_\Sigma$ is a monad and $g\cdot f$ is, in fact, the Kleisli
composition  for $T_\Sigma$ (see Example~\ref{example:tree_functor} for
details).

Finally, $T^{\ast}_{\Sigma}X\subseteq T_{\Sigma}X$ and
$T^\omega_{\Sigma}X\subseteq T_{\Sigma}X$ are sets of finite and infinite trees
from $T_{\Sigma}X$ respectively. Note that trees in $T^\omega_{\Sigma}X$ have no
leaves, hence $T_{\Sigma}^\omega X =T_\Sigma^\omega \varnothing$ for any set
$X$.
\subsubsection{B\"uchi tree automata and their languages}
Let $\mathcal{Q}=(Q,\Sigma,\delta,\mathfrak{F})$ be a tree automaton. A
\emph{run} of the automaton $\mathcal{Q}$ on a finite tree $t\in T_\Sigma 1$
starting at the state $s\in Q$ is a map $\mathfrak{r}:\mathsf{dom}^+(t)\to Q$
such that $\mathfrak{r}(\varepsilon)=s$ and for any $x\in
\mathsf{dom}(t)\setminus \mathsf{fr}(t)$ we have
\[
(\mathfrak{r}(xl),\mathfrak{r}(xr))\in \delta(\mathfrak{r}(x),t(x)).
\]
We say that the run $\mathfrak{r}$ is \emph{successful} if $\mathfrak{r}(w)\in
\mathfrak{F}$ for any $w\in \mathsf{fr}^+(t)$ for the tree $t$. The set of
finite trees recognized by a state $s$ in $\mathcal{Q}$ is defined as the set of
finite trees  $t\in T_\Sigma^\ast 1$ for which there is a run  in
$\mathcal{Q}$ starting at $s$ which accepts the tree $t$.

Finally, let $t\in T_\Sigma^\omega \varnothing$ be an infinite tree with nodes
in $\Sigma$. An \emph{infinite run} for $t$ starting at $s\in Q$ is a map
$\mathfrak{r}:\{l,r\}^\ast\to Q$ such that $\mathfrak{r}(\varepsilon) = s$ and:

\[
(\mathfrak{r}(xl),\mathfrak{r}(xr))\in \delta(\mathfrak{r}(x),t(x)) \text{ for
all }x\in \{l,r\}^\ast.
\]
The tree $t$ is said to be \emph{recognized by} the state $s$ in $\mathcal{Q}$
if there is a run $\mathfrak{r}$ for $t$ which start at $s$ and for each path in
$t$ some final state occurs infinitely often~\cite{pin:automata}.

\subsubsection{Rational tree languages} Rational tree languages are
 analogues of rational languages for non-deterministic automata. Akin to the standard case,
 they are defined as sets of trees obtained from trees of height $\leq 1$ by a sequence of applications of:
 finite union, composition and Kleene star closure. However,
 the non-sequential nature of trees requires us to consider composition of rational trees with
 more than one variable.

 Formally, for any subset $T\subseteq X\to T_\Sigma X$ we define $T^\ast$ by $T^\ast \defeq
\bigcup_{n} T^n$, where $T^0 = \{ \mathsf{id} \}$ and
$
T^{n} = T^{n-1} \cup \{ t'\cdot t \mid t'\in T^{n-1} \text{ and }t \in T\}.
$
For any natural number $n\in \{0,1,\ldots \}$ we slightly abuse the notation
and put $n\defeq\{1,\ldots,n\}$ and
define $\mathfrak{Rat}(1,n)$ to be the smallest family of
subsets of $T_\Sigma n$ which satisfies:
\begin{itemize}
\item $\varnothing \in \mathfrak{Rat}(1,n)$,
\item $\{t\}\in \mathfrak{Rat}(1,n)$, where $t$ is of height less than or equal
to $1$,
\item if $T\in \mathfrak{Rat}(1,n)$ and $T_1,\ldots,T_m\in \mathfrak{Rat}(1,m)$
then:
\[
\{ [t_1,\ldots, t_n]\cdot t \mid t\in T, t_i \in T_i\}\in \mathfrak{Rat}(1,m),
\]
\item if $T\in \mathfrak{Rat}(1,n)$ then for any $i\in n$:
\[
\left \{f(i) \mid f\in \{[t_1,\ldots,t_n]:n \to T_\Sigma n \mid t_i \in
T\}^\ast\right \}  \in \mathfrak{Rat}(1,n).
\]
\end{itemize}
It is easy to check that if we extend the definition of $\mathfrak{Rat}$ and
put \[\mathfrak{Rat}(m,n) \defeq m\to \mathfrak{Rat}(1,n)\] then the last
item in the above list implies that for any $T\in
\mathfrak{Rat}(n,n)$ we have $T^\ast \in \mathfrak{Rat}(n,n)$.

Now for $T\subseteq n\to T^\ast_\Sigma n$ we define $T^\omega$ as the subset
of $n \to T^\omega_\Sigma \varnothing$ consisting of common extensions of
functions in $T^k$ for any $k$.
Finally, the $\omega$-rational subset of trees is defined by~\cite{pin:automata}:
\[
\omega\mathfrak{Rat} \defeq \{ T^\omega \cdot T'\mid T\in \mathfrak{Rat}(n,n) \text{
and } T'\in \mathfrak{Rat}(1,n)\},
\]
where  $T\cdot T' \defeq \{t\cdot t'\mid t\in T, t'\in T'\}$.

\subsubsection{Kleene theorems}\label{subsection:kleene_tree}
Let $\mathfrak{Reg}$ be the set of subsets of trees from $T_\Sigma^\ast 1$ for which
there is an automaton accepting the given set of trees. Similarly, we define the set
$\omega\mathfrak{Reg}$ of infinite trees accepted by the tree automata. In this case the Kleene
theorems for regular and $\omega$-regular input are respectively given by~\cite{pin:automata}:
\[
\mathfrak{Reg} = \mathfrak{Rat}(1,1) \text{ and }
\omega\mathfrak{Reg}=\omega\mathfrak{Rat}.
\]

In Section~\ref{section:buchi_coalgebraically} we will show that Kleene theorems
for non-deterministic automata and tree automata are instances of a generic pair
 of theorems formulated on a categorical level.

\subsection{Algebras and coalgebras}
Let $F:\mathsf{C}\to \mathsf{C}$ be a functor. An \emph{$F$-coalgebra}
(\emph{$F$-algebra}) is a morphism $\alpha:A\to FA$ (resp. $a:FA\to A$). The
object $A$ is called a \emph{carrier} of the underlying $F$-(co)algebra. Given
two coalgebras $\alpha:A\to FA$ and $\beta:B\to FB$ a morphism $h:A\to B$ is
\emph{homomorphism} from $\alpha$ to $\beta$ provided that $\beta\circ h=
F(h)\circ \alpha$. For two algebras $a:FA\to A$ and $b:FB\to B$ a morphism
$h:A\to B$ is called \emph{homomorphism} from $a$ to $b$ if $b\circ F(h) =
h\circ a$. The category of all $F$-coalgebras ($F$-algebras) and homomorphisms
between them is denoted by $\mathsf{CoAlg}(F)$ (resp. $\mathsf{Alg}(F)$). We say
that a coalgebra $\zeta:Z\to FZ$ is \emph{final} or \emph{terminal} if for any
$F$-coalgebra $\alpha:A\to FA$ there is a unique homomorphism $[[\alpha]]:A\to
Z$ from $\alpha$ to $\zeta$.

\begin{exa}%
  \label{example:lts_non_deterministic_automata}
Let $\Sigma$ be a set of \emph{labels}.  \emph{Labelled transition systems} (see e.g.~\cite{sangiorgi2011:bis}) can be
  viewed as coalgebras of the type $\mathcal{P}(\Sigma\times
  \mathcal{I}d):\mathsf{Set}\to \mathsf{Set}$~\cite{rutten:universal}. Here,
  $\mathcal{P}:\mathsf{Set}\to \mathsf{Set}$ is the {powerset functor} which maps
  any $X$ to the set $\mathcal{P}X=\{ A\mid A\subseteq X\}$ and any $f:X\to Y$ to
  $\mathcal{P}f:\mathcal{P}X\to \mathcal{P}Y;A\mapsto f(A)$.

  \emph{Non-deterministic automata} as defined in Subsection~\ref{subsection:non-deterministic_automata}  are modelled as coalgebras of the
  type  $\mathcal{P}(\Sigma \times \mathcal{I}d+1)$, where $1= \{\checked\}$ (e.g.~\cite{hasuo07:trace}). Indeed, any non-deterministic automaton
  $(Q,\Sigma,\delta,\mathfrak{F})$ is modelled by  $\alpha:Q\to
  \mathcal{P}(\Sigma \times Q +1)$ where:
  \[
  \alpha(q) = \{(a,q')\mid q'\in \delta(a,q)\} \cup \chi(q),
  \]
  where $\chi(q) = \left \{
  \begin{array}{cc}
      \{\checked\} & q\in \mathfrak{F},\\
      \varnothing & \text{ otherwise.}
  \end{array}\right.$
  In a similar manner, we can model \emph{tree automata} coalgebraically, i.e.\ as
  coalgebras of the type $Q\to \mathcal{P}(\Sigma\times Q\times   Q+1)$.
\end{exa}

\begin{exa} \label{example:fully_prob}\label{example:distr_monad}
  \emph{Fully probabilistic processes}~\cite{baier97:cav} sometimes referred to as
  \emph{fully probabilistic systems}~\cite{sokolova11}
   are modelled as
  $\mathcal{D}(\Sigma \times \mathcal{I}d)$-coalgebras~\cite{sokolova11}.  Here,
  $\mathcal{D}$ denotes the subdistribution functor assigning to any set $X$ the
  set $\{\mu:X\to [0,1]\mid \sum_x \mu(x) \leq 1\}$ of subdistributions with
  countable support and to any map $f:X\to Y$ the map
  $\mathcal{D}f:\mathcal{D}X\to \mathcal{D}Y; \mu \mapsto \mathcal{D}f(\mu)$ with % chktex 13
  \[\mathcal{D}f(\mu)(y) = \sum \{\mu(x) \mid x\in X \text{ such that }f(x)=y\}.\]
\end{exa}

\subsection{Monads}%
\label{subsection:monads}
For a general introduction to the theory of monads the reader is referred to e.g.~\cite{barrwells:ttt,maclane:cwm}.
A \emph{monad} on $\mathsf{C}$ is a triple $(T,\mu,\eta)$, where
$T:\mathsf{C}\to \mathsf{C}$ is an endofunctor and $\mu:T^2\implies T$,
$\eta:\mathcal{I}d\implies T$ are two natural transformations for which the
following diagrams commute:
\[
  \begin{tikzpicture}
    \node (t2) at (0,0) {$T^2$};
    \node (t) at (2,0) {$T$};
    \node (t3) at (0,1) {$T^3$};
    \node (t22) at (2,1) {$T^2$};

    \draw[->] (t3) -- (t22) node[pos=.5,above] {$\mu$};
    \draw[->] (t3) -- (t2) node[pos=.5,left] {$T\mu$ };
    \draw[->] (t2) -- (t) node[pos=.5,below] {$\mu$};
    \draw[->] (t22) -- (t) node[pos=.5,right] {$\mu$};

    \node (at2) at (4,0) {$T^2$};
    \node (at) at (6,0) {$T$};
    \node (at3) at (4,1) {$T$};
    \node (at22) at (6,1) {$T^2$};

    \draw[->] (at3) -- (at22) node[pos=.5,above] {$T\eta$};
    \draw[->] (at3) -- (at2) node[pos=.5,left] {$\eta_T$ };
    \draw[->] (at3) -- (at) node[pos=.3,below] {$\mathsf{id}$ };
    \draw[->] (at2) -- (at) node[pos=.5,below] {$\mu$};
    \draw[->] (at22) -- (at) node[pos=.5,right] {$\mu$};
  \end{tikzpicture}
\]
The transformation $\mu$ is called  \emph{multiplication} and the transformation
$\eta$ is called \emph{unit}.

For any monad $(T:\mathsf{C}\to \mathsf{C},\mu,\eta)$ we define the \emph{Kleisli category}
$\mathcal{K}l(T)$ for $T$ has whose class of objects is the class of objects
of $\mathsf{C}$ and for two objects $X,Y$ in $\mathcal{K}l(T)$ we put
${\mathcal{K}l(T)}(X,Y) = {\mathsf{C}}(X,TY)$
with the composition $\cdot$ in $\mathcal{K}l(T)$ defined between two morphisms
$f:X\to TY$ and $g:Y\to TZ$ by
$g\cdot f := \mu_Z \circ T(g) \circ f$. Since most of the
time we work with two categories at once, namely $\mathsf{C}$ and
$\mathcal{K}l(T)$, morphisms in $\mathsf{C}$ will be denoted using standard
arrows $\to$, whereas for morphisms in $\mathcal{K}l(T)$ we will use the symbol $\rightdcirc$.
Hence, $f:X\rightdcirc Y = X\to TY$ and the composition is given by
\[X\stackrel{f}{\rightdcirc } Y\stackrel{g}{\rightdcirc Z} =
X\stackrel{f}{\to}TY \stackrel{Tg}{\to } TTZ\stackrel{\mu_Z}{\to} TZ.\]

We define a functor
$^{\sharp}:\mathsf{C}\to\mathcal{K}l(T)$
which sends each object $X\in \mathsf{C}$ to itself and each morphism $f:X\to Y$ in
$\mathsf{C}$ to the morphism
$f^{\sharp}:X\to TY; f^{\sharp} \defeq \eta_Y\circ f$.
Maps in $\mathcal{K}l(T)$ of the form $f^\sharp$ for some $f:X\to Y\in
\mathsf{C}$ are referred to as \emph{base morphisms}.

Every monad $(T,\mu,\eta)$ on a category $\mathsf{C}$ arises from the
composition of a left and a right adjoint:
$\mathsf{C}\leftrightarrows \mathcal{K}l(T)$,
where the left adjoint is $^\sharp:\mathsf{C}\to \mathcal{K}l(T)$ and the right
adjoint $U_T:\mathcal{K}l(T)\to \mathsf{C}$ is defined as follows: for any
object $X\in \mathcal{K}l(T)$ (i.e.\ $X\in \mathsf{C}$) the object $U_T X$ is
given by $U_T X := TX$ and for any morphism $f:X\to TY$ in $\mathcal{K}l(T)$ the
morphism $U_T f:TX\to TY$ is given by $U_T f = \mu_Y\circ Tf$.

We say that a functor $F:\mathsf{C}\to\mathsf{C}$ \emph{lifts to} a functor
$\overline{F}:\mathcal{K}l(T)\to\mathcal{K}l(T)$ provided that the following
diagram commutes:
\begin{center}
  \begin{tikzpicture}
    \node (C1) at (0,0) {$\mathsf{C}$};
    \node (C2) at (2,0) {$\mathsf{C}$};
    \node (K1) at (0,1) {$\mathcal{K}l(T)$};
    \node (K2) at (2,1) {$\mathcal{K}l(T)$};

    \draw[->] (C1) -- (C2) node[pos=.5,below] {\tiny $F$};
    \draw[->] (K1) -- (K2) node[pos=.5,above] {\tiny $\overline{F}$};
    \draw[->] (C1) -- (K1) node[pos=.5,left] {\tiny $\sharp$};
    \draw[->] (C2) -- (K2) node[pos=.5,right] {\tiny $\sharp$};
  \end{tikzpicture}
\end{center}
\noindent There is a one-to-one correspondence between liftings $\overline{F}$
and \emph{distributive laws}
$\lambda: FT\implies TF$ between the functor $F$
and the monad $T$\footnote{A distributive law between
  a functor $F$ and a monad $T$ is a
  natural transformation  $FT\implies TF$
  which additionally satisfies extra conditions listed in e.g.
~\cite{jacobssilvasokolova2012:cmcs,mulry:mfps1993}.
}.
Indeed, any lifting $\overline{F}:\mathcal{K}l(T)\to \mathcal{K}l(T)$ induces
the transformation $\lambda$ whose $X$-component  $\lambda_X:FTX\to TFX$ is
$\lambda_X= \overline{F}(\mathsf{id}_{TX}:TX\to TX)$ and any distributive law
$\lambda:FT\implies TF$ gives rise to a lifting $\overline{F}:\mathcal{K}l(T)\to
\mathcal{K}l(T)$ given by:
\begin{align*}
&\overline{F}X=FX \text{ and }
\overline{F}(X\stackrel{f}{\to} TY)=FX\stackrel{Ff}{\to}
FTY\stackrel{\lambda_Y}{\to} TFY.
\end{align*}

A monad $(T,\mu,\eta)$ on a category $\mathsf{C}$ with finite products is called
\emph{strong} if there is a natural transformation $t_{X,Y}:X\times TY\to
T(X\times Y)$ called \emph{tensorial strength} satisfying the strength laws
listed in e.g.~\cite{kock1972strong}. Existence of strength guarantees that for
any object $\Sigma$ the functor $\Sigma\times \mathcal{I}d:\mathsf{C}\to
\mathsf{C}$ admits a lifting
$\overline{\Sigma}:\mathcal{K}l(T)\to\mathcal{K}l(T)$ defined as follows. For
any $X\in \mathcal{K}l(T)$ we put
$
\overline{\Sigma}X := \Sigma\times X,
$
and for any  $f:X\rightdcirc Y = X\to TY$ we define
$
\overline{\Sigma}f\defeq t_{\Sigma,Y}\circ ( id_{\Sigma}\times f).
$
Existence of the transformation ${t}$ is not a strong requirement. For instance
all monads on $\mathsf{Set}$ are strong.

\begin{exa}\label{example:powerset_monad}
The powerset endofunctor $\mathcal{P}:\mathsf{Set}\to \mathsf{Set}$, used in the
definition of labelled transition systems, non-deterministic automata and tree
automata, carries a monadic structure $(\mathcal{P},\bigcup,\{-\})$  for which
the multiplication and the unit are given by:
\[
\bigcup:\mathcal{PP}X\to \mathcal{P}X; S\mapsto \bigcup S, \qquad \{-\}:X\to
\mathcal{P}X; x\mapsto \{x\}.
\]
The Kleisli category $\mathcal{K}l(\mathcal{P})$ consists of sets as objects and
morphisms given by
the
maps $f:X\rightdcirc Y = X\to \mathcal{P}Y$ and $g:Y\rightdcirc Z=Y\to \mathcal{P}Z$ with the composition
$g\cdot f:X\rightdcirc Z = X\to \mathcal{P}Z$ given by
\[
  (g\cdot f)(x) = \bigcup_{y\in f(x)} g(y).
\]
The identity morphisms $\mathsf{id}:X\rightdcirc X = X\to \mathcal{P}X$ are given for any $x\in
X$ by $\mathsf{id}(x) =\{x\}$. The Kleisli category for $\mathcal{P}$ is
isomorphic to $\mathsf{Rel}$ --- the category of sets as objects, and relations as
morphisms. The $X$-component of the distributive law $\lambda:\Sigma \times
\mathcal{P}X\to \mathcal{P}(\Sigma\times X)$ induced by strength of
$\mathcal{P}$ is:
\[
\lambda(a,X') = \{(a,x)\mid x\in X'\}.
\]
\end{exa}

\begin{exa}
The subdistribution functor $\mathcal{D}:\mathsf{Set}\to\mathsf{Set}$ from
Example~\ref{example:fully_prob} carries a monadic structure
$(\mathcal{D},\mu,\eta)$, where $\mu_X:\mathcal{D}\mathcal{D}X\to \mathcal{D}X$
is
\begin{align*}
\mu(\psi)(x) = \sum_{\phi\in \mathcal{D}X}\psi(\phi)\cdot \phi(x)
\end{align*}
and $\eta_X:X\to \mathcal{D}X$ assigns to any $x$ the Dirac delta distribution
$\delta_x:X\to [0,1]$.
\end{exa}

\begin{exa}\label{example:monoid_monad}
For any monoid $(M,\cdot,1)$ the $\mathsf{Set}$-functor $M\times \mathcal{I}d$
carries a monadic structure $(M\times \mathcal{I}d,m,e)$, where $m_X:M\times
M\times X\to M\times X; (m,n,x)\mapsto (m\cdot n,x)$ and $e_X:X\to M\times X;
x\mapsto (1,x)$.
\end{exa}

From the perspective of this paper, the most imporant instance of the family of
monads from Example~\ref{example:monoid_monad} is the monad $(\Sigma^\ast\times
\mathcal{I}d,m,e)$, where $(\Sigma^\ast,\cdot, \varepsilon)$ is the free monoid
over $\Sigma$. The reason is that $\Sigma^\ast\times \mathcal{I}d$ is the free
monad over the functor $\Sigma\times \mathcal{I}d$ and hence,  since
$\Sigma\times \mathcal{I}d$ lifts to the Kleisli category for any
$\mathsf{Set}$-based monad $T$ (since all $\mathsf{Set}$-based monads are
strong), then so does $\Sigma^\ast\times \mathcal{I}d$ whose lifting is the free
monad over the lifting of $\Sigma\times \mathcal{I}d$~\cite{brengos2015:lmcs}.
In practice, this yields a monadic structure on $T(\Sigma^\ast\times
\mathcal{I}d)$ for any monad $T$ on the category of sets~\cite{brengos2015:lmcs}.

\begin{exa}\label{example:lts_free}
If $T=\mathcal{P}$ then the Kleisli category for $\mathcal{P}(\Sigma^\ast\times
\mathcal{I}d)$  has the composition given as follows~\cite{brengos2015:lmcs}.
For two morphisms $f:X\rightdcirc Y=X\to \mathcal{P}(\Sigma^{*}\times Y)$ and $g:Y\rightdcirc Z=Y\to
\mathcal{P}(\Sigma^{*}\times Z)$ we have
\[
g\cdot f (x) = \{(\sigma_1\sigma_2,z)\mid x\stackrel{\sigma_1}{\to}_f y
\stackrel{\sigma_2}{\to}_g z \text{ for some }y\in Y\}.
\]
The identity morphisms in this category are $\mathsf{id}:X\rightdcirc X =X\to
\mathcal{P}(\Sigma^\ast\times X)$ given by $\mathsf{id}(x) =
\{(\varepsilon,x)\}$.

In a similar manner, using the remark above, we show that
$\mathcal{D}(\Sigma^\ast\times \mathcal{I}d)$ carries a monadic structure.
\end{exa}

\subsection{Coalgebras with internal moves}%
\label{subsection:coalgebras_with_internal_moves}

Coalgebras with internal moves were  first introduced in
the context of coalgebraic trace semantics as coalgebras of the type
$TF_\varepsilon$ for a monad $T$ and an endofunctor $F$ on $\mathsf{C}$ with $F_\varepsilon$
defined by  $F_\varepsilon \defeq F+\mathcal{I}d$~\cite{hasuo06,silva2013:calco}. If we take  $F=\Sigma\times \mathcal{I}d$ then
we have
$TF_\varepsilon =T(\Sigma\times \mathcal{I}d + \mathcal{I}d)\cong
T(\Sigma_\varepsilon\times \mathcal{I}d)$, where $\Sigma_\varepsilon \defeq
\Sigma +\{\varepsilon\}$.
In~\cite{brengos2015:lmcs} we showed that given certain assumptions on $T$ and
$F$ we may embed the functor $TF_\varepsilon$ into the monad $TF^{*}$, where
$F^{*}$ is the free monad over $F$. In particular, if we apply this construction
to $T=\mathcal{P}$ and $F=\Sigma\times \mathcal{I}d$ we obtain the monad
$\mathcal{P}(\Sigma^\ast\times \mathcal{I}d)$ from Example~\ref{example:lts_free}. The construction of $TF^\ast$ is revisited in this paper in
Section~\ref{section:monads}.
The trick of modelling  the invisible steps via a monadic structure allows us
\emph{not} to specify the internal moves explicitly. Instead of considering
$TF_\varepsilon$-coalgebras we consider $T'$-coalgebras for a monad $T'$ on an
arbitrary category.

The strategy of finding a suitable monad (for modelling the behaviour taken into
consideration) will also be applied in this paper. Unfortunately, from the point
of view of the infinite behaviour of coalgebras, considering systems of the type
$TF^\ast$ is not sufficient (see Section~\ref{section:buchi_coalgebraically} for
a discussion). Hence, in Section~\ref{section:monads} we show how to obtain
a monad suitable for modelling infinite behaviour. Intuitively, the new monad
extends $TF^\ast$ by adding an ingredient associated with the terminal $F$-coalgebra
$\zeta:F^\omega\to FF^\omega$. The construction presented in Section~\ref{section:monads}
yields the monad $TF^\infty=T(F^\ast \oplus F^\omega)$ suitable to capture both: finite and infinite behaviour
of systems. Below we give two examples of such monad.

\begin{exa}\label{example:kleisli_powerset}\label{example:lts_omega_monad}
Although the monad $\mathcal{P}(\Sigma^\ast\times \mathcal{I}d)$ from Example~\ref{example:lts_free} proves to be sufficient to model finite behaviours of
non-deterministic automata (see~\cite{brengos2015:lmcs,brengos2015:jlamp}), it  will not be suitable to model
their infinite behaviour (see Section~\ref{section:buchi_coalgebraically} for
details). Hence, we extend $\mathcal{P}(\Sigma^\ast\times \mathcal{I}d)$
and consider the following. Let $\Sigma^\omega$ be the set of all infinite
sequences of elements from $\Sigma$. As it will be shown in sections to come,
the functor $\mathcal{P}(\Sigma^\ast \times \mathcal{I}d + \Sigma^\omega)$
carries a monadic structure whose Kleisli composition is as follows. For
$f:X\to \mathcal{P}(\Sigma^\ast \times Y + \Sigma^\omega)$ and $g:Y\to
\mathcal{P}(\Sigma^\ast \times Z+\Sigma^\omega)$ the map $g\cdot f:X\rightdcirc Z = X\to
\mathcal{P}(\Sigma^\ast \times Z+\Sigma^\omega)$ satisfies:
%\vspace{-0.2cm}
\begin{align*}
&x\stackrel{\sigma}{\to}_{g\cdot f} z \iff \exists y \text{ such that }
x\stackrel{\sigma_1}{\to}_{f} y \text{ and }y\stackrel{\sigma_2}{\to}_{g} z,
\text{ where } \sigma=\sigma_1\sigma_2\in \Sigma^\ast,\\
&x\downarrow_{g\cdot f} v \iff x\downarrow_f v \text{ or } x\stackrel{\sigma
}{\to}_f y \text{ with } y\downarrow_g v' \text{ and } v=\sigma v'\in \Sigma^\omega.
\end{align*}
In the above we write $x\stackrel{\sigma}{\to}_f y$ whenever $(\sigma,y)\in
f(x)$ and $x\downarrow_f v$ if $v\in f(x)$ for $\sigma\in \Sigma^\ast$, $v\in
\Sigma^\omega$.  The identity morphisms in this category are the same as in the
Kleisli category for the monad $\mathcal{P}(\Sigma^\ast\times \mathcal{I}d)$.
The monadic structure of $\mathcal{P}(\Sigma^\ast \times \mathcal{I}d +
\Sigma^\omega)$ arises as a consequence of a general construction of monads
modelling (in)finite behaviour described in detail in Section~\ref{section:monads}.
\end{exa}

\begin{exa}%
  \label{example:in_finite_tree_monad}
  If we move from non-deterministic automata  towards
  tree automata we have to find a suitable monadic
  setting to  talk about their (in)finite behaviour.
  It turns out that a good candidate for this monad can be built
  from the ingredients already presented in this paper. Indeed, if we take the
  powerset monad and the monad  $T_\Sigma$  from Subsection~\ref{subsection:trees},
  then their composition $\mathcal{P}T_\Sigma$ carries a monadic
  structure\footnote{The proof of this claim can be found in Section~\ref{section:monads}. See Example~\ref{example:tree_functor} for details.}.
  The formula for the composition in the Kleisli category for the monad
  $\mathcal{P}T_\Sigma$ is given for $f:X\to \mathcal{P}T_\Sigma Y$ and $g:Y\to
  \mathcal{P}T_\Sigma Z$  by  $g\cdot f:X\rightdcirc Z = X\to \mathcal{P}T_\Sigma Z$ with $g\cdot
  f(x)$ being a set of trees obtained from trees in $f(x)\subseteq T_\Sigma Y$ by
  replacing any occurence of the leaf $y\in Y$ with a tree from $g(y)\subseteq
  T_\Sigma Z$. As will be witnessed in Section~\ref{section:monads}, this monad
  and the monad $\mathcal{P}(\Sigma^\ast\times \mathcal{I}d+\Sigma^\omega)$
  arise from the same categorical construction.
\end{exa}
\noindent The list of examples of monads used in the paper will be extended in the
upcoming sections.

\subsection{Categorical order enrichment}
Our main ingredients for defining
(in)finite behaviours of automata will turn out to be two fixpoint operators:
$(-)^\ast$ and $(-)^\omega$. In order to establish them on a categorical level
we require the category under consideration to be suitably
order enriched. A category is said to be \emph{order enriched}, or simply
\emph{ordered}, if each hom-set is a poset with the order preserved by the
composition. It is \emph{$\vee$-ordered} if all hom-posets admit arbitrary
finite (possibly empty) suprema. Note that, given such suprema exist,
\begin{wrapfigure}[2]{r}{0.35\textwidth}
\vspace{-0.3cm}
\resizebox{.35\textwidth}{!}{
\begin{tikzpicture}
\usetikzlibrary{patterns};
\pgfmathsetmacro{\przesuniecieX}{0};

\pgfmathsetmacro{\x}{-3};
\pgfmathsetmacro{\y}{0};
\pgfmathsetmacro{\shadow}{0.03};

\draw[-] (\x+1.4,\y+.35) -- (\x+1.7,\y+0.35);
\draw[fill = white] (\x+1.7,\y+0.1) rectangle (\x+2.2,\y+0.6);
\node (v) at (\x+1.95,\y+0.35) {$f$};
\draw[-] (\x+2.2,\y+0.35) -- (\x+2.5,\y+0.35);

\draw[fill = white] (\x+2.72,\y+0.3) circle (0.15);
\node (v) at (\x+2.72,\y+0.3) {$\vee$};

\pgfmathsetmacro{\x}{-1.45};
\pgfmathsetmacro{\y}{0};
\pgfmathsetmacro{\shadow}{0.03};

\draw[-] (\x+1.4,\y+.35) -- (\x+1.7,\y+0.35);
\draw[fill = white] (\x+1.7,\y+0.1) rectangle (\x+2.2,\y+0.6);
\node (v) at (\x+1.95,\y+0.35) {$g$};
\draw[-] (\x+2.2,\y+0.35) -- (\x+2.5,\y+0.35);

\node (v) at (\x+2.7,\y+0.3) {\small $=$};
\draw[-] (1.4,0.35) -- (1.7,0.35);
\draw[fill = white] (1.7,0.1) rectangle (2.7,0.6);
\node (v) at (2.2,0.35) {$f\vee g$};
\draw[-] (2.7,0.35) -- (3,0.35);
\end{tikzpicture}
}
\end{wrapfigure}
the composition in $\mathsf{C}$ does not have to distribute over them in
general. We call a category \emph{left distributive} (or \emph{LD} in short)
if it is $\vee$-ordered and
$g\cdot (\bigvee_{i\in I} f_i) = \bigvee_{i\in I} g\cdot f_i$ for any finite set $I$.
We define \emph{right distributivity}
analogously. In this paper we  come across many left distributive categories that do not
necessarily satisfy  right distributivity. Still, however, all examples of Kleisli categories
taken into consideration satisfy
its weaker form. To be more precise, we say that the Kleisli category
$\mathcal{K}l(T)$ for a monad $T$ on $\mathsf{C}$ is
\emph{right distributive w.r.t.\ base morphisms}
provided that $(\bigvee_{i\in I} f_i)\cdot j^\sharp = \bigvee_{i\in I} f_i\cdot j^\sharp$
for any  $f_i:Y\rightdcirc Z=Y\to TZ\in \mathcal{K}l(T)(Y,Z)$, any $j:X\to Y\in \mathsf{C}(X,Y)$ and
any finite set $I$.
We say that an order enriched category is $\cpo$-enriched
if any countable ascending chain of morphisms $f_1\leq f_2\leq \ldots$ with common
domain and codomain admits a supremum which is preserved by the morphism
composition. Finally, in an ordered category with finite coproducts we say that
\emph{cotupling preserves order} if $[f_1,f_2]\leq [g_1,g_2] \iff f_1\leq g_1
\text{ and }f_2\leq g_2$ for any $f_i,g_i$ with suitable domains and codomains.

\begin{rem}\label{remark:right_distributivity}
  Right distributivity w.r.t.\ the base morphisms and cotupling order preservation
  are properties we often get as a consequence of other general assumptions.
  Indeed, any  $\mathsf{Set}$-based monad $T$ whose order enrichment of the Kleisli category
  is given by $f\leq g \iff \forall x. f(x)\leq g(x)$, for $f,g:X\to TY$
  where $TY$ is a poset for any $Y$\footnote{In this case, we say that the order enrichement of $\mathcal{K}l(T)$
  is \emph{pointwise induced}.}, satisfies these conditions.
  Note that, in this case, the Kleisli composition over any suprema or infima that exist is right distributive
  w.r.t.\ morphisms of the form $j^\sharp = \eta_Y \circ j:X\rightdcirc Y = X\to TY$ for any set map
  $j:X\to Y$. A similar argument applies to
  cotupling order preservation.
\end{rem}

\begin{exa}%
\label{example:lts_ordered_categories}%
\label{example:conditions}
The next section of this paper focuses on three categories, namely:
$\mathcal{K}l(\mathcal{P}(\Sigma^\ast\times \mathcal{I}d))$,
$\mathcal{K}l(\mathcal{P}(\Sigma^\ast\times \mathcal{I}d+\Sigma^\omega))$ and
$\mathcal{K}l(\mathcal{P}T_\Sigma )$.
These categories are order-enriched with the hom-set ordering given by $f\leq g
\iff f(x) \subseteq g(x) \text{ for any } x$.
The base morphisms of the first two examples  are of the form \[X \to
\mathcal{P}(\{\varepsilon\}\times Y);x\mapsto \{(\varepsilon,j(x))\}\] for a set
map $j:X\to Y$. The base morphisms of the third example are given by
$X\to \mathcal{P}T_\Sigma Y; x\mapsto \{j(x)\}$.
We leave it as an exercise to the reader to verify
that all these examples satisfy the following conditions: the order enrichment is pointwise induced;
 they are $\cpo$-enriched; their hom-sets are complete lattices; they
 are left distributive\footnote{We refer the reader to~\cite{brengos2015:lmcs} for a
 proof that $\mathcal{K}l(\mathcal{P}(\Sigma^\ast\times \mathcal{I}d))$
 satisfies these conditions.}.
 These conditions
  play a central role in defining (in)finite behaviours on a coalgebraic level.
We will elaborate more on them in Section~\ref{section:automata}.
\end{exa}
\subsection{Lawvere theories}%
\label{subsection:lawvere_theories}

The primary interest of the theory of automata and formal languages focuses on
automata over a \emph{finite} state space. Hence, since we are interested in
systems with internal moves (i.e.\ coalgebras $X\to TX$ for a monad $T$), without
loss of generality we may focus our attention on coalgebras of the form $n\to
Tn$, where $n = \{1,\ldots,n\}$ with $n=0,1,\ldots$.
These morphisms are endomorphisms in a full
subcategory of the Kleisli category for $T$, we will later refer to as
{(Lawvere) theory}. Restricting the scope to this category
instead of considering the whole Kleisli category for a given monad
plays an important role in Kleene theorems characterizing
regular and $\omega$-regular behaviour
(see e.g.~\cite{Hopcroft:2000:IAT:557657,pin:automata}).

 Formally, a \emph{Lawvere theory}, or simply
 \emph{theory}, is a category whose
objects are natural numbers $n\geq 0$ such that each $n$ is an $n$-fold
coproduct of $1$.  The definition used here is dual to the classical notion~\cite{lawvere:1963} and can be found in e.g.~\cite{esik2011,esik2013,esikhajgato2009:ai}. The reason why
we use our version of the definition is the following:
we want the connection between Lawvere theories and Kleisli categories for
$\mathsf{Set}$-based monads to be as direct as possible. Indeed, in our case,
any monad $T$ on $\mathsf{Set}$ induces a theory $\mathbb{T}$ associated with
it by restricting the Kleisli category $\mathcal{K}l(T)$ to objects  $n$ for
any $n\geq 0$.  Conversely, for any theory $\mathbb{T}$ there is a
$\mathsf{Set}$ based monad the theory is associated with  (see e.g.~\cite{hyland:power:2007} for details). This remark also motivates us to use the notation introduced
before and denote morphisms from a theory by $\rightdcirc$.

For any element $i \in n$ let $i_n:1\rightdcirc n$
denote the $i$-th coproduct injection
 \begin{wrapfigure}[4]{r}{0.15\textwidth}
\vspace{-0.4cm}
\resizebox{.137\textwidth}{!}{
\begin{tikzpicture}
\draw[-] (-0.3,0.25) -- (-0,0.25)node[pos=0.5,below]{\tiny $n_k$};
\draw[-] (-0.3,0.95) -- (-0,0.95)node[pos=0.5,above]{\tiny $n_1$};
\draw[fill = white] (0,0) rectangle (0.5,0.5);
\draw[fill = white] (0,0.7) rectangle (0.5,1.2);
\node (dots) at (0.1,0.57) {$\ldots$};
\draw[rounded corners=4pt,-]
(0.5,0.85)--(0.7,0.85)--(0.7,0.6)--(0.91,0.6)--(1,0.6);
\draw[rounded corners=4pt,-]
(0.5,0.25)--(0.7,0.25)--(0.7,0.6)--(0.91,0.6)--(1,0.6);

\draw[rounded corners=4pt,-] (0.91,0.6)--(1,0.6) node[pos=0.5,above]{\tiny $n$};

\node (v) at (0.23,0.2) {$f_k$};
\node (v) at (0.23,0.9) {$f_1$};
\end{tikzpicture}
}
\end{wrapfigure} and $[ f_1, \ldots ,f_k ]:n_1+ \cdots + n_k\rightdcirc n$ the cotuple of
$\{f_l:n_l\rightdcirc n\}_{l}$  depicted in the diagram on the right. The
coprojection morphism $n_i\rightdcirc n_1+\cdots +n_k$ into the $i$-th component of the coproduct
will be denoted by  $\mathsf{in}^{n_i}_{n_1+\cdots +n_k}$. Any morphism $k\rightdcirc n$
of the form $[(i^1)_n,\ldots,(i^k)_n]:k\rightdcirc n$ for $i^j\in n$ is called \emph{base
morphism} or \emph{base map}. If $\mathbb{T}$ is associated with a
monad $T$ then the base morphisms in $\mathbb{T}$ are exactly
given by $f^\sharp:m\rightdcirc n =  m\to T n$ for $\mathsf{Set}$-maps $f: m\to n$.
Finally, let $!:n\rightdcirc 1$ be defined by $!\defeq
[1_1,1_1,\ldots,1_1]$.  We say that a theory $\mathbb{T}'$ is a \emph{subtheory}
of $\mathbb{T}$ if there is a faithful functor $\mathbb{T}'\to \mathbb{T}$ which
maps any object $n$ onto itself.

\begin{exa}\label{example:lts_ordered_theories}
  By $\mathsf{LTS}$,  $\mathsf{LTS}^\omega$ and $\mathsf{TTS}^\omega$
  we denote the theories associated
  with the monads $\mathcal{P}(\Sigma^\ast \times \mathcal{I}d)$,
  $\mathcal{P}(\Sigma^\ast\times \mathcal{I}d+\Sigma^\omega)$ and
  $\mathcal{P}T_\Sigma$ respectively.
\end{exa}

\section{Non-deterministic automata, coalgebraically}\label{section:buchi_coalgebraically}
The purpose of this section is to give motivations for the abstract theory
presented in the remainder of the paper. In the first part of this section
we focus on finite
non-deterministic (B\"uchi) automata and their (in)finite behaviour from the
perspective of the categories $\mathcal{K}l(\mathcal{P}(\Sigma^\ast\times
\mathcal{I}d))$ and  $\mathcal{K}l(\mathcal{P}(\Sigma^\ast\times \mathcal{I}d+\Sigma^\omega))$.
Afterwards, we deal with tree automata and their behaviour. Finally, we give
a categorical perspective on Kleene theorems for automata taken into consideration.

\subsection{Non-deterministic automata}%
\label{subsection:non-deterministic-automata}
Without any loss of generality we may only consider automata over the state
space $n=\{1,\ldots,n\}$ for some natural number $n$.  As mentioned in
Example~\ref{example:lts_non_deterministic_automata} any non-deterministic automaton
 $(n,\Sigma, \delta, \mathfrak{F})$ may be
modelled as a $\mathcal{P}(\Sigma\times \mathcal{I}d+1)-$coalgebra
$n\to \mathcal{P}(\Sigma \times n + 1)$~\cite{rutten:universal}.
However, as it has been already noted in~\cite{urabe_et_al:LIPIcs:2016:6186},
from the point of view of infinite behaviour with BAC it is more useful to
extract the information about the final states of the automaton and not to
encode it into the transition map as above. Instead, given an automaton $(n,\Sigma,\delta,\mathfrak{F})$
we encode it as a pair $(\alpha,\mathfrak{F})$ where $\alpha: n\to
\mathcal{P}(\Sigma \times n)$  is defined by
$\alpha(i)=\{(a,j)\mid j\in \delta(a,i)\}$ and consider the map:
\begin{align*}
\mathfrak{f_F}:n \to \mathcal{P}(\{\varepsilon\} \times n );
i \mapsto
\left \{ \begin{array}{cc}  \{(\varepsilon,i)\} & \text{ if } i\in
\mathfrak{F},\\ \varnothing & \text{
otherwise.}\end{array}\right.%\tag{FIN}\label{form:final_states}
\end{align*}
Note that by extending the codomain of $\alpha$ and $\mathfrak{f_F}$ both maps
can be viewed as endomorphisms in $\mathsf{LTS}$ and $\mathsf{LTS}^\omega$.
The purpose of $\mathfrak{f_F}$ is to encode the set of accepting states with an
endomorphism in the same Kleisli category  in which the transition $\alpha$ is
an endomorphism. Now, we have all the necessary ingredients to revisit finite
and infinite behaviour (with BAC) of non-deterministic automata from the
perspective of the theories $\mathsf{LTS}$ and $\mathsf{LTS}^\omega$.

\subsubsection{Finite behaviour}
Consider $\alpha^\ast:n\rightdcirc n$ to be an endomorphism in $\mathsf{LTS}$ (or
$\mathsf{LTS}^\omega$) given by $\alpha^\ast = \mu x.(\mathsf{id}\vee x\cdot
\alpha) = \bigvee_{n\in \omega } \alpha^n$, where the order is as in
 Example~\ref{example:lts_ordered_theories}. We have~\cite{brengos2015:lmcs}:
\[
\alpha^\ast (i) = \{(\sigma,j) \mid i \stackrel{\sigma}{\implies } j \},
\]
where $\stackrel{\sigma}{\implies} \defeq (\stackrel{\varepsilon}{\to} )^\ast
\circ\stackrel{a_1}{\to} \circ (\stackrel{\varepsilon}{\to} )^\ast \circ \ldots
(\stackrel{\varepsilon}{\to} )^\ast \circ \stackrel{a_n}{\to}
(\stackrel{\varepsilon}{\to} )^\ast$ for $\sigma = a_1\ldots a_n$, $a_i\in
\Sigma$ and  $\stackrel{\varepsilon}{\implies }\defeq
(\stackrel{\varepsilon}{\to})^\ast$.  Let us observe that the theory
morphism $!:n\rightdcirc 1$
is explicitly given in the case of theories
$\mathsf{LTS}$ and $\mathsf{LTS}^\omega$ by $!(i) =
\{(\varepsilon,1)\}$ for any $i\in n$. Finally, consider the morphism
$!\cdot \mathfrak{f_F} \cdot \alpha^\ast:n\rightdcirc 1$ in $\mathsf{LTS}$ (or
$\mathsf{LTS}^\omega$) which is:
\[!\cdot \mathfrak{f_F} \cdot \alpha^\ast (i) = \{ (\sigma,1) \mid \sigma \in
\Sigma^\ast \text{ such that } i\stackrel{\sigma}{\implies} j \text{ and }j\in
\mathfrak{F}\}.\]
Since $\mathcal{P}(\Sigma^\ast \times 1 )\cong \mathcal{P}(\Sigma^\ast)$, the
set $!\cdot \mathfrak{f_F} \cdot \alpha^\ast (i)$ represents the set of all
finite words accepted by the state $i$ in the automaton
$(n,\Sigma, \delta, \mathfrak{F})$.

\subsubsection{Infinite behaviour}
Note that the hom-posets of theories $\mathsf{LTS}$ and $\mathsf{LTS}^\omega$
are complete lattices and, hence (by the Tarski-Knaster theorem), come
equipped with an operator which assigns to any endomorphism $\beta:n\rightdcirc n$ the
morphism $\beta^\omega:n\rightdcirc 0$ defined as the greatest fixpoint of the
assignment  $x\mapsto x\cdot \beta$. Now, if $\alpha$ is given as in the previous
subsections and considered as an endomorphism in the theory $\mathsf{LTS}$
then the map $\alpha^\omega:n \rightdcirc 0 =n \to
\mathcal{P}(\Sigma^\ast \times \varnothing)$ in
$\mathsf{LTS}$ satifies
$\alpha^\omega (i) = \varnothing$. However, if we consider $\alpha$ to be an
endomorphism in $\mathsf{LTS}^\omega$ and compute
$\alpha^\omega: n\rightdcirc 0=n\to \mathcal{P}(\Sigma^\ast\times \varnothing + \Sigma^\omega) = n\to \mathcal{P}(\Sigma^\omega)$ in
$\mathsf{LTS}^\omega$ the result will be different. Indeed, we have the
following.
%\vspace{-0.3cm}
\begin{prop}\label{theorem:omega_iteration} Let $\beta: n\to
\mathcal{P}( \Sigma^\ast \times n )$ and, since $\mathcal{P}(\Sigma^\ast \times
n) \subseteq \mathcal{P}(\Sigma^\ast \times n+\Sigma^\omega)$, it can be considered as an endomorphism in
$\mathsf{LTS}^\omega$.
In this case, the explicit formula for the greatest fixpoint
$\beta^\omega:n\rightdcirc 0= n\to \mathcal{P}(\Sigma^\omega)$ of the
assignment $x\mapsto x\cdot \beta$ calculated in  $\mathsf{LTS}^\omega$ is given
by:
{\small
\begin{align}
\beta^\omega(i) =\bigcup \{ |\sigma_1, \sigma_2,\ldots | \subseteq  \Sigma^\omega\mid i
\stackrel{\sigma_1}{\to}_\beta i_1\stackrel{\sigma_2}{\to}_\beta i_2  \ldots  \text{ for
some } i_k \in n \text{ and }\sigma_k\in \Sigma^\ast
\}, \label{formula:omega_sat}
\end{align}
}
where $|-|: (\Sigma^\ast)^\omega \to \mathcal{P}(\Sigma^\omega)$ assigns to any sequence $\sigma_1,\sigma_2,\ldots$
of words over $\Sigma$ the set
\[
|\sigma_1,\sigma_2,\ldots | \defeq \left \{
\begin{array}{cc}
\{\sigma_1\sigma_2\ldots \}\times \Sigma^\omega & \text{ if } \sigma_1\sigma_2\ldots \text{ is a finite word,}\\
\{\sigma_1\sigma_2\ldots\} & \text{ otherwise.}
\end{array}
\right.
\]
\end{prop}
\begin{proof}

At first let us note that we may assume $\beta = \beta^\ast\cdot \beta$. This is a consequence
of the fact that $\beta^\omega = (\beta^\ast\cdot \beta)^\omega$\footnote{This
  identity is proven in Lemma~\ref{theorem:least_fixpoint}
  in the general setting of ordered theories equipped with $(-)^\ast=
  \mu x.(\mathsf{id} \vee x\cdot (-))$ and $(-)^\omega = \nu x. x\cdot (-)$
  satisfying additional conditions that, in particular, hold for
  $\mathsf{LTS}^\omega$. Hence, we refer the reader to
  Section~\ref{section:automata} for the general proof of the statement.}.
Restated, this condition means that for any $i,j,k\in n$ we have:
\begin{align}
i\stackrel{\sigma_1}{\to}_\beta j \stackrel{\sigma_2}{\to}_\beta k\implies i\stackrel{\sigma_1\sigma_2}{\to}_\beta k.
\label{formula:transitive}
\end{align}
Let $\beta^o:n\rightdcirc 0=n\to \mathcal{P}(\Sigma^\omega)$ be a map whose value $\beta^o(i)$
is given in terms of the right hand side of the equality~\ref{formula:omega_sat}.
Observe that this map satisfies $\beta^o = \beta^o\cdot \beta$
in $\mathsf{LTS}^\omega$. This follows directly from the definition of $\beta^o$ and the
formula for the composition in
$\mathcal{K}l(\mathcal{P}(\Sigma^\ast \times \mathcal{I}d+\Sigma^\omega))$. Thus,
$\beta^o \leq \beta^\omega$. Now, by contradiction, if $\beta^o < \beta^\omega$ then there is
$i$ and $v\in \Sigma^\omega$ such that $v\in \beta^\omega(i)$ and $v\notin \beta^o(i)$.
Hence, in particular, this means that there is an infinite sequence of transitions $i\stackrel{\sigma_1}{\to} i_1\stackrel{\sigma_2}{\to} i_2\ldots$
in $\beta$ which starts at $i$. If there was no such sequence, this would mean that $\beta^o(i) = \varnothing=\beta^\omega(i)$ which cannot hold.
Since $\beta^\omega = \beta^\omega\cdot \beta$,
there is a state $i_1$ and  $\sigma_1\in \Sigma^\ast$, $v_1\in \Sigma^\omega$
such that $i\stackrel{\sigma_1}{\to}_\beta i_1$ and $v_1\in \beta^\omega(i_1)$ with $v=\sigma_1v_1$.
Note that we may assume $\sigma_1\neq \varepsilon$ as there has to be a prefix $\sigma_1\neq \varepsilon$ of
$v=\sigma_1v_1$
with $i\stackrel{\sigma_1}{\to} i_1$ for some state $i_1$ with $v_1\in \beta^\omega(i_1)$.
If it was otherwise,
then by~\ref{formula:transitive} we would have an infinite
sequence $i\stackrel{\varepsilon}{\to}i_1\stackrel{\varepsilon}{\to }i_2\stackrel{\varepsilon}{\to}\ldots$ yielding
$\beta^\omega(i)= \Sigma^\omega=\beta^o(i)$ which contradicts our assumptions.
Hence, if $i\stackrel{\sigma_1}{\to} i_1$ for $\sigma_1\neq \varepsilon$ and $v_1\in \beta^\omega(i_1)$ then
 we also have $v_1\notin \beta^o(i_1)$. By inductively repeating this argument
we get a sequence $i\stackrel{\sigma_1}{\to} i_1 \stackrel{\sigma_2}{\to} i_2 \ldots $  in $\beta$ such that $\sigma_k\neq \varepsilon$ and $v= \sigma_1\sigma_2\ldots$. Thus, by the definition of $\beta^o$ we also get
$v\in \beta^o(i)$ which is a contradiction.
\end{proof}

\subsubsection{B\"uchi acceptance condition} Before we spell out the recipe of how to extract
$\omega$-language of any state in the  automaton $(\alpha,\mathfrak{F})$ in terms of $(-)^\ast$, $(-)^\omega$
and the composition in $\mathsf{LTS}^\omega$, we
need one last ingredient. Let us define $\alpha^+ \defeq \alpha^\ast \cdot
\alpha$ and note
\[
\alpha^+(i) = \{(\sigma,j)\mid i\stackrel{a_1}{\to}i_1\ldots
\stackrel{a_k}{\to}i_k \text{ in }\alpha \text{ and } \sigma=a_1\ldots a_k
\text{ for } k\geq 1\}.
\]
Hence, $\mathfrak{f_F}\cdot \alpha^+$ viewed as an endomorphism in $\mathsf{LTS}^\omega$ is given by
$\mathfrak{f_F}\cdot \alpha^+:n\rightdcirc n= n \to \mathcal{P}(\Sigma^*\times n+\Sigma^\omega)$
where:
\[\mathfrak{f_F}\cdot \alpha^{+}(i) = \{(\sigma,j)\mid
i\stackrel{\sigma}{\to} j \text{ in } \alpha^+ \text{ and }j\in \mathfrak{F}
\}.\]

\noindent Finally,  consider the following map in $\mathsf{LTS}^\omega$:
 \[(\mathfrak{f_F}\cdot \alpha^+)^\omega: n\rightdcirc 0=n\to \mathcal{P}(\Sigma^\omega).\]
By Proposition~\ref{theorem:omega_iteration}, the map
$(\mathfrak{f_F}\cdot \alpha^+)^\omega$
satisfies:
\[
(\mathfrak{f_F}\cdot \alpha^+)^\omega(i) = \text{ the $\omega$-language of $i$
in the B\"uchi automaton represented by $(\alpha,\mathfrak{F})$}.
\]
The above statement suggests a general approach towards modelling
($\omega$-)behaviours of abstract (coalgebraic) automata which we will develop in
the sections to come.
\begin{figure}
\begin{center}
%\vspace{-0.6cm}
\resizebox{0.9\textwidth}{!}{
\begin{tikzpicture}[shorten >=1pt,node distance=2cm,on grid]
  \node[state]   (q_0)                {$s_0$};
  \node[state]           (q_1) [right=of q_0] {$s_1$};
  \node[state,accepting] (q_2) [right=of q_1] {$s_2$};
  \path[->] (q_0) edge                node [above] {0} (q_1)
                  edge [bend right]   node [below] {$0+1$} (q_2)
            (q_1) edge                node [above] {1} (q_2);
  \path[->] (q_2) edge [loop above]   node         {$1$} ();
\end{tikzpicture}

\hspace{0.5cm}
\begin{tikzpicture}[shorten >=1pt,node distance=2cm,on grid]
  \node[state]   (q_0)                {$s_0$};
  \node[state]           (q_1) [right=of q_0] {$s_1$};
  \node[state] (q_2) [right=of q_1] {$s_2$};
  \path[->] (q_0) edge                node [above] {0} (q_1)
                  edge [bend right]   node [below] {$(0+1)1^\ast$} (q_2)
            (q_1) edge                node [above] {$1 1^\ast$} (q_2);
  \path[->] (q_2) edge [loop above]   node         {$11^\ast$} ();
\end{tikzpicture}
\hspace{0.5cm}
\begin{tikzpicture}[shorten >=1pt,node distance=2cm,on grid]
  \node[state]   (q_0)                {$s_0$};
  \node[state]           (q_1) [right=of q_0] {$s_1$};
  \node[state] (q_2) [right=of q_1] {$s_2$};
  \path[->] (q_0)

                  edge [bend right]   node [below] {$(0+1)1^\ast$} (q_2)
            (q_1) edge                node [above] {$1 1^\ast$} (q_2);
  \path[->] (q_2) edge [loop above]   node         {$1 1^\ast$} ();
\end{tikzpicture}
}
\end{center}
\caption{An automaton $(\alpha,\mathfrak{F})$ and the maps $\alpha^+$ and
$\mathfrak{f_F}\cdot \alpha^+$.}\label{na:example}\label{fig:automata}
\end{figure}

\subsection{Tree automata}
Let us now focus our attention on tree automata and their behaviour. Just like
in the previous subsection we may consider automata over the state space $n$.
Moreover, as before, we also encode any tree automaton $(n,\Sigma,\delta,\mathfrak{F})$
as a pair $(\alpha: n\to \mathcal{P}(\Sigma\times n\times n), \mathfrak{F})$.
Since $\mathcal{P}(\Sigma\times n \times n) \subseteq \mathcal{P}T_\Sigma n$,
the transition map $\alpha$ can be viewed as $\alpha:n\to \mathcal{P}T_\Sigma n$ (i.e.\ as
an endomorphism in the Kleisli category for the monad $\mathcal{P}T_\Sigma$ or, equivalently,
as an endomorphism in the theory $\mathsf{TTS}^\omega$). The hom-sets of the
 Kleisli category for $\mathcal{P}T_\Sigma$ and its full subcategory $\mathsf{TTS}^\omega$
admit ordering in which we can define $\beta^\ast$, $\beta^+$
and $\beta^\omega$ for any $\beta:n\rightdcirc n$ as in the previous subsection.
Not surpisingly, if we now compute
$!\cdot \mathfrak{f_F} \cdot \alpha^\ast$
and $(\mathfrak{f_F}\cdot \alpha^+)^\omega$ in $\mathsf{TTS}^\omega$
we exactly get the following\footnote{The proof of Proposition~\ref{proposition:tree_languages} is
intensionally omitted as it goes along the lines of the series of statements made in Subsection~\ref{subsection:non-deterministic-automata} for non-deterministic automata.  }.
\begin{prop}\label{proposition:tree_languages}
For any $i\in n$ we have:
\begin{align*}
    !\cdot\mathfrak{f_F}\cdot \alpha^\ast (i) & =  \text{ the set of finite trees recognized by }
     i \text{ in } (n,\Sigma,\delta,\mathfrak{F}), \\
    (\mathfrak{f_F}\cdot \alpha^+)^\omega (i) & = \text{ the set of inifinite trees recognized by }
    i  \text{ in } (n,\Sigma,\delta,\mathfrak{F}).
\end{align*}
\end{prop}

\subsection{Kleene theorems, categorically} The purpose of this subsection is
to restate classical Kleene theorems from Subsection~\ref{subsection:non_determinist_automata}
on the categorical level for $\mathsf{LTS}^\omega$ and $\mathsf{TTS}^\omega$. Before we do this let
us elaborate more on why we choose our setting to be systems whose type is a monad.

\begin{rem}
As the examples of non-deterministic (tree) automata studied in the previous subsection
do not admit silent moves, the reader may get an impression that the need for categorical
modelling of infinite behaviour for systems with silent steps is \emph{not}
sufficiently justified. To add to this, although $\varepsilon$-moves are a standard feature of automata
whenever it comes to their finitary languages, invisible moves
rarely occur in practice in the classical literature on the \emph{infinite}
behaviour (with BAC) (see e.g.~\cite{pin:automata}).
However, as already mentioned in the introduction, incorporation of silent
moves should be viewed as a by-product of our paper's framework, not its main
purpose. The main aim is to build a simple bridge between syntax and
semantics of regular and $\omega$-regular behaviours in the form of generic Kleene theorems. Once we
embed our systems into systems whose type is a monad $T$, the syntax
arises from the algebraicity of $T$ and the semantics is provided by automata whose
transition maps are certain $T$-coalgebras. This also allows us to
abstract away from several ``unnecessary'' details and focus on
core properties.
\end{rem}

As witnessed in Subsection~\ref{subsection:non-deterministic_automata},
Kleene theorems for tree automata were slightly more involved than their
classical counterparts for non-deterministic automata. The reason for this is
simple: non-deterministic automata accept sequential data types.
Whenever we deal with non-sequential data, e.g.\ trees, the set of ($\omega$-)regular languages
is expected to be closed under a more complex type of composition, i.e.\ the
composition of regular languages with multiple variables~\cite{Gradel:2002:ALI:938135,pin:automata}.
Hence, if we aim at  categorical statements generalizing theory
from Subsection~\ref{subsection:non_determinist_automata} then we should expect
a slightly more involved
formulation to be our point of reference. Hence, we start with presenting a categorical perspective
of Kleene theorems for tree automata first.

\subsubsection{Tree automata}  The Kleene theorems for regular and
$\omega$-regular input from Subsection~\ref{subsection:kleene_tree}
are equivalent to the following proposition.
\begin{prop}\label{prop:kleene_tree}
    Let $\mathfrak{Reg}$ and $\omega\mathfrak{Reg}$ be defined as in Subsection~\ref{subsection:kleene_tree}.
    Let $\mathfrak{Rat}$ be the smallest subtheory of $\mathsf{TTS}^\omega$ such that:
    \begin{enumerate}[(a)]
        \item it contains all maps of the form
        $n\to \mathcal{P}(\Sigma_\varepsilon \times n \times n)\hookrightarrow \mathcal{P}T_\Sigma n$,
        \item is closed under finite suprema,
        \item its endomorphisms are closed under $(-)^\ast$.
    \end{enumerate}
    Then the hom-set $\mathfrak{Rat}(1,1)$ of the theory $\mathfrak{Rat}$ equals to $\mathfrak{Reg}$.
    Moreover, the set
    $\omega\mathfrak{Reg}$ of $\omega$-regular
    languages for tree automata satisfies $\{ r^\omega \cdot  s\mid r\in \mathfrak{Rat}(n,n), s\in \mathfrak{Rat}(1,n)  \}=\omega\mathfrak{Reg}$.
\end{prop}

\subsubsection{Non-deterministic automata}
The formulation of Proposition~\ref{prop:kleene_tree} allows us to instantiate it for non-deterministic automata.
A simple verification proves that the following  holds.

\begin{prop}\label{prop:kleene_lts}
Let $\mathfrak{Rat}$ be the smallest subtheory of $\mathsf{LTS}^\omega$ such that:
\begin{enumerate}[(a)]
\item it contains all maps of the form $n\to \mathcal{P}(\Sigma_\varepsilon\times n)\hookrightarrow \mathcal{P}(\Sigma^\ast \times n +\Sigma^\omega)$,
\item it is closed under finite suprema,
\item its endomorphisms are closed under $(-)^\ast$.
\end{enumerate}
Then the hom-set $\mathfrak{Rat}(1,1)$ of the theory $\mathfrak{Rat}$ equals to:
\begin{align*}
\{r:1\to \mathcal{P}(\Sigma^\ast\times 1)\subseteq \mathcal{P}(\Sigma^\ast\times 1+\Sigma^\omega) \mid r(1)=R\times \{1\} \text{ where }
R\subseteq \Sigma^\ast\text{ is regular}\}.
\end{align*}
Additionally, the set $\omega\mathfrak{Reg}\defeq \{ r:1\rightdcirc 0=1\to \mathcal{P}(\Sigma^\omega)\mid r(1) \text{ is $\omega$-regular}\}$ satisfies:
\begin{align*}
& \{r^\omega\cdot s:1\rightdcirc 0=1\to \mathcal{P}(\Sigma^\omega) \mid r\in \mathfrak{Rat}(n,n), s\in \mathfrak{Rat}(1,n)\} = \omega\mathfrak{Reg}.
\end{align*}
\end{prop}

\subsection{Beyond tree automata}
There are variants of non-deterministic (B\"uchi) automata that accept
other types of input (e.g.\ arbitrary finitely-branching trees, see e.g.~\cite{urabe_et_al:LIPIcs:2016:6186}).
In general, given a functor $F:\mathsf{Set}\to \mathsf{Set}$ we
define a non-deterministic (B\"uchi) $F$-automaton as a pair
$(\alpha,\mathfrak{F})$, where $\alpha:n \to \mathcal{P}F n$ and
$\mathfrak{F}\subseteq n$. A natural question that arises is the following:
are we able to build a general categorical setting in which we can reason about the (in)finite
behaviour of systems for arbitrary non-deterministic B\"uchi $F$-automata (or
even more generally, for systems of the type $TF$ for a
monad $T$)? If so, then is it possible to generalize the Kleene theorem for
$(\omega-)$regular languages to a coalgebraic level?
We will answer these questions positively in the next sections.

\section{Monads for (in)finite behaviour}\label{section:monads}

Given a monad $T$ and an endofunctor $F$ on a common category, the purpose
of this section is to provide a construction
of a monad $TF^\infty$ which extends the functor $TF$.
The monad $TF^\infty$ will prove itself sufficient to model the combination of finite
and infinite behaviour (akin to the monad $\mathcal{P}(\Sigma^\ast\times
\mathcal{I}d+\Sigma^\omega)$ for the functor
$\mathcal{P}(\Sigma\times \mathcal{I}d)$,
or $\mathcal{P}T_\Sigma$ for $\mathcal{P}(\Sigma \times \mathcal{I}d^2)$).

At first we list all assumptions required in the remainder of this section. Later, in
Subsection~\ref{subsection:free_monad}, we revisit the construction of the monad
$TF^\ast$ from~\cite{brengos2015:lmcs}. Finally, we give a
description of $TF^\infty$.

\begin{asm}\label{subsection:assumptions_monads}
Let $\mathsf{C}$ be a category which admits binary coproducts. We denote the
coproduct operator by $+$ and the coprojection into the first and the second
component of a coproduct by $\mathsf{inl}$ and $\mathsf{inr}$ respectively.
Moreover, let $F:\mathsf{C}\to \mathsf{C}$ be a functor. In what follows, in this section
we additionally assume:
\begin{enumerate}[(i)]
\item  $(T,\mu,\eta)$ is a monad on $\mathsf{C}$ and $F:\mathsf{C}\to
\mathsf{C}$ lifts to $\mathcal{K}l(T)$ via a distributive law $\lambda:FT\implies TF$,
\item there is an initial $F(-)+X$-algebra for any object $X$ and a terminal
$F$-coalgebra $\zeta:F^\omega\to F F^\omega$,
\item the category $\mathsf{Alg}(F)$ of $F$-algebras admits binary coproducts
(with the coproduct operator denoted by $\oplus$).
\end{enumerate}
\end{asm}

\subsection{Preliminaries}\label{subsection:monads_preliminaries}
 The initial $F(-)+X$-algebra $i_X:FF^\ast X +X\to F^\ast X$  yields
 the free $F$-algebra over $X$  given by $i_X\circ \mathsf{inl}: FF^\ast X \to F^\ast X$.
Hence, by our assumptions we have an adjoint situation $\mathsf{C}\rightleftarrows \mathsf{Alg}(F)$, where
the left adjoint is the free algebra functor which assigns to any object $X$ the
free algebra $i_X\circ \mathsf{inl}$ over it.
The right adjoint is the forgetful functor which assigns to any $F$-algebra its
carrier and is the identity on morphisms. The adjunction yields the  monad
$F^\ast:\mathsf{C}\to \mathsf{C}$ which assigns to any object $X$ the carrier of
the free $F$-algebra over $X$.

\begin{exa}
For any set $\Sigma$ and $X$ the initial $\Sigma\times \mathcal{I}d+X$-algebra
is given by the morphism  $i_X:\Sigma \times \Sigma^\ast \times X+X\to
\Sigma^\ast \times X$, where \[i_X(a,(\sigma,x)) = (a\sigma,x) \text{ and }
i_X(x) = (\varepsilon,x).\]
\end{exa}

\subsubsection{Bloom algebras}
The purpose of this subsection is to recall basic definitions and properties of Bloom $F$-algebras~\cite{AdamekHM14} whose free algebras yield a monad $F^\infty$ on $\mathsf{C}$ which extends the functor $F$.
This will  allow us to embed systems of the type $X\to TFX$ to systems
of the type $X\to TF^\infty X$ and discuss their (in)finite behaviour in the latter context.

A pair $(a:FA\to A, (-)^\dagger)$ is called a \emph{Bloom
$F$-algebra} provided that any $F$-coalgebra $e:X\to FX$ yields the map
$e^\dagger:X\to A$ which satisfies:
\begin{center}
\resizebox{.7\textwidth}{!}{
\begin{tikzpicture}
\node(x1) at (-1,1) {$X$};
\node (a1) at (0,1) {$A$};
\node(fx1) at (-1,0) {$FX$};
\node (fa1) at (0,0) {$FA$};
\draw[->] (x1) -- (fx1) node[pos=.5,left] {$e$};
\draw[->] (fa1) -- (a1) node[pos=.5,right] {$a$};
\draw[->] (x1) -- (a1) node[pos=.5,above] {\tiny $e^\dagger$};
\draw[->] (fx1) -- (fa1) node[pos=.5,below] {\tiny $Fe^\dagger$};
\node (and) at (2,0.5) {\text{and}};
\node(x2) at (4,1) {$X$};
\node (y2) at (5,1) {$Y$};
\node(fx2) at (4,0) {$FX$};
\node (fy2) at (5,0) {$FY$};
\draw[->] (x2) -- (fx2) node[pos=.5,left] {$e$};
\draw[->] (y2) -- (fy2) node[pos=.5,right] {$f$};
\draw[->] (x2) -- (y2) node[pos=.5,above] {\tiny $h$};
\draw[->] (fx2) -- (fy2) node[pos=.5,below] {\tiny $Fh$};
\node (implies) at (6.5,0.5) {\text{implies}};
\node(x3) at (8,1) {$X$};
\node (y3) at (9,1) {$Y$};
\node(a3) at (8.5,0) {$A$};
;
\draw[->] (x3) -- (a3) node[pos=.5,left] {$e^\dagger$};
\draw[->] (y3) -- (a3) node[pos=.5,right] {$f^\dagger$};
\draw[->] (x3) -- (y3) node[pos=.5,above] {\tiny $h$};
\end{tikzpicture}
}

\end{center}
\noindent A \emph{homomorphism} from a Bloom algebra $(a:FA\to A,
(-)^\dagger)$ to a Bloom algebra $(b:FB\to B,(-)^\ddagger)$ is a map $h:A\to B$ which is
an $F$-algebra homomorphism from $a$ to $b$, which additionally preserves the dagger, i.e.
$e^\dagger \circ h = e^\ddagger$.  The category of Bloom algebras and
homomorphisms between them is denoted by $\mathsf{Alg}_B(F)$. We have the
following theorem.

\begin{thmC}[\cite{AdamekHM14}]\label{theorem:adamek}
The pair $(\zeta^{-1}:FF^\omega \to F^\omega,[[-]])$, where $[[-]]$ assigns to
$e:~X\to FX$ the unique coalgebra homomorphism $[[e]]:X\to F^\omega$ between $e$
and $\zeta$, is an initial object in $\mathsf{Alg}_B(F)$.  Moreover, the $F$-algebra
 coproduct \[(i_X\circ \mathsf{inl}:FF^\ast X\to F^\ast X )\oplus (\zeta^{-1}:FF^\omega\to F^\omega)\]
 is the free Bloom algebra over $X$.
\end{thmC}

\begin{rem}
Let $F^\infty:\mathsf{C}\to \mathsf{C}$ be defined as the composition of the
left and right adjoints $\mathsf{C}\rightleftarrows \mathsf{Alg}_B(F)$,
where the left adjoint is the free Bloom algebra functor  and the
right adjoint is the forgetful functor. The functor $F^\infty$ carries a monadic
structure which extends $F^\ast$. Indeed, by Theorem~\ref{theorem:adamek}, the monad
$F^\ast$ is a submonad of $F^\infty$ (via the transformation induced by the
coprojection into the first component of $i_X\circ \mathsf{inl}\oplus
\zeta^{-1}$ in $\mathsf{Alg}(F)$). The formula for the free Bloom algebra
from the above theorem indicates that $F^\infty$ is a natural extension of
$F^\ast$ encompassing infinite behaviours of the final $F$-coalgebra. By abusing
the notation slightly, we can write \[F^\infty = F^\ast \oplus F^\omega.\]
The functor $F_\varepsilon$ is a subfunctor of $F^\ast$~\cite[Lemma
4.12]{brengos2015:lmcs} and hence, by the above, also of $F^\infty$. In the
following sections this will let us  turn any coalgebra $X\to TFX$ or $X\to
TF_\varepsilon X$ into a system $X\to TF^\infty X$ and, by doing so, allow us to
model their (in)finite behaviour.
\end{rem}

\begin{exa}\label{example:bloom_algebras}
The terminal $\Sigma\times \mathcal{I}d$-coalgebra is \[\zeta:\Sigma^\omega \to
\Sigma\times \Sigma^\omega;a_1 a_2\ldots\mapsto (a_1,a_2a_3\ldots).\]
The coproduct of $a:\Sigma\times A\to A$ and $b:\Sigma\times B\to B$ in
$\mathsf{Alg}(F)$ is
\[a\oplus b:\Sigma\times (A+B)\to A+B; (\sigma,x) \mapsto \left \{
\begin{array}{cc} a(\sigma,x)  &\text{if }x\in A,\\ b(\sigma,x) &
\text{otherwise.} \end{array}\right.\]
 Hence, the free Bloom algebra over $X$ is a map $\Sigma\times (\Sigma^\ast \times X+
\Sigma^\omega) \to \Sigma^\ast \times X + \Sigma^\omega$ explicitly given by:
$(a,(\sigma,x))\mapsto (a\sigma,x) \text{ and } (a,a_1a_2\ldots) \mapsto
aa_1a_2\ldots$.
\end{exa}
\begin{wrapfigure}[5]{r}{0.22\textwidth}
\resizebox{.22\textwidth}{!}{
\begin{tikzpicture}
\node(x) at (0,1) {$X$};
\node (a) at (1.5,1) {$A$};
\node(fx) at (0,0) {$FX$};
\node (fa) at (1.5,0) {$FA$};
\node(b) at (3,1) {$B$};
\node (fb) at (3,0) {$FB$};

\draw[->] (x) ..controls (1.5,1.3).. (b) node[pos=.5,above] {\small
$e^\ddagger$};
\draw[->] (x) -- (fx) node[pos=.5,left] {$e$};
\draw[->] (fa) -- (a) node[pos=.5,right] {$a$};
\draw[->] (x) -- (a) node[pos=.5,below] {$e^\dagger$};
\draw[->] (fx) -- (fa) node[pos=.5,above] { $Fe^\dagger$};
\draw[->] (a) -- (b) node[pos=.5,below] {$h$};
\draw[->] (fb) -- (b) node[pos=.5,right] {$b$};
\draw[->] (fa) -- (fb) node[pos=.5,above] { $Fh$};
\draw[->] (fx) ..controls (1.5,-0.3).. (fb) node[pos=.5,below] {\small
$Fe^\ddagger$};
\end{tikzpicture}
}
%\vspace{-0.8cm}
\end{wrapfigure}
 Let $(a:FA\to A, (-)^\dagger)$ be a Bloom algebra, $b:FB\to B$ an $F$-algebra and
 $h:A\to B$ a homomorphism between $F$-algebras $a$ and $b$. Then there is a
unique assignment $(-)^\ddagger$ which turns $(b:FB\to B,(-)^\ddagger)$ into a
Bloom algebra and $h$ into a Bloom algebra homomorphism and it is defined as
follows~\cite{AdamekHM14}: for $e:X\to FX$ the map $e^\ddagger:X\to B$ is
$e^\ddagger \defeq h\circ e^\dagger$.

\subsection{Lifting monads to algebras}\label{subsection:free_monad}
Take an $F$-algebra $a:FA\to A$ and define
$\bar{T}(a)\defeq FTA \stackrel{\lambda_A}{\to} TFA\stackrel{Ta}{\to} TA$. If
$h:A\to B$ is a homomorphism of algebras $a$ and $b:FB\to B$ we put $\bar{T}(h)
= T(h)$.
In this case $\bar{T}:\mathsf{Alg}(F)\to \mathsf{Alg}(F)$ is a functor for which
the morphism $\eta_A:A\to TA$ is an $F$-algebra homomorphism from $a:FA\to A$ to
$\bar{T}(a):FTA\to TA$. Moreover, $\mu_A:T^2 A\to TA$ is a homomorphism from
$\bar{T}^2(a)$ to $\bar{T}(a)$ (see~\cite{10.1007/BFb0083084} for details). A
direct consequence of this construction is the following.

\begin{thmC}[\cite{10.1007/BFb0083084}]
The triple $(\bar{T},\bar{\mu},\bar{\eta})$, where for  $a:FA\to A$ we put
\[\bar{\mu}_a:\bar{T}^2(a)\to~\bar{T}(a); \bar{\mu}_a = \mu_A\text{ and
}\bar{\eta}_a: a\to \bar{T}(a); \bar{\eta}_a = \eta_A\] is a monad on
$\mathsf{Alg}(F)$.
\end{thmC}

 The above theorem together with the assumption of existence of an arbitrary
\begin{wrapfigure}[3]{r}{0.3\textwidth}

\resizebox{.3\textwidth}{!}{
\begin{tikzpicture}
\node(C) at (0,0) {$\mathsf{C}$};
\node (A) at (2,0) {$\mathsf{Alg}(F)$};
\node (K) at (4,0) {$\mathcal{K}l(\bar{T})$};
\node(P) at (0.8,0) {$\perp$};
\node(P2) at (3,0) {$\perp$};
\draw[->] (C) ..controls (0.7,0.3).. (A);
\draw[->] (A) ..controls (0.7,-0.3)..  (C);
\draw[->] (A) ..controls (3,0.3).. (K);
\draw[->] (K) ..controls (3,-0.3)..  (A);
\end{tikzpicture}%
\label{form:adjoint_pair}
}
\end{wrapfigure}
free $F$-algebra  in $\mathsf{Alg}(F)$ leads to a pair of adjoint situations
captured by the diagram on the right. Since the composition of adjunctions is an
adjunction this yields a monadic structure on the functor $T
F^\ast:\mathsf{C}\to \mathsf{C}$.

\begin{exa}
An example of this phenomenon is given by the monad
$\mathcal{P}(\Sigma^\ast\times \mathcal{I}d)$ from
Example~\ref{example:lts_free} where in the above we set $T=\mathcal{P}$ and
$F=\Sigma\times \mathcal{I}d$. This monad has already been described e.g.\ in~\cite{brengos2015:lmcs}, but it arose as a consequence of the composition of a
different pair of adjunctions.
\end{exa}

\subsubsection{Monads on Bloom algebras}\label{subsection:cofree_monad}
Above we gave a recipe for a general construction of a monadic structure on the
functor $TF^\ast$. As witnessed in~\cite{brengos2015:lmcs,bonchi2015killing},
this monad is suitable to model coalgebras and their weak bisimulations and weak
finite trace semantics (i.e.\ their finite behaviour). Our primary interest is in
modelling infinite behaviour and this monad proves itself insufficient. The
purpose of this subsection is to show how to tweak the middle category from the
pair of adjunctions in pictured in the diagram above so that the monad obtained
from the composition of two adjunctions is suitable for our needs.

Let $(a:FA\to A, (-)^\dagger)$ be a Bloom algebra and define
\[\bar{T}_B( (a:FA\to A, (-)^\dagger)) \defeq ( \bar{T}(a):FTA\to
TA,(-)^\ddagger ),\]
where for any $e:X\to FX$ the map $e^\ddagger$ is given by $\eta_A \circ
e^\dagger$.
Since $\eta_A:A\to TA$ is a homomorphism between $a:FA\to A$ and
$\bar{T}(a):FTA\to TA$ the pair $(\bar{T}(a),(-)^\ddagger)$ is a Bloom algebra.
For a pair of Bloom algebras $(a:FA\to A, (-)^\dagger)$ and $(b:FB\to
B,(-)^\ddagger)$ and a Bloom algebra homomorphism $h:A\to B$ between them put
$\bar{T}_B(h) = T(h)$.  This defines a functor $\bar{T}_B:\mathsf{Alg}_B(F)\to
\mathsf{Alg}_B(F)$. Analogously to the previous subsection we have the following
direct consequence of the construction.

\begin{thm}
The triple $(\bar{T}_B,\bar{\mu}^B,\bar{\eta}^B)$ is a monad on
$\mathsf{Alg}_B(F)$, where for any Bloom algebra $(a:FA\to A, (-)^\dagger)$ the
$(a,(-)^\dagger)$-components of the  transformations $\bar{\mu}^B$ and
$\bar{\eta}^B$ are
\begin{align*}
&\bar{\mu}^B_{(a,(-)^\dagger)}:\bar{T}^2_B(a,(-)^\dagger) \to
\bar{T}_B(a,(-)^\dagger); \quad \bar{\mu}^B_{(a,(-)^\dagger)} = \mu_A\text{ and
}\\
&\bar{\eta}^B_{(a,(-)^\dagger)}:(a,(-)^\dagger) \to
\bar{T}_B(a,(-)^\dagger)\text{ with }\bar{\eta}^B_{(a,(-)^\dagger)} = \eta_A.
\end{align*}
\end{thm}

Hence, we obtain two adjoint situations captured in the diagram below. These
\begin{wrapfigure}[3]{r}{0.33\textwidth}
\resizebox{.33\textwidth}{!}{
\begin{tikzpicture}
\node(C) at (0,0) {$\mathsf{C}$};
\node (A) at (2,0) {$\mathsf{Alg}_B(F)$};
\node (K) at (4,0) {$\mathcal{K}l(\bar{T}_B)$};
\node(P) at (0.8,0) {$\perp$};
\node(P2) at (3,0) {$\perp$};
\draw[->] (C) ..controls (0.7,0.3).. (A);
\draw[->] (A) ..controls (0.7,-0.3)..  (C);
\draw[->] (A) ..controls (3,0.3).. (K);
\draw[->] (K) ..controls (3,-0.3)..  (A);
\end{tikzpicture}
}
\end{wrapfigure}
adjunctions impose a monadic structure on the functor $TF^\infty:\mathsf{C}\to
\mathsf{C}$.
The monad $\mathcal{P}(\Sigma^\ast\times \mathcal{I}d+\Sigma^\omega)$ from
Example~\ref{example:lts_omega_monad} arises from the composition of the above
adjoint situations (see also Example~\ref{example:bloom_algebras}).

\begin{exa}\label{example:tree_functor}
Let $F=\Sigma\times \mathcal{I}d^2$. This functor lifts to $\mathcal{K}l(\mathcal{P})$~\cite{hasuo07:trace}
and, up to isomorphism,
$F^{\infty} = T_{\Sigma}$ is a
functor which assigns to any set $X$ the set of complete binary trees  (i.e.
every node has either two children or no children) with inner nodes taking
values in $\Sigma$ and finitely many leaves, all taken from $X$~\cite{AdamekHM14} (see Subsection~\ref{subsec:buchi_tree}).  This yields a
monadic structure on $\mathcal{P}F^\infty=\mathcal{P}T_\Sigma$ defined in
Example~\ref{example:in_finite_tree_monad}.
\end{exa}

The above example can be easily generalized. Indeed, if $T$ is a commutative $\mathsf{Set}$-based monad
then any {polynomial functor}\footnote{A \emph{polynomial functor} is a functor defined by
the grammar $F \defeq  \Sigma \in \mathsf{Set} \mid  \mathcal{I}d \mid  F \times
F \mid  \sum F$~\cite{hasuo07:trace}.}
$F:\mathsf{Set}\to \mathsf{Set}$ lifts to $\mathcal{K}l(T)$~\cite{hasuo07:trace}. If it admits all free $F$-algebras and the final
$F$-coalgebra then Assumption~\ref{subsection:assumptions_monads} holds for $T$ and $F$ yielding
the monad $TF^\infty$.

\section{Abstract automata and their behaviour}%
\label{section:automata}

The purpose of this section is to generalize the concepts from Section~\ref{section:buchi_coalgebraically} to an arbitrary Kleisli category with a suitable
ordering. In other words, given a $\mathsf{Set}$-monad $T$,  we define a $T$-automaton,
its finite behaviour, its infinite behaviour with BAC and provide generic Kleene theorems
for $T$.

Let $(T,\mu,\eta)$ be a $\mathsf{Set}$-based monad. Since
we will often consider the Lawvere theory $\mathbb{T}$ associated with it, recall that its
objects are sets given by $n=\{1,\ldots,n\}$ for $n=0,1,\ldots$. We start with the definition of a $T$-automaton.

\begin{defi}
A \emph{$T$-automaton} or simply \emph{automaton} is a pair $(\alpha,
\mathfrak{F})$, where $\alpha:X\rightdcirc X = X\to TX$ is a $T$-coalgebra called
\emph{transition morphism} and $\mathfrak{F}\subseteq X$.
\end{defi}

\begin{asm}\label{assumptions:kleisli_cat}
In order to define finite and infinite behaviour of $(\alpha,\mathfrak{F})$ and reason about it we
require the Kleisli category for $T$ to satisfy more assumptions. In this section we assume that:
\begin{enumerate}[(A)]
    \item  $\mathcal{K}l(T)$ is order enriched with a pointwise induced order,\label{condition:0}
    \item it is $\cpo$-enriched,\label{condition:1}
    \item it is left distributive,\label{condition:2}\label{condition:4}
    \item its hom-sets are complete lattices.\label{condition:3}
\end{enumerate}
\end{asm}

\noindent
 At first let us note that, in the light of
 Remark~\ref{remark:right_distributivity}, by~\ref{condition:0} we get:
\begin{itemize}
\item right distributivity w.r.t.\ the base morphisms and
\item  cotupling order preservation.
\end{itemize}
 Additionally, since the Kleisli category for $T$ is left distributive we have:
\begin{itemize}
\item   the bottoms $\perp_{X,Y}\in \mathcal{K}l(T)(X,Y)$ satisfy
  $g\cdot \perp_{X,Y} = \perp_{X,Z}$ for any $g:Y\rightdcirc Z$.
\end{itemize}

\noindent
The  axioms~\ref{condition:0}-\ref{condition:2}  guarantee that the saturation
 $\alpha\mapsto \alpha^\ast$ is computable in $\omega$-steps and is expressive enough.
For a given  $\alpha:X\rightdcirc X = X\to T X$  we can define maps $\alpha^\ast,\alpha^+:X\rightdcirc X= X\to TX$:
\[
\alpha^\ast \defeq \mu x. (\mathsf{id} \vee x\cdot \alpha) \text{ and } \alpha^+ \defeq \alpha^\ast\cdot \alpha.
\]
 The operator $\alpha\mapsto \alpha^\ast$ was thoroughly studied in~\cite{brengos2014:cmcs,brengos2015:lmcs,brengos2015:jlamp,brengos2016:concur} in
the context of coalgebraic weak bisimulation. Its definition does \emph{not}
require a complete lattice order.  See \emph{loc.\ cit.} for a discussion.

 Assumption~\ref{condition:3} allows us to define the greatest
 fixpoint of the map $x\mapsto x\cdot \alpha$.
Indeed for any $\alpha:X\rightdcirc X=X\to TX$ put $\alpha^\omega:X\rightdcirc 0=X\to T0$ to be:
 \[\alpha^\omega \defeq \bigwedge_{\kappa \in \mathsf{Ord} }(x\mapsto x\cdot \alpha)^\kappa
(\top),\]
where $\top:X\rightdcirc 0 = X\to T0$ is the greatest element of $\mathcal{K}l(T)(X,0)$ and
\[(x\mapsto x\cdot \alpha)^\kappa:(X\rightdcirc 0)\to (X\rightdcirc 0)\]
is defined in terms of the transfinite induction by
\[
(x\mapsto  x\cdot \alpha)^{\kappa +1} \defeq  ( x\mapsto x\cdot \alpha)\circ ( x\mapsto x\cdot
\alpha)^{\kappa}
\]
for a successor ordinal $\kappa+1$ and the ordinary map composition operator $\circ$,
and $( x\mapsto x\cdot \alpha)^{\kappa}\defeq \bigwedge_{\lambda <\kappa }
( x\mapsto x\cdot \alpha)^{\lambda}$ for a limit ordinal $\kappa$.
By the Tarski-Knaster theorem, the map $\alpha^\omega:X\rightdcirc 0=X\to T0$
is the greatest fixpoint of the assignment $x\mapsto  x\cdot \alpha$.

\begin{exa}
The Kleisli categories for the monads from Example~\ref{example:conditions}
satisfy~\ref{condition:0}-\ref{condition:3}.
Section~\ref{section:probabilistic_systems} presents one more example
of a Kleisli category that fits the setting in the context of probabilistic
automata.
\end{exa}

Before we present the definition of finite and infinite behaviour of automata we
need one more technical result.

\begin{lem}\label{theorem:least_fixpoint}
 For any $\alpha,\beta:X\rightdcirc X = X\to TX$ we have:
\begin{enumerate}
  \item\label{prop:0} $\alpha^\ast = \bigvee_n (\mathsf{id} \vee \alpha)^n$,
  \item\label{prop:1} $\mathsf{id}^\ast = \mathsf{id}$, $\mathsf{id}\leq \alpha^\ast$ and
  $\alpha^\ast \cdot \alpha^\ast  = \alpha^\ast$,
\item\label{prop:4} $(\alpha\cdot\beta)^\omega = (\beta\cdot \alpha)^\omega \cdot \beta$,
\item\label{prop:5} $(\alpha^n)^\omega = \alpha^\omega$ for any $n>0$,
\item\label{prop:6} $\alpha^\omega = (\alpha^+)^\omega$.
\end{enumerate}
\end{lem}
\begin{proof}
  The proof of~\ref{prop:0} and~\ref{prop:1} can be found in~\cite[Lemma 5.1 and Theorem 5.7]{brengos2015:lmcs}.
  To see~\ref{prop:4} holds, i.e.\  $(\alpha\cdot\beta)^\omega = (\beta\cdot \alpha)^\omega
\cdot \beta$ note that $(\beta\cdot \alpha)^\omega \cdot \beta $ is a fixpoint
of $x\mapsto  x\cdot \alpha\cdot \beta$ and hence $(\beta\cdot \alpha)^\omega
\cdot \beta \leq (\alpha\cdot\beta)^\omega$. By a similar argument we show
$(\alpha\cdot \beta)^\omega \cdot \alpha \leq (\beta\cdot \alpha)^\omega$.
Thus,
\[
(\beta\cdot \alpha)^\omega=(\beta\cdot \alpha)^\omega \cdot \beta\cdot \alpha
\leq (\alpha\cdot\beta)^\omega\cdot \alpha\leq (\beta\cdot \alpha)^\omega.
\]

To prove $(\alpha^n)^\omega = \alpha^\omega$ note that by~\ref{prop:4} we have
$(\alpha^n)^\omega = (\alpha^{n-1}\cdot \alpha)^\omega = (\alpha\cdot
\alpha^{n-1})^\omega\cdot \alpha$. Hence, $(\alpha^{n})^\omega \leq
\alpha^\omega$. Moreover, since $\alpha^\omega\cdot \alpha^n = \alpha^\omega
\cdot \alpha^{n-1}= \cdots =\alpha^\omega$ we get the converse inequality, i.e.
$\alpha^\omega \leq (\alpha^n)^\omega$. This proves the assertion.

Finally, note that by monotonicity of $(-)^\omega$ since $\alpha \leq
\alpha^\ast \cdot \alpha = \alpha^+$ we have $\alpha^\omega \leq
(\alpha^+)^\omega$. Moreover,
\begin{align*}
&(\alpha^+)^\omega = (\alpha^\ast\cdot \alpha)^\omega\leq (\alpha^\ast\cdot
\alpha)^\omega \cdot (\mathsf{id}\vee \alpha) = (\alpha^\ast\cdot
\alpha)^\omega\vee (\alpha^\ast\cdot \alpha)^\omega\cdot \alpha \leq \\
&(\alpha^\ast\cdot \alpha)^\omega\vee (\alpha^\ast\cdot \alpha)^\omega\cdot
\alpha^\ast \cdot \alpha = (\alpha^\ast\cdot \alpha)^\omega\vee
(\alpha^\ast\cdot \alpha)^\omega= (\alpha^+)^\omega
\end{align*}
 Hence, by induction we prove that $(\alpha^+)^\omega \cdot (\mathsf{id}\vee
\alpha)^i = (\alpha^+)^\omega$. By the fact that our theory is \cpo -enriched we
get:
\[
(\alpha^+)^\omega \cdot \alpha^\ast = (\alpha^+)^\omega \cdot \bigvee_i
(\mathsf{id}\vee \alpha)^i = \bigvee_i (\alpha^+)^\omega \cdot (\mathsf{id}\vee
\alpha)^i = (\alpha^+)^\omega.
\]
This proves that $(\alpha^+)^\omega$ satisfies $(\alpha^+)^\omega \cdot
\alpha^\ast \cdot \alpha = (\alpha^+)^\omega \cdot \alpha$. Thus,
$(\alpha^+)^\omega \leq \alpha^\omega$ which completes the proof.
\end{proof}

\subsection{Finite and infinite behaviour}
\noindent The purpose of this subsection is to present
the definitions of the finite and infinite behaviour with BAC
for $T$-automata. Let $(\alpha:X\to TX, \mathfrak{F}\subseteq X)$ be a $T$-automaton. Before we start,
let us first encode the set $\mathfrak{F}$
of accepting states in terms of an endomorphism $\mathfrak{f}_{\mathfrak{F}}:X\rightdcirc X=X\to TX$
by:
\[
\mathfrak{f_F}(x)  =
\left\{
    \begin{array}{cc}
        \eta_X(x) & \text{ if }x\in \mathfrak{F},\\
        \perp & \text{ otherwise}
    \end{array}
\right.
\text{ for any }x\in X,
\]
where $\perp$ denotes the bottom element of the poset $TX$.

\begin{defi}\label{definition:behaviours}
 \emph{Finite} and  \emph{$\omega$-behaviour} of the automaton $(\alpha,\mathfrak{F})$
 are given respectively in terms of morphisms in $\mathcal{K}l(T)$ by:
$||\alpha,\mathfrak{F}||:X\rightdcirc 1$ and $|| \alpha,\mathfrak{F}
||_\omega:X\rightdcirc 0$, where
\[|| \alpha,\mathfrak{F} || \defeq !\cdot \mathfrak{f_F}\cdot
\alpha^\ast\text{ and } || \alpha,\mathfrak{F} ||_\omega \defeq (
\mathfrak{f_F} \cdot \alpha^+)^\omega.\]
\emph{Finite behaviour of a state  $x\in X$} of $(\alpha,\mathfrak{F})$ is the
map $||\alpha,\mathfrak{F}|| \cdot x_{X}:1\rightdcirc 1$, and its \emph{$\omega$-behaviour}
is given by $||\alpha,\mathfrak{F} ||_{\omega} \cdot x_X:1\rightdcirc 0$.
Here, \[x_X:1\rightdcirc X=1\to TX; 1\mapsto \eta_X(x).\]
\end{defi}

\begin{exa}
As we have already seen in Section~\ref{section:buchi_coalgebraically}, the
finite and $\omega$-behaviour of
$\mathcal{P}(\Sigma^\ast\times \mathcal{I}d+\Sigma^\omega)$-automata coincides with
the classical notions whenever the tuple is given by $(\alpha:n\to \mathcal{P}(\Sigma\times
n),\mathfrak{F}\subseteq n)$. The same applies to tree automata
(see Proposition~\ref{proposition:tree_languages}).
\end{exa}

\subsection{Additional remarks}\label{remark:trace_equivalence}
Our approach
to defining semantics for coalgebras seems to diverge slightly
from the established coalgebraic takes known from e.g.~\cite{hasuo07:trace,silva2013:calco,bonchi2015killing,urabe_et_al:LIPIcs:2016:6186}.
The purpose of this subsection is to compare our setting with the frameworks presented
in the literature and try to justify the (slight) differences.

Our approach builds on top of two fixpoint operators, namely $(-)^\ast$ and $(-)^\omega$.
The choice of these two operators, and not other (e.g.\ the \emph{dagger} operator from~\cite{esik2011,bonchi2015killing})
follows from the premise that we wanted to make the connection with the classical
results in regular and $\omega$-regular languages as clear and as direct as possible.
As we witness here, the classical Kleene
star operation and $(-)^\omega$~\cite{pin:automata}
prove to have their general categorical counterparts.

It may not be clear to the reader why the finite
and $\omega$-behaviour maps have different codomains, i.e.\ the former is a map $X\rightdcirc 1$
and the latter $X\rightdcirc 0$. Let us focus on the finite behaviour first.
So far in the coalgebraic literature, finite behaviour of systems was introduced
in terms of finite trace~\cite{silva2013:calco,bonchi2015killing,jacobssilvasokolova2012:cmcs}.
In the setting of systems $X\to TFX$ it is obtained in terms of the
initial algebra-final coalgebra coincidence~\cite{hasuo07:trace,bonchi2015killing}.
When translated to the setting of systems with internal moves, the
finite trace is given by $\mu
x.x\cdot \alpha: X\rightdcirc 0$ and is calculated
in the Kleisli category for the monad $TF^\ast$~\cite{brengos2014:cmcs,brengos2015:lmcs}.
However, this holds for coalgebras whose type monad encodes accepting states.
From the point of our
setting, the accepting states are not part of the transition and are
encoded in terms of $\mathfrak{F}\subseteq X$ instead. The direct use of
initial algebra-final coalgebra coincidence makes no sense here, as the initial algebra
would simply be degenerate. Luckily, there is a simple formal argument showing that
our approach from this paper and the aforementioned
approach established in the coalgebraic literature coincide. For the monad $T$ and a
 $T$-automaton $(\alpha,\mathfrak{F})$ consider the
 monad $T(\mathcal{I}d+1)$\footnote{It can be easily verified that for
 any monad $T$ the functor $T(\mathcal{I}d+1)$
 carries a monadic structure. It follows from the fact that the exception monad $\mathcal{I}d+E$ induces
 an exception monad transfomer $T\mapsto T(\mathcal{I}d+E)$.} and the map $X\to T(X+1)$ defined
 for any $x\in X$ by $\alpha(x) \vee \chi_\mathfrak{F}(x)$,
 where $\chi_\mathfrak{F}:X\to T(X+1); x\mapsto
 \left \{
     \begin{array}{cc}
       \eta_{X+1}(1)   & \text{ if } x\in \mathfrak{F},\\
      \perp & \text{ otherwise}.
     \end{array}
 \right.$
It is a simple exercise
to prove that the least fixpoint
$\mu x. x\cdot (\alpha\vee \chi_\mathfrak{F}):X\rightdcirc 0=X\to T(0+1)=X\to T1$ % chktex 12
calculated in $\mathcal{K}l(T(\mathcal{I}d+1))$ is the same as the finite
behaviour map $||\alpha,\mathfrak{F}||:X\rightdcirc 1=X\to T1$ calculated in
$\mathcal{K}l(T)$. Therefore, our definition of finite behaviour via
$(-)^\ast$ coincides with the coalgebraic finite trace semantics via
$\mu x.x\cdot \alpha$.

The finite behaviour of a state of an automaton from Definition~\ref{definition:behaviours}
is of type $1\rightdcirc 1$. We argue that this map
should be viewed as a generalization of a finitary language. Classically,
these languages have been considered in some algebraic context, e.g. with the familiar
algebraic operations of concatenation, Kleene start closure and finite union. These operations
considered on our abstract categorical level directly translate
into morphism composition, saturation and finite joins of endomorphisms $1\rightdcirc 1$
respectively.

As far as the infinite behaviour is concerned, it should be noted here that our prototypical
example of a monad is $TF^\infty = T(F^\ast \oplus F^\omega)$ from Section~\ref{section:monads}.
By Theorem~\ref{theorem:adamek} the object $(F^\ast \oplus F^\omega)(0)=F^\omega$ is the carrier
of the terminal $F$-coalgebra $\zeta:F^\omega\to F F^\omega$ making $TF^\infty 0 =TF^\omega$. This is exactly what
we expect to have as a codomain of an infinite trace map
(see also for comparison~\cite{urabe_et_al:LIPIcs:2016:6186,DBLP:journals/entcs/Cirstea10}).

Additionally, the type of the infinite behaviour of a state of an automaton is $1\rightdcirc 0$
and it reflects the \emph{partial}
algebraic nature of (in)finitary languages. In particular, it makes sense to compose (concatenate)
a finitary langugage $1\rightdcirc 1$
with an infinitary language $1\rightdcirc 0$ and get an infinitary one
($1\rightdcirc 1\rightdcirc 0 = 1\rightdcirc 0$), but \emph{not} vice versa. Moreover, it does \emph{not} necessarily make
sense to compose two infinitary languages ($1\rightdcirc 0$ and $1\rightdcirc 0$) with each other.

%%%%%%%%%%%%%%%%%%%
%%%%%%%%%%%%%%%%%%%%
%%%%%%%%%%%%%%%%%%%%%
%%%%%%%%%%%%%%%%%%%%%%
%%%%%%%%%%%%%%%%%%%%%%%
\subsection{Kleene theorems}\label{subsection:kleene_omega_reg}
The purpose of this part of the paper
is to state and prove Kleene theorems akin
to Proposition~\ref{prop:kleene_tree} and ~\ref{prop:kleene_lts}.
These theorems require us to work with \emph{finite}
automata and their behaviour, so we will restrict the setting of this subsection
to the Lawvere theory $\mathbb{T}$ associated with the monad $T$.

In this
subsection we consider a set $\mathcal{A}$ of endomorphisms from $\mathbb{T}$ such that:
\begin{itemize}
\item $\mathcal{A}$ contains all base map endomorphisms,
\item $ \perp_{n,n} :n\rightdcirc n\in \mathcal{A}$ % chktex 26
for any $n<\omega$
\footnote{In order to simplify the notation we will often omit the subscript and write $\perp$
to denote $\perp_{n,m}$ if the domain and codomain of $\perp$ can be deduced from the context.},
%\item for any $\alpha:n\rightdcirc n \in \mathcal{A}$ we have $\perp \cdot \alpha = \perp$,
\item if $\{\alpha_k:n_i\rightdcirc n_i\}_{k=1,\ldots,k}\subseteq  \mathcal{A}$
  then $\alpha_1+\ldots +\alpha_k\in \mathcal{A}$,
\item $\mathcal{A}$ is closed under taking finite suprema.
\end{itemize}
The set $\mathcal{A}$ plays a role of a set of admissible transition functions for automata taken into
consideration.
A $T$-automaton $(\alpha,\mathfrak{F})$ whose transition $\alpha:n\rightdcirc n$
is an arrow in $\mathbb{T}$
is called \emph{$\mathcal{A}$-automaton} if $\alpha\in \mathcal{A}$.

\begin{exa}
In the case of our leading examples of theories, namely $\mathsf{LTS}^\omega$ and $\mathsf{TTS}^\omega$,
the prototypical choice for $\mathcal{A}$ was given in condition (a)
in Proposition~\ref{prop:kleene_lts} and~\ref{prop:kleene_tree}
respectively.
\end{exa}

\begin{defi}
The set of \emph{regular
morphisms} $m\rightdcirc p\in \mathbb{T}$ is defined by:
\begin{align*}
\mathfrak{Reg}(m,p) \defeq & \{ j'\cdot \mathfrak{f_F} \cdot \alpha^\ast \cdot
j\mid  (\alpha:n\rightdcirc  n,\mathfrak{F}) \text{ is a }\mathcal{A}\text{-automaton
and } \\
&j:m\rightdcirc n, j':n\rightdcirc p \text{ are base maps}\}.
\end{align*}
\end{defi}
\noindent The set of regular morphisms $\mathfrak{Reg}(1,p)$ will be often
referred to as the set of \emph{regular trees with variables in $p$}.
Note that $\mathfrak{Reg}(1,1)$ is  exactly the set of finite behaviours of
states in $\mathcal{A}$-automata.

We list the statements without proofs which we later provide in Subsection~\ref{subsubsection:omitted}.
\begin{lem}\label{lemma:regular_category}
The identity maps in $\mathbb{T}$ are regular morphisms. Moreover,
regular morphisms are closed under  composition from
$\mathbb{T}$.
\end{lem}
\noindent The above lemma precisely says that the collection of objects $n=0,1,\ldots$
with morphisms $\mathfrak{Reg}(m,n)$ forms a category with the composition taken
from $\mathbb{T}$. We denote this category by $\mathfrak{Reg}(\mathcal{A})$.
We have the following.
\begin{thm}[Kleene theorem for regular behaviour]\label{theorem:kleene_regular}
The category $\mathfrak{Reg}(\mathcal{A})$ is a subtheory of $\mathbb{T}$ such that:
%\vspace{-0.2cm}
\begin{enumerate}[(a)]
\item it contains all maps from $\mathcal{A}$,\label{item:aa}
\item it admits finite suprema,\label{item:bb}
\item its endomorphisms are closed under $(-)^\ast$.\label{item:cc}
\end{enumerate}
Moreover, if $\mathfrak{Rat}(\mathcal{A})$ denotes the smallest subtheory of $\mathbb{T}$
which satsfies~\ref{item:aa}-\ref{item:cc} then
\[\mathfrak{Rat}(\mathcal{A}) = \mathfrak{Reg}(\mathcal{A}).\]
\end{thm}

 Finally, we define
 \begin{align*}
\omega\mathfrak{Rat}(\mathcal{A}) & \defeq \{ [r_1,\ldots,
r_{m}]^\omega  \cdot r \mid r, r_i \in
\mathfrak{Rat}(1,m)   \text{ for } m  <\omega   \} \\
\omega\mathfrak{Reg}(\mathcal{A}) & \defeq  \{ ||\alpha,\mathfrak{F}||_{\omega}\cdot
i_n:1\rightdcirc 0 \mid (\alpha,\mathfrak{F}) \text{ is an }
\mathcal{A}\text{-automaton}\}.
\end{align*}

\begin{thm}[Kleene theorem for $\omega$-regular
behaviour]\label{theorem:kleene_omega_regular} We have
\[\omega\mathfrak{Rat}(\mathcal{A}) = \omega\mathfrak{Reg}(\mathcal{A}).\]
\end{thm}

%%%%%%%%%%%%%%%%%%%%%%%%%%%%
% Omitted proofs
%%%%%%%%%%%%%%%%%%%%%%%%%%%%
\subsubsection{Proofs}\label{subsubsection:omitted}

The purpose of this subsection is to present the proofs of the statements above.
Before we proceed we would like to make a remark concerning the material presented here
and its originality.
Several Kleene theorems (akin to Theorem~\ref{theorem:kleene_regular})
have been presented and proven in the literature on the level of
\emph{iteration theories}
(see e.g.~\cite{DBLP:journals/tcs/Esik97,bloomesik:93,esikhajgato2009:ai,esik2011,esik2013}).
According to our knowledge, due to minor differences in the formulation,
the theorems presented in our paper do not fit directly into any existing setting.
However, the classical proof
techniques used in \emph{loc.\ cit.} are still applicable here.
These methods are based on using well
know properties satisfied by a fixpoint operator.
We recall them here and present detailed proofs of our statements.
We additionally use string diagrams
as a visual aid to help the reader understand the techniques better.

  In order to proceed with the proofs we need to introduce some new notions
and define a notation used below. We start off by defining $[\mathcal{A}]$ to
be the set of morphisms from $\mathbb{T}$ obtained by (pre- and post-)composing
maps from $\mathcal{A}$ with base morphisms with suitable domains and codomains:

\[
  [\mathcal{A}]\defeq \{ i\cdot \alpha \cdot j \mid \alpha:n\rightdcirc n\in \mathcal{A}\text{ and }
  i:n\rightdcirc p, j:m\rightdcirc n \text{ are base maps}\}.
\]
Moreover, since the proofs presented below use the identity~\ref{equation:gspi}
which requires an extended definition of the saturation operator,
for any morphism $\alpha: n\rightdcirc n+p$ we define:
\begin{align}
  \alpha^\otimes \defeq [\alpha,\mathsf{in}_{n+p}^p]^\ast \cdot \mathsf{in}^n_{n+p}.
  \label{equation:extended_saturation}
\end{align}
Note that $[\alpha^\otimes,\mathsf{in}^p_{n+p}] = [\alpha,\mathsf{in}^{p}_{n+p}]^\ast$. Hence,
if $p=0$ then $\alpha^\otimes = \alpha^\ast$
for $\alpha : n\rightdcirc n+p$.
%%%%%%%%%%%%%%%%%%%%
% subsection
%%%%%%%%%%%%%%%%%%%%
\subsubsection*{String diagram notation} Let us now develop a string diagram notation which will
clarify the proofs considerably.
\begin{rem}
  It is important to emphasize that the purpose of the new notation is to
  build a visual aid to the technical statements made below. The reader should note that
  all proofs presented here are written so that
  the knowledge of the string diagram calculus is not required.
  However, given the complexity of some of
  the (in)equalities used, we strongly believe that the diagrammatic notation improves their readability
  (\emph{conf.} e.g.~\ref{equation:gspi} and its diagrammatic representation). Hence, we decide to proceed with
  its introduction.
\end{rem}

We adopt the standard string diagram calculus for monoidal
categories~\cite{selinger:2011:survey_graphical,fong:spivak:seven:2018} which
will be tailored to our purposes.
A  morphism $f:m\rightdcirc n$ is depicted by
\resizebox{.1\textwidth}{!}{
\begin{tikzpicture}
\pgfmathsetmacro{\shadow}{0.0};
\pgfmathsetmacro{\przesuniecieX}{0}
\pgfmathsetmacro{\x}{\przesuniecieX};
\pgfmathsetmacro{\y}{0};

\draw[-] (\x-0.25,\y+0.25) -- (\x+0.75,\y+0.25);
%\draw[-] (\x,\y+0.15) -- (\x+0.75,\y+0.15);
\draw[fill = white] (\x+0,\y+0) rectangle (\x+0.5,\y+0.5);
%\draw[fill = black] (\x+0.4,\y+0.4) rectangle (\x+0.5,\y+0.5);

\node (alpha) at (\x+0.25,\y+0.25) {$f$};
\node (aa) at (\x-0.25,\y+0.4) {\small $m$};
\node (bb) at (\x+0.75,\y+0.4) {\small $n$};
\end{tikzpicture}
}.
We will often drop the (co)domain types from the notation and depict $f$ simply by
\resizebox{.07\textwidth}{!}{
\begin{tikzpicture}
\pgfmathsetmacro{\shadow}{0.0};
\pgfmathsetmacro{\przesuniecieX}{0}
\pgfmathsetmacro{\x}{\przesuniecieX};
\pgfmathsetmacro{\y}{0};

\draw[-] (\x-0.25,\y+0.25) -- (\x+0.75,\y+0.25);
%\draw[-] (\x,\y+0.15) -- (\x+0.75,\y+0.15);
\draw[fill = white] (\x+0,\y+0) rectangle (\x+0.5,\y+0.5);
%\draw[fill = black] (\x+0.4,\y+0.4) rectangle (\x+0.5,\y+0.5);

\node (alpha) at (\x+0.25,\y+0.25) {$f$};
%\node (aa) at (\x-0.25,\y+0.4) {\small $m$};
%\node (bb) at (\x+0.75,\y+0.4) {\small $n$};
\end{tikzpicture}
}.   If $f:m\rightdcirc n+p$ and the coproduct codomain needs to be emphasized
by the diagram notation then we depict $f$ by
\resizebox{.07\textwidth}{!}{
\begin{tikzpicture}
\pgfmathsetmacro{\shadow}{0.0};
\pgfmathsetmacro{\przesuniecieX}{0}
\pgfmathsetmacro{\x}{\przesuniecieX};
\pgfmathsetmacro{\y}{0};

\draw[-] (\x,\y+0.3) -- (\x+0.75,\y+0.3);
\draw[-] (\x-0.25,\y+0.25) -- (\x,\y+0.25);
\draw[-] (\x,\y+0.15) -- (\x+0.75,\y+0.15);
\draw[fill = white] (\x+0,\y+0) rectangle (\x+0.5,\y+0.5);
%\draw[fill = black] (\x+0.4,\y+0.4) rectangle (\x+0.5,\y+0.5);

\node (alpha) at (\x+0.25,\y+0.25) {$f$};
%\node (aa) at (\x-0.25,\y+0.4) {\small $m$};
%\node (bb) at (\x+0.75,\y+0.4) {\small $n$};
\end{tikzpicture}
}. This generalizes to $m_1+\cdots+ m_k \rightdcirc n_1+\cdots +n_l$ in an obvious manner.
Whenever $f:m\rightdcirc n$ and $g:n\rightdcirc p$ then the composition $g\cdot f:m\rightdcirc p$
is
\resizebox{.12\textwidth}{!}{
\begin{tikzpicture}
\pgfmathsetmacro{\shadow}{0.0};
\pgfmathsetmacro{\przesuniecieX}{0}
\pgfmathsetmacro{\x}{\przesuniecieX};
\pgfmathsetmacro{\y}{0};

\draw[-] (\x-0.25,\y+0.25) -- (\x+0.75,\y+0.25);
%\draw[-] (\x,\y+0.15) -- (\x+0.75,\y+0.15);
\draw[fill = white] (\x+0,\y+0) rectangle (\x+0.5,\y+0.5);
%\draw[fill = black] (\x+0.4,\y+0.4) rectangle (\x+0.5,\y+0.5);

\node (alpha) at (\x+0.25,\y+0.25) {$f$};

\pgfmathsetmacro{\x}{1};
\draw[-] (\x-0.25,\y+0.25) -- (\x+0.75,\y+0.25);
%\draw[-] (\x,\y+0.15) -- (\x+0.75,\y+0.15);
\draw[fill = white] (\x+0,\y+0) rectangle (\x+0.5,\y+0.5);
%\draw[fill = black] (\x+0.4,\y+0.4) rectangle (\x+0.5,\y+0.5);

\node (alpha) at (\x+0.25,\y+0.25) {$g$};
%\node (aa) at (\x-0.25,\y+0.4) {\small $m$};
%\node (bb) at (\x+0.75,\y+0.4) {\small $n$};
\end{tikzpicture}
}.
Given two maps $f:m_1\rightdcirc n_1$ and $g:m_2\rightdcirc n_2$
the coproduct $f+g:m_1+m_2\rightdcirc n_1+n_2$
is depicted by
\begin{center}
\resizebox{.1\textwidth}{!}{
\begin{tikzpicture}
\pgfmathsetmacro{\shadow}{0.0};
\pgfmathsetmacro{\przesuniecieX}{0}
\pgfmathsetmacro{\x}{\przesuniecieX};
\pgfmathsetmacro{\y}{0};

\draw[-] (\x-0.25,\y+0.25) -- (\x+0.75,\y+0.25);
%\draw[-] (\x,\y+0.15) -- (\x+0.75,\y+0.15);
\draw[fill = white] (\x+0,\y+0) rectangle (\x+0.5,\y+0.5);
%\draw[fill = black] (\x+0.4,\y+0.4) rectangle (\x+0.5,\y+0.5);

\node (alpha) at (\x+0.25,\y+0.25) {$f$};

\pgfmathsetmacro{\y}{-0.6};
\draw[-] (\x-0.25,\y+0.25) -- (\x+0.75,\y+0.25);
%\draw[-] (\x,\y+0.15) -- (\x+0.75,\y+0.15);
\draw[fill = white] (\x+0,\y+0) rectangle (\x+0.5,\y+0.5);
%\draw[fill = black] (\x+0.4,\y+0.4) rectangle (\x+0.5,\y+0.5);

\node (alpha) at (\x+0.25,\y+0.25) {$g$};
%\node (aa) at (\x-0.25,\y+0.4) {\small $m$};
%\node (bb) at (\x+0.75,\y+0.4) {\small $n$};
\end{tikzpicture}
}.
\end{center}
Given any endomorphism $\alpha:n\rightdcirc n$ we depict
the saturated map $\alpha^\ast$
by
\begin{center}
\resizebox{.11\textwidth}{!}{
\begin{tikzpicture}
\pgfmathsetmacro{\shadow}{0.0};
\pgfmathsetmacro{\przesuniecieX}{0}
\pgfmathsetmacro{\x}{\przesuniecieX};
\pgfmathsetmacro{\y}{0};

\draw[-] (\x-0.25,\y+0.25) -- (\x+0.75,\y+0.25);
\draw[fill = white] (\x+0,\y+0) rectangle (\x+0.5,\y+0.5);
\draw[fill = black] (\x+0.4,\y+0.4) rectangle (\x+0.5,\y+0.5);

\node (alpha) at (\x+0.25,\y+0.25) {$\alpha$};
\end{tikzpicture}
}.
\end{center}
%%%%%
% Extended notation
%%%%%
By slightly abusing the notation we extend it to the generalized saturation $(-)^\otimes$ operator and for any
$\alpha:n\rightdcirc n+p$ denote $\alpha^\otimes :n\rightdcirc n+p$ diagrammatically by:
\[
\resizebox{.15\textwidth}{!}{
\begin{tikzpicture}
\pgfmathsetmacro{\przesuniecieX}{0};
\pgfmathsetmacro{\x}{-0.5};
\pgfmathsetmacro{\y}{0};
\pgfmathsetmacro{\shadow}{0.0};

\draw[fill = gray] (\x+0+\shadow,\y+0+\shadow) rectangle
(\x+0.5+\shadow,\y+0.8+\shadow);
\draw[fill = white] (\x+0,\y+0) rectangle (\x+0.5,\y+0.8);

\node (alpha) at (\x+0.25,\y+0.4) {$\alpha$};
\draw[-] (\x-0.5,\y+0.6) -- (\x,\y+0.6) node[pos=.5,above] {\tiny $n$};
\draw[-] (\x+0.5,\y+0.6) -- (\x+1,\y+0.6) node[pos=.5,above] {\tiny $n$};
\draw[-] (\x+0.5,\y+0.1) -- (\x+1,\y+0.1) node[pos=.5,below] {\tiny $p$};

\draw[fill = white] (\x+0,\y+0) rectangle (\x+0.5,\y+0.8);
\draw[fill = black] (\x+0.3,\y+0.6) rectangle (\x+0.5,\y+0.8);
\end{tikzpicture}
}.
\]
We will use a separate notation to denote special morphisms.
The identity map $\mathsf{id}:n\rightdcirc n$ is depicted by
\resizebox{.05\textwidth}{!}{
\begin{tikzpicture}
\pgfmathsetmacro{\przesuniecieX}{0};

\pgfmathsetmacro{\shadow}{0.0};
\pgfmathsetmacro{\x}{\przesuniecieX};

\pgfmathsetmacro{\y}{0};
\draw[-] (\x,\y+0.25) -- (\x+1,\y+0.25);
\draw[-] (\x,\y) -- (\x,\y);
\end{tikzpicture}
},
the maps ${\perp_{0,m}}:0\rightdcirc m$ and ${\perp_{m,0}}:m\rightdcirc 0$ by
\resizebox{.05\textwidth}{!}{
\begin{tikzpicture}
\pgfmathsetmacro{\przesuniecieX}{0};
\pgfmathsetmacro{\shadow}{0.0};
\pgfmathsetmacro{\x}{\przesuniecieX};
\pgfmathsetmacro{\y}{0};
\draw[-] [rounded corners=4pt,-] (\x+0.1,\y+0.25) -- (\x+1,\y+.25);
\draw[fill = white] (\x+0.1,\y+0.25) circle (0.1);
\draw[-] (\x,\y) -- (\x,\y);
%\draw[-] [rounded corners=4pt,-] (\x,\y+0.25)
%--(\x+0.3,\y+0.25)--(\x+0.7,\y-0.45)-- (\x+1,\y-0.45);
%
%\pgfmathsetmacro{\y}{0};
%\draw[-] (\x,\y+0.25) -- (\x+1,\y+0.25);
\end{tikzpicture}
} and
\resizebox{.05\textwidth}{!}{
\begin{tikzpicture}
\pgfmathsetmacro{\przesuniecieX}{0};
\pgfmathsetmacro{\shadow}{0.0};
\pgfmathsetmacro{\x}{\przesuniecieX};
\pgfmathsetmacro{\y}{0};
\draw[-] [rounded corners=4pt,-] (\x+0.1,\y+0.25) -- (\x+1,\y+.25);
\draw[fill = black] (\x+1,\y+0.25) circle (0.1);
\draw[-] (\x,\y) -- (\x,\y);
%\draw[-] [rounded corners=4pt,-] (\x,\y+0.25)
%--(\x+0.3,\y+0.25)--(\x+0.7,\y-0.45)-- (\x+1,\y-0.45);
%
%\pgfmathsetmacro{\y}{0};
%\draw[-] (\x,\y+0.25) -- (\x+1,\y+0.25);
\end{tikzpicture}
} respectively.

Since by Assumption~\ref{condition:4} we have
$\perp_{m,n} = \perp_{0,n} \cdot \perp_{m,0}$ we depict $\perp_{m,n}$ by

\begin{center}
\resizebox{.15\textwidth}{!}{
\begin{tikzpicture}
\pgfmathsetmacro{\przesuniecieX}{0};
\pgfmathsetmacro{\shadow}{0.0};
\pgfmathsetmacro{\x}{\przesuniecieX};
\pgfmathsetmacro{\y}{0};
\draw[-] [rounded corners=4pt,-] (\x+0.3,\y+0.25) -- (\x+1,\y+.25);
\draw[fill = black] (\x+1,\y+0.25) circle (0.1);
\draw[-] [rounded corners=4pt,-] (\x+1.4,\y+0.25) -- (\x+2,\y+.25);
\draw[fill = white] (\x+1.4,\y+0.25) circle (0.1);
\
\draw[-] (\x,\y) -- (\x,\y);
%\draw[-] [rounded corners=4pt,-] (\x,\y+0.25)
%--(\x+0.3,\y+0.25)--(\x+0.7,\y-0.45)-- (\x+1,\y-0.45);
%
%\pgfmathsetmacro{\y}{0};
%\draw[-] (\x,\y+0.25) -- (\x+1,\y+0.25);
\end{tikzpicture}
}.
\end{center}
Additionally, since the map $\mathsf{in}^{m}_{m+n}:m\rightdcirc m+n$ satisfies $\mathsf{in}^{m}_{m+n} = \mathsf{id}_m + \perp_{0,n}$,
its diagrammatic representation is
\begin{center}
\resizebox{.1\textwidth}{!}{
\begin{tikzpicture}
\pgfmathsetmacro{\przesuniecieX}{0};

\pgfmathsetmacro{\shadow}{0.0};
\pgfmathsetmacro{\x}{\przesuniecieX};
\pgfmathsetmacro{\y}{0};
\pgfmathsetmacro{\y}{-0.3};
\draw[-] [rounded corners=4pt,-] (\x+0.1,\y+0.25) -- (\x+1,\y+.25);
\draw[fill = white] (\x+0.1,\y+0.25) circle (0.1);
%\draw[-] [rounded corners=4pt,-] (\x,\y+0.25)
%--(\x+0.3,\y+0.25)--(\x+0.7,\y-0.45)-- (\x+1,\y-0.45);
%
\pgfmathsetmacro{\y}{0};
\draw[-] (\x,\y+0.25) -- (\x+1,\y+0.25);
\end{tikzpicture}
}.
\end{center}
The cotuple string diagram notation has already been
presented in Subsection~\ref{subsection:lawvere_theories}.
However, since the the cotuple
\[[\perp_{m,n}:m\rightdcirc n,f :n\rightdcirc n ]:m+n\rightdcirc n\] satisfies
\[[\perp_{m,n}:m\rightdcirc n,f :n\rightdcirc n ] = \perp_{m,0} + f\]
the following diagram depicts it:
\begin{center}
\resizebox{.1\textwidth}{!}{
\begin{tikzpicture}
\pgfmathsetmacro{\przesuniecieX}{0};

\pgfmathsetmacro{\shadow}{0.0};
\pgfmathsetmacro{\x}{\przesuniecieX};
\pgfmathsetmacro{\y}{0};

\pgfmathsetmacro{\y}{0};
\draw[-] [rounded corners=4pt,-] (\x,\y+0.25) -- (\x+1,\y+.25);
\draw[fill = black] (\x+0.9,\y+0.25) circle (0.1);
%\draw[-] [rounded corners=4pt,-] (\x,\y+0.25)
%--(\x+0.3,\y+0.25)--(\x+0.7,\y-0.45)-- (\x+1,\y-0.45);
%
\pgfmathsetmacro{\y}{-0.3};
\draw[-] (\x,\y+0.25) -- (\x+1,\y+0.25);
\draw[fill = white] (\x+0.3,\y+0) rectangle (\x+0.7,\y+0.4);

\end{tikzpicture}
}.
\end{center}
Moreover, the morphism
$[\alpha,\mathsf{in}^{p}_{n+p}]:n+p\rightdcirc n+p$ satisfies the identity
\[[\alpha,\mathsf{in}^{p}_{n+p}] = (\mathsf{id}_n+[\mathsf{id}_p,\mathsf{id}_p])\cdot (\alpha+\mathsf{id}_p)\]
and its diagrammatic representation is
\begin{center}
\resizebox{.2\textwidth}{!}{
\begin{tikzpicture}
\pgfmathsetmacro{\przesuniecieX}{0};
\pgfmathsetmacro{\x}{-0.5};
\pgfmathsetmacro{\y}{0};
\pgfmathsetmacro{\shadow}{0.0};

\draw[fill = gray] (\x+0+\shadow,\y+0+\shadow) rectangle
(\x+0.5+\shadow,\y+0.8+\shadow);
\draw[fill = white] (\x+0,\y+0) rectangle (\x+0.5,\y+0.8);
%\draw[fill = gray] (\x+-0.7+\shadow,\y+0.2+\shadow) rectangle
%(\x+-0.3+\shadow,\y+0.6+\shadow);
%\draw[fill = white] (\x+-0.7,\y+0.2) rectangle (\x+-0.3,\y+0.6);
%
\node (alpha) at (\x+0.25,\y+0.4) {$\alpha$};
%\node (in) at (\x+-0.5,\y+0.4) {\tiny $\mathsf{in}$};
\draw[-] (\x-0.5,\y+0.6) -- (\x,\y+0.6) node[pos=.5,above] {\tiny $n$};
%\draw[-] (\x-0.5,\y+0.3) -- (\x,\y+0.3) node[pos=.1,below] {\tiny $n-m$};
%\draw[->] (\x-1.3,\y+0.4) -- (\x-0.7,\y+0.4);
\draw[-] (\x+0.5,\y+0.6) -- (\x+1.3,\y+0.6) node[pos=.5,above] {\tiny $n$};
%\draw[rounded corners=4pt,->] (\x+-0.7,\y+-0.2) --
%(\x+1.5,\y+-0.2)--(\x+1.5,\y+0.4)--(\x+2.2,\y+0.4);
%\draw[rounded corners=4pt,-] (\x+-0.3,\y+-0.2) -- (\x+0.5,\y+-0.2) ;
\draw[rounded corners=4pt,-] (\x-0.5,\y+-0.2) -- (\x+1.3,\y+-0.2)
node[pos=.5,below] {\tiny $p$} ;
\draw[rounded corners=4pt,-] (\x+0.5,\y+0.1) -- (\x+0.8,\y+0.1)
--(\x+0.8,\y+-0.2)--(\x+1,\y+-0.2);
%\draw[rounded corners=4pt,-] (\x+0.5,\y+0.3) --
%(\x+1.3,\y+0.3)node[pos=.8,below] {\tiny $n-m$};

%\pgfmathsetmacro{\shadow}{0.0};
%\pgfmathsetmacro{\x}{\przesuniecieX};
%\pgfmathsetmacro{\y}{0};
%
%
%\draw[fill = gray] (\x+0+\shadow,\y+0+\shadow) rectangle
%(\x+0.5+\shadow,\y+0.8+\shadow);
%\draw[fill = white] (\x+0,\y+0) rectangle (\x+0.5,\y+0.8);
%\draw[fill = black] (\x+0.3,\y+0.6) rectangle (\x+0.5,\y+0.8);
%\draw[-] (\x-0.5,\y+0.6) -- (\x,\y+0.6);
%
%\draw[-] (\x-0.4,\y+0.3) -- (\x,\y+0.3);
%
%\draw[-] (\x+0.5,\y+0.3) -- (\x+1,\y+0.3);
%\draw[fill = black] (\x+1,\y+0.3) circle (0.1);
%\draw[rounded corners=4pt,-] (\x-0.4,\y+-0.2) -- (\x+1,\y+-0.2) ;
%\draw[-] (\x+0.5,\y+0.6) -- (\x+1,\y+0.6);
%\draw[fill = black] (\x+1,\y+0.6) circle (0.1);
%\node (alpha) at (\x+0.25,\y+0.4) {$\alpha$};
%\draw[rounded corners=4pt,-] (\x+-0.2,\y+-0.2) -- (\x+0.5,\y+-0.2) ;
%\draw[rounded corners=2pt,-] (\x+0.5,\y+0.1) -- (\x+0.7,\y+0.1)
%--(\x+0.7,\y+-0.2)--(\x+1.2,\y+-0.2);
%%\draw[rounded corners=2pt,-] (\x+1,\y-0.2) -- (\x+1.2,\y-0.2)
%%--(\x+1.2,\y-0.7)--(\x+1.4,\y-0.7);
%%\draw[rounded corners=4pt,-] (\x+0.5,\y+0.2) -- (\x+0.8,\y+0.2) ;
%%
%\draw[fill = white] (\x-0.3,\y+0.3) circle (0.1);
%\draw[fill = white] (\x-0.3,\y-0.2) circle (0.1);

\end{tikzpicture}
}.
\end{center}

We are now ready to list some basic observations and remarks about the diagram calculus introduced
above. First of all note that by the properties of saturation the following diagram (in)equalities hold:
\begin{center}
\resizebox{.6\textwidth}{!}{
\begin{tikzpicture}
\pgfmathsetmacro{\shadow}{0.0};
\pgfmathsetmacro{\przesuniecieX}{0}
\pgfmathsetmacro{\x}{\przesuniecieX};
\pgfmathsetmacro{\y}{0};

\pgfmathsetmacro{\x}{-4};

\draw[-] (\x-0.25,\y+0.25) -- (\x+0.75,\y+0.25);

\node (alpha) at (\x+1,\y+0.25) {$\leq $};

\pgfmathsetmacro{\x}{\x+1.5};

\draw[-] (\x-0.25,\y+0.25) -- (\x+0.75,\y+0.25);
\draw[fill = white] (\x+0,\y+0) rectangle (\x+0.5,\y+0.5);
\draw[fill = black] (\x+0.4,\y+0.4) rectangle (\x+0.5,\y+0.5);

\node (alpha) at (\x+0.25,\y+0.25) {$\alpha$};
\node (alpha) at (\x+1.5,\y+0.25) {and};
\pgfmathsetmacro{\x}{0};

\draw[-] (\x-0.25,\y+0.25) -- (\x+0.75,\y+0.25);
\draw[fill = white] (\x+0,\y+0) rectangle (\x+0.5,\y+0.5);
\draw[fill = black] (\x+0.4,\y+0.4) rectangle (\x+0.5,\y+0.5);

\node (alpha) at (\x+0.25,\y+0.25) {$\alpha$};
\pgfmathsetmacro{\x}{1};

\draw[-] (\x-0.25,\y+0.25) -- (\x+0.75,\y+0.25);
\draw[fill = white] (\x+0,\y+0) rectangle (\x+0.5,\y+0.5);
\draw[fill = black] (\x+0.4,\y+0.4) rectangle (\x+0.5,\y+0.5);

\node (alpha) at (\x+0.25,\y+0.25) {$\alpha$};
\node (alpha) at (\x+1,\y+0.25) {$=$};

\pgfmathsetmacro{\x}{2.5};

\draw[-] (\x-0.25,\y+0.25) -- (\x+0.75,\y+0.25);
\draw[fill = white] (\x+0,\y+0) rectangle (\x+0.5,\y+0.5);
\draw[fill = black] (\x+0.4,\y+0.4) rectangle (\x+0.5,\y+0.5);

\node (alpha) at (\x+0.25,\y+0.25) {$\alpha$};

\end{tikzpicture}
}.
\end{center}
Moreover, since $\perp$ satisfies $f\cdot \perp=\perp$ (Assumption~\ref{condition:4})
for any $f$, we have:
\begin{align}
  \label{equation:composition_with_perp}  f\cdot [\perp,\mathsf{id}] = [f\cdot \perp, f\cdot \mathsf{id}] = [\perp,f],
\end{align}
which diagrammatically is represented in terms of the
following identity:
\begin{center}
\resizebox{.3\textwidth}{!}{
\begin{tikzpicture}
\pgfmathsetmacro{\przesuniecieX}{0};

\pgfmathsetmacro{\shadow}{0.0};
\pgfmathsetmacro{\x}{\przesuniecieX};
\pgfmathsetmacro{\y}{0};
\pgfmathsetmacro{\y}{0};
\draw[-] [rounded corners=4pt,-] (\x,\y+0.25) -- (\x+1,\y+.25);
\draw[fill = black] (\x+0.9,\y+0.25) circle (0.1);
%\draw[-] [rounded corners=4pt,-] (\x,\y+0.25)
%--(\x+0.3,\y+0.25)--(\x+0.7,\y-0.45)-- (\x+1,\y-0.45);
%
\pgfmathsetmacro{\y}{-0.3};
\draw[-] (\x,\y+0.2) -- (\x+1,\y+0.2);
\draw[fill = white] (\x+0.3,\y+0) rectangle (\x+0.7,\y+0.4);

\node (eq) at (\x+1.3,\y+0.25) {\Large $=$};

\pgfmathsetmacro{\przesuniecieX}{\przesuniecieX+1.6};
\pgfmathsetmacro{\x}{\przesuniecieX};

\pgfmathsetmacro{\y}{0};
\draw[-] [rounded corners=4pt,-] (\x,\y+0.25) -- (\x+1,\y+.25);
\draw[fill = black] (\x+0.9,\y+0.25) circle (0.1);
%\draw[-] [rounded corners=4pt,-] (\x,\y+0.25)
%--(\x+0.3,\y+0.25)--(\x+0.7,\y-0.45)-- (\x+1,\y-0.45);
%
\pgfmathsetmacro{\y}{-0.3};
\draw[-] (\x,\y+0.2) -- (\x+2,\y+0.2);
\draw[fill = white] (\x+1.3,\y+0) rectangle (\x+1.7,\y+0.4);

%\node (eq) at (\x+1.2,\y+0.25) {\Large $=$};
\end{tikzpicture}
}.
\end{center}
In the above, the right hand side of the equality, namely
\resizebox{.07\textwidth}{!}{
\begin{tikzpicture}
\pgfmathsetmacro{\przesuniecieX}{0};

\pgfmathsetmacro{\shadow}{0.0};
\pgfmathsetmacro{\x}{\przesuniecieX};
\pgfmathsetmacro{\y}{0};
\draw[-] [rounded corners=4pt,-] (\x,\y+0.25) -- (\x+1,\y+.25);
\draw[fill = black] (\x+0.9,\y+0.25) circle (0.1);
%\draw[-] [rounded corners=4pt,-] (\x,\y+0.25)
%--(\x+0.3,\y+0.25)--(\x+0.7,\y-0.45)-- (\x+1,\y-0.45);
%
\pgfmathsetmacro{\y}{-0.3};
\draw[-] (\x,\y+0.2) -- (\x+2,\y+0.2);
\draw[fill = white] (\x+1.3,\y+0) rectangle (\x+1.7,\y+0.4);

%\node (eq) at (\x+1.2,\y+0.25) {\Large $=$};
\end{tikzpicture}
},
is the composition of
\resizebox{.05\textwidth}{!}{
\begin{tikzpicture}
\pgfmathsetmacro{\przesuniecieX}{0};

\pgfmathsetmacro{\shadow}{0.0};
\pgfmathsetmacro{\x}{\przesuniecieX};
\pgfmathsetmacro{\y}{0};
\draw[-] [rounded corners=4pt,-] (\x,\y+0.25) -- (\x+1,\y+.25);
\draw[fill = black] (\x+0.9,\y+0.25) circle (0.1);
%\draw[-] [rounded corners=4pt,-] (\x,\y+0.25)
%--(\x+0.3,\y+0.25)--(\x+0.7,\y-0.45)-- (\x+1,\y-0.45);
%
\pgfmathsetmacro{\y}{-0.3};
\draw[-] (\x,\y+0.2) -- (\x+1,\y+0.2);
%\draw[fill = white] (\x+1.3,\y+0) rectangle (\x+1.7,\y+0.4);

%\node (eq) at (\x+1.2,\y+0.25) {\Large $=$};
\end{tikzpicture}
} and
\resizebox{.06\textwidth}{!}{
\begin{tikzpicture}
\pgfmathsetmacro{\przesuniecieX}{0};

\pgfmathsetmacro{\shadow}{0.0};
\pgfmathsetmacro{\x}{\przesuniecieX};
\pgfmathsetmacro{\y}{0};
%\draw[-] [rounded corners=4pt,-] (\x,\y+0.25) -- (\x+1,\y+.25);
%\draw[fill = black] (\x+0.9,\y+0.25) circle (0.1);
%%\draw[-] [rounded corners=4pt,-] (\x,\y+0.25)
%--(\x+0.3,\y+0.25)--(\x+0.7,\y-0.45)-- (\x+1,\y-0.45);
%
\pgfmathsetmacro{\y}{-0.3};
\draw[-] (\x+1,\y+0.2) -- (\x+2,\y+0.2);
\draw[fill = white] (\x+1.3,\y+0) rectangle (\x+1.7,\y+0.4);

%\node (eq) at (\x+1.2,\y+0.25) {\Large $=$};
\end{tikzpicture}
}.

\subsubsection*{Generalized star pairing identity}
Here, we present the so-called generalized star pairing identity described in any Lawvere theory
equipped with an operator $(-)^\otimes$ assigning to each morphism $f:n\rightdcirc n+p$
a morphism $f^\otimes : n\rightdcirc n+p$~\cite{esikhajgato2009:ai}.
This identity will hold in our setting and will be used in
the proof of Theorem~\ref{theorem:kleene_regular} and lemmas that precede it.

For $f:n\rightdcirc n+m+p$ and $g:m\rightdcirc n+m+p$ the
\emph{generalized star pairing identity}
is~\cite{esikhajgato2009:ai}:
\begin{align}
  [f,g]^\otimes =
  [ [\mathsf{in}^{n}_{n+m+p},(\pi^{-1}+\mathsf{id}_p)\cdot
  k^\otimes,\mathsf{in}^{p}_{n+m+p} ] \cdot f^\otimes,
  (\pi^{-1}+\mathsf{id}_p) \cdot k^\otimes], \tag{GSPI} \label{equation:gspi}
\end{align}
where $\pi:m+n\rightdcirc n+m$ is given by $\pi \defeq [\mathsf{in}^{m}_{n+m}, \mathsf{in}^{n}_{n+m}]$ and
\[
  k \defeq [(\pi+\mathsf{id}_p)\cdot f^\otimes,  [\mathsf{in}^m_{m+n+p},\mathsf{in}^p_{m+n+p}] ] \cdot g.
\]
The generalized star pairing identity is depicted by the string diagram:
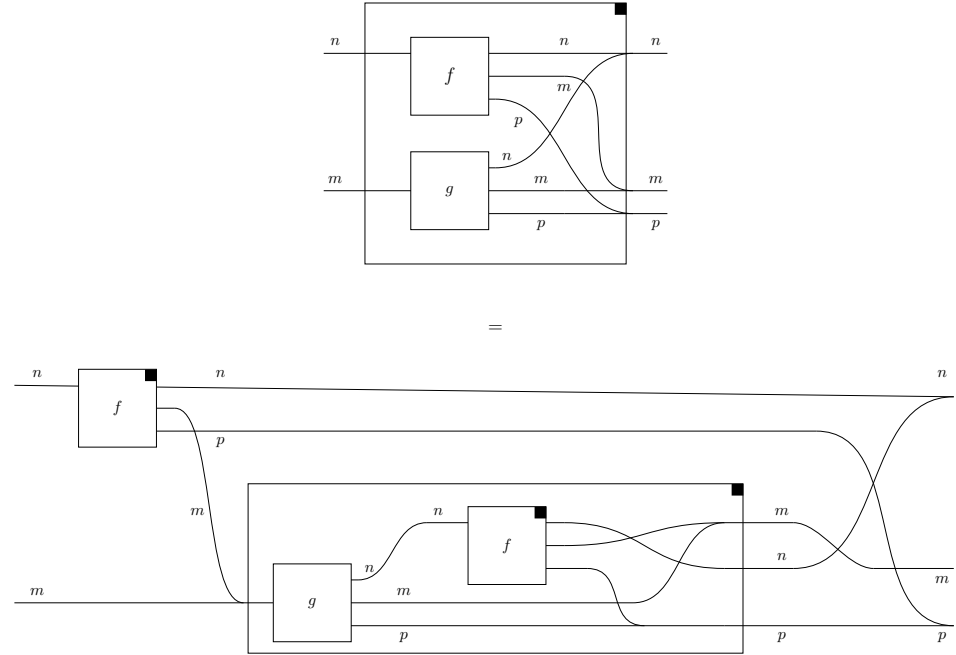
\begin{center}
\resizebox{\textwidth}{!}{
\begin{tikzpicture}
	\begin{pgfonlayer}{nodelayer}
		\node [style=none] (0) at (-12.75, 10.75) {};
		\node [style=none] (1) at (-6, 10.75) {};
		\node [style=new style 0] (4) at (-10, 10.25) {\Large $f$};
		\node [style=none] (5) at (-9, 9.75) {};
		\node [style=none] (6) at (-10, 9.75) {};
		\node [style=none] (8) at (-7.5, 11) {{$n$}};
		\node [style=none] (10) at (-12.5, 11) {{$n$}};
		\node [style=none] (12) at (-8.5, 9.25) {{$p$}};
		\node [style=none] (13) at (-12.75, 7.75) {};
		\node [style=none] (14) at (-7.5, 7.75) {};
		\node [style=new style 0] (15) at (-10, 7.75) {$g$};
		\node [style=none] (16) at (-7.5, 7.25) {};
		\node [style=none] (17) at (-10, 7.25) {};
		\node [style=none] (18) at (-8.75, 8.5) {{$n$}};
		\node [style=none] (20) at (-8, 7) {{$p$}};
		\node [style=none] (22) at (-6, 7.75) {};
		\node [style=none] (24) at (-6, 7.25) {};
		\node [style=none] (26) at (-10, 10.25) {};
		\node [style=none] (27) at (-7.5, 10.25) {};
		\node [style=none] (28) at (-7.5, 10) {{$m$}};
		\node [style=none] (30) at (-9, 8.25) {};
		\node [style=none] (31) at (-12.5, 8) {{$m$}};
		\node [style=none] (32) at (-8, 8) {{$m$}};
		\node [style=empty square] (33) at (-9, 9) {};
		\node [style=none] (34) at (-5.25, 10.75) {};
		\node [style=none] (35) at (-5.25, 7.75) {};
		\node [style=none] (36) at (-5.25, 7.25) {};
		\node [style=none] (37) at (-5.5, 11) {{$n$}};
		\node [style=none] (38) at (-5.5, 8) {{$m$}};
		\node [style=none] (39) at (-5.5, 7) {{$p$}};
		\node [style=none] (43) at (-9, 4.75) {\Large $=$};
		\node [style=none] (44) at (-19.5, 3.5) {};
		\node [style=none] (45) at (1, 3.25) {};
		\node [style=new style 0] (46) at (-17.25, 3) {$f$};
		\node [style=none] (47) at (-2, 2.5) {};
		\node [style=none] (48) at (-17.25, 2.5) {};
		\node [style=none] (49) at (-19, 3.75) {{$n$}};
		\node [style=none] (50) at (-17.25, 3) {};
		\node [style=none] (51) at (-16, 3) {};
		\node [style=none] (53) at (-10.5, 0.5) {};
		\node [style=none] (54) at (-19.5, -1.25) {};
		\node [style=none] (55) at (-6, -1.25) {};
		\node [style=new style 0] (56) at (-13, -1.25) {$g$};
		\node [style=none] (57) at (-4, -1.75) {};
		\node [style=none] (58) at (-13, -1.75) {};
		\node [style=none] (59) at (-11.75, -0.5) {{$n$}};
		\node [style=none] (60) at (-11, -2) {{$p$}};
		\node [style=none] (61) at (-12, -0.75) {};
		\node [style=none] (62) at (-19, -1) {{$m$}};
		\node [style=none] (63) at (-11, -1) {{$m$}};
		\node [style=none] (64) at (-10.5, 0.5) {};
		\node [style=none] (65) at (-7.5, 0.5) {};
		\node [style=new style 0] (66) at (-8.75, 0) {$f$};
		\node [style=none] (67) at (-7, -0.5) {};
		\node [style=none] (68) at (-8.75, -0.5) {};
		\node [style=none] (69) at (-10.25, 0.75) {{$n$}};
		\node [style=none] (70) at (-8.75, 0) {};
		\node [style=none] (71) at (-7.5, 0) {};
		\node [style=none] (73) at (-4, -0.5) {};
		\node [style=none] (74) at (-4, 0.5) {};
		\node [style=none] (75) at (-15, 3.75) {{$n$}};
		\node [style=none] (76) at (-2.75, -0.25) {{$n$}};
		\node [style=none] (77) at (-2.75, 0.75) {{$m$}};
		\node [style=none] (78) at (-5.75, -1.75) {};
		\node [style=none] (79) at (-2.75, -2) {{$p$}};
		\node [style=none] (80) at (1, -1.75) {};
		\node [style=none] (81) at (-2.5, -0.5) {};
		\node [style=none] (82) at (-2.5, 0.5) {};
		\node [style=new style 1] (83) at (-9, -0.5) {};
		\node [style=none] (86) at (-0.75, -0.5) {};
		\node [style=none] (87) at (1, -0.5) {};
		\node [style=none] (88) at (0.75, 3.75) {{$n$}};
		\node [style=none] (89) at (0.75, -0.75) {{$m$}};
		\node [style=none] (90) at (0.75, -2) {{$p$}};
		\node [style=none] (91) at (-15.5, 0.75) {{$m$}};
		\node [style=none] (92) at (-15, 2.25) {{$p$}};
		\node [style=none] (93) at (-14.5, -1.25) {};
		\node [style={czarny ma?y kwadrat}] (94) at (-6.27, 11.72) {};
		\node [style={czarny ma?y kwadrat}] (95) at (-16.52, 3.72) {};
		\node [style={czarny ma?y kwadrat}] (96) at (-3.72, 1.22) {};
		\node [style={czarny ma?y kwadrat}] (97) at (-8.02, 0.72) {};
		\node [style=none] (98) at (-13, -0.75) {};
		\node [style=none] (99) at (-10, 8.25) {};
	\end{pgfonlayer}
	\begin{pgfonlayer}{edgelayer}
		\draw (0.center) to (1.center);
		\draw (6.center) to (5.center);
		\draw (13.center) to (14.center);
		\draw (17.center) to (16.center);
		\draw (22.center) to (14.center);
		\draw (16.center) to (24.center);
		\draw [in=180, out=0] (5.center) to (24.center);
		\draw (26.center) to (27.center);
		\draw [in=180, out=0] (30.center) to (1.center);
		\draw [in=-180, out=0, looseness=1.25] (27.center) to (22.center);
		\draw (1.center) to (34.center);
		\draw (22.center) to (35.center);
		\draw (24.center) to (36.center);
		\draw (44.center) to (45.center);
		\draw (48.center) to (47.center);
		\draw (50.center) to (51.center);
		\draw (54.center) to (55.center);
		\draw (58.center) to (57.center);
		\draw [in=180, out=0] (61.center) to (53.center);
		\draw (64.center) to (65.center);
		\draw (68.center) to (67.center);
		\draw (70.center) to (71.center);
		\draw [in=-180, out=0, looseness=1.25] (65.center) to (73.center);
		\draw [in=180, out=0, looseness=1.25] (71.center) to (74.center);
		\draw [in=-180, out=0, looseness=1.25] (67.center) to (78.center);
		\draw [in=180, out=0] (55.center) to (74.center);
		\draw (74.center) to (82.center);
		\draw (73.center) to (81.center);
		\draw (57.center) to (80.center);
		\draw [in=-180, out=0, looseness=0.75] (82.center) to (86.center);
		\draw (86.center) to (87.center);
		\draw [in=180, out=0] (47.center) to (80.center);
		\draw [in=180, out=0] (81.center) to (45.center);
		\draw [in=180, out=0, looseness=0.50] (51.center) to (93.center);
		\draw (99.center) to (30.center);
		\draw (98.center) to (61.center);
	\end{pgfonlayer}
\end{tikzpicture}
}
\end{center}

\subsubsection*{Rational morphisms.}
%
%
%%%%%%%%%%%%%%%%%%%%%%%%%%%%%%%%%%%%%%%
Let $\mathfrak{Rat}(A)$ be the theory defined in Theorem~\ref{theorem:kleene_regular}.
Since $(-)^\ast$ (with its extension $(-)^\otimes$ given in~\ref{equation:extended_saturation})
is defined in $\mathfrak{Rat}(\mathcal{A})$ in terms of a least
fixpoint operator in a more general setting of the Kleisli category for $T$
which satisfies Assumption~\ref{assumptions:kleisli_cat}, we have the following~\cite{DBLP:journals/tcs/Esik97,esikhajgato2009:ai}
\footnote{
  Here, we sketch a proof of
  Lemma~\ref{lemma:rat_star_pairing_identity}. The  generalized star pairing identity is equivalent to
  the so-called \emph{pairing identity} given in \emph{dagger theories} which are also
  \emph{grove theories},
  where the dagger operator is compatible with the star operator~\cite{esikhajgato2009:ai}.
  The pairing identity for dagger theories holds in any \emph{$\omega$-continuous theory}~\cite{DBLP:journals/tcs/Esik97}. Our theory $\mathfrak{Rat}(\mathcal{A})$ satisfies
  the assumptions of an $\omega$-continuous grove theory where the dagger operator is given
  by $(\alpha:n\rightdcirc n+p)  \mapsto  (\mu x.x\cdot \alpha : n\rightdcirc p)$ and the
  extended saturation operator~\ref{equation:extended_saturation} is compatible
  with it. This completes the proof.
  We skip the definitions of
  the theories and new notions introduced in the footnote
  and refer the reader to \emph{loc.\ cit.} for details.
}
(see also~\cite{bloomesik:93,esik2011,esik2013}):

\begin{lem}\label{lemma:rat_star_pairing_identity}
  The theory $\mathfrak{Rat}(A)$ satisfies the generalized star pairing identity for the operator $(-)^\otimes$.
\end{lem}

\subsubsection*{Regular morphisms and normal form.}

Note that any regular map is a morphism in $\mathfrak{Rat}(A)$.
In particular, this means that
 % by Lemma~\ref{lemma:rational_identity}
 % all regular maps satisfy the identity
 % (\ref{equation:identity_perp}) and that
regular morphisms satisfy the generalized
star pairing identity.

Let us first introduce a new notion.

\begin{defi}\label{definition:normal_form}
A morphism $r:m\rightdcirc p$
is said to be in \emph{normal form} if
\[
   r=[\perp_{n,p}, \mathsf{id}_p]\cdot \alpha^\otimes\cdot \mathsf{in}_{n}^m
\]
for  some  $\alpha:n\rightdcirc n+p\in [\mathcal{A}]$  and $m\leq n$.
\end{defi}

\noindent Let the family of all maps in normal form be denoted by $NF(\mathcal{A})$.
The map $\alpha$ and $r$ in $NF(\mathcal{A})$ from Definition~\ref{definition:normal_form} are depicted
by the string diagrams below:
%%%%%%%%%%%%%%%%%%%%%%
% diagram
%%%%%%%%%%%%%%%%%%%%%%
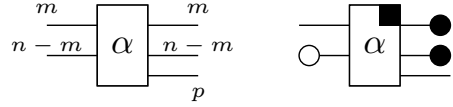
\begin{center}
\resizebox{.4\textwidth}{!}{
\begin{tikzpicture}
%%% set move variables
\pgfmathsetmacro{\przesuniecieX}{0};
\pgfmathsetmacro{\x}{-0.5};
\pgfmathsetmacro{\y}{0};
\pgfmathsetmacro{\shadow}{0.0};
%%% picture 1
\draw[fill = gray] (\x+0+\shadow,\y+0+\shadow) rectangle
(\x+0.5+\shadow,\y+0.8+\shadow);
\draw[fill = white] (\x+0,\y+0) rectangle (\x+0.5,\y+0.8);
%\draw[fill = gray] (\x+-0.7+\shadow,\y+0.2+\shadow) rectangle
%(\x+-0.3+\shadow,\y+0.6+\shadow);
%\draw[fill = white] (\x+-0.7,\y+0.2) rectangle (\x+-0.3,\y+0.6);
%
\node (alpha) at (\x+0.25,\y+0.4) {$\alpha$};
%\node (in) at (\x+-0.5,\y+0.4) {\tiny $\mathsf{in}$};
\draw[-] (\x-0.5,\y+0.6) -- (\x,\y+0.6) node[at start,above] {\tiny $m$};
\draw[-] (\x-0.5,\y+0.3) -- (\x,\y+0.3) node[at start,above,yshift=-3] {\tiny $n-m$};
%\draw[->] (\x-1.3,\y+0.4) -- (\x-0.7,\y+0.4);
\draw[-] (\x+0.5,\y+0.6) -- (\x+1,\y+0.6) node[at end,above] {\tiny $m$};
%\draw[rounded corners=4pt,->] (\x+-0.7,\y+-0.2) --
%(\x+1.5,\y+-0.2)--(\x+1.5,\y+0.4)--(\x+2.2,\y+0.4);
%\draw[rounded corners=4pt,-] (\x+-0.3,\y+-0.2) -- (\x+0.5,\y+-0.2) ;
%\draw[rounded corners=4pt,-] (\x-0.5,\y+-0.2) -- (\x+1.3,\y+-0.2)
%node[pos=.5,below] {\tiny $p$} ;
\draw[rounded corners=4pt,-] (\x+0.5,\y+0.1) -- (\x+1,\y+0.1) node[at end,below] {\tiny $p$};
%--(\x+0.8,\y+-0.2)--(\x+1,\y+-0.2);
\draw[-] (\x+0.5,\y+0.3) -- (\x+1,\y+0.3) node[at end,above,yshift=-3] {\tiny $n-m$};

%%%%%%%%%%%%%%%%%%%%%%%%%%
% variables for picture 2
%%%%%%%%%%%%%%%%%%%%%%%%%%
\pgfmathsetmacro{\przesuniecieX}{2};
\pgfmathsetmacro{\shadow}{0.0};
\pgfmathsetmacro{\x}{\przesuniecieX};
\pgfmathsetmacro{\y}{0};
%%%%%%%%%%%%%%%%%%%%%
% picture 2
%%%%%%%%%%%%%%%%%%%%%
\draw[fill = gray] (\x+0+\shadow,\y+0+\shadow) rectangle
(\x+0.5+\shadow,\y+0.8+\shadow);
\draw[fill = white] (\x+0,\y+0) rectangle (\x+0.5,\y+0.8);

\node (alpha) at (\x+0.25,\y+0.4) {$\alpha$};
\draw[-] (\x-0.5,\y+0.6) -- (\x,\y+0.6);
\draw[-] (\x-0.5,\y+0.3) -- (\x,\y+0.3);
\draw[-] (\x+0.5,\y+0.6) -- (\x+1,\y+0.6);
\draw[rounded corners=4pt,-] (\x+0.5,\y+0.1) -- (\x+1,\y+0.1);
\draw[-] (\x+0.5,\y+0.3) -- (\x+1,\y+0.3);

\draw[fill = black] (\x+0.3,\y+0.6) rectangle (\x+0.5,\y+0.8);
\draw[fill = black] (\x+0.9,\y+0.3) circle (0.1);
\draw[fill = black] (\x+0.9,\y+0.6) circle (0.1);
\draw[fill = white] (\x-0.4,\y+0.3) circle (0.1);
\end{tikzpicture}
}
%\caption{\tiny The map $[\alpha,\mathsf{in}^p]$ and $r$ in NF}
\end{center}
The right-hand-side diagram is a correct representation of $r$
as it is the result of the composition of three maps:
%%%%%%%%%%%%%%%%%%%%%%%%%%%%%%%%%%%%%%%%%%%%%%%%%%%%%%%%%%%
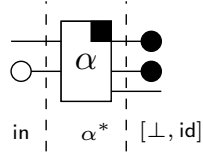
\begin{center}
\resizebox{.25\textwidth}{!}{
\begin{tikzpicture}
\pgfmathsetmacro{\przesuniecieX}{0};
\pgfmathsetmacro{\x}{-0.5};
\pgfmathsetmacro{\y}{0};
\pgfmathsetmacro{\shadow}{0.0};

\pgfmathsetmacro{\przesuniecieX}{2};

\pgfmathsetmacro{\shadow}{0.0};
\pgfmathsetmacro{\x}{\przesuniecieX};
\pgfmathsetmacro{\y}{0};

\draw[fill = gray] (\x+0+\shadow,\y+0+\shadow) rectangle
(\x+0.5+\shadow,\y+0.8+\shadow);
\draw[fill = white] (\x+0,\y+0) rectangle (\x+0.5,\y+0.8);
\draw[fill = black] (\x+0.3,\y+0.6) rectangle (\x+0.5,\y+0.8);
\draw[-] (\x-0.5,\y+0.6) -- (\x,\y+0.6);

\draw[-] (\x-0.4,\y+0.3) -- (\x,\y+0.3);

\draw[-] (\x+0.5,\y+0.3) -- (\x+1,\y+0.3);
\draw[fill = black] (\x+0.9,\y+0.3) circle (0.1);
%\draw[rounded corners=4pt,-] (\x-0.4,\y+-0.2) -- (\x+1,\y+-0.2) ;
\draw[-] (\x+0.5,\y+0.6) -- (\x+1,\y+0.6);
\draw[fill = black] (\x+0.9,\y+0.6) circle (0.1);
\node (alpha) at (\x+0.25,\y+0.4) {$\alpha$};
\draw[rounded corners=2pt,-] (\x+0.5,\y+0.1) -- (\x+1,\y+0.1);
%--(\x+0.7,\y+-0.2)--(\x+1.2,\y+-0.2);
%\draw[rounded corners=2pt,-] (\x+1,\y-0.2) -- (\x+1.2,\y-0.2)
%--(\x+1.2,\y-0.7)--(\x+1.4,\y-0.7);
%\draw[rounded corners=4pt,-] (\x+0.5,\y+0.2) -- (\x+0.8,\y+0.2) ;
%
\draw[fill = white] (\x-0.4,\y+0.3) circle (0.1);

\node (alpha) at (\x-0.4,\y-0.3) {\tiny $\mathsf{in}$};
\draw[dashed, rounded corners=4pt,-] (\x-0.15,\y-0.5) -- (\x-0.15,\y+1) ;
\node (alpha) at (\x+0.35,\y-0.3) {\tiny $\alpha^\otimes$};
\draw[dashed, rounded corners=4pt,-] (\x+0.65,\y-0.5) -- (\x+0.65,\y+1) ;
\node (alpha) at (\x+1.7,\y-0.3) {\tiny $[\perp,\mathsf{id}]=\perp+\mathsf{id}$};
\end{tikzpicture}
}
%\caption{\tiny The map $[\alpha,\mathsf{in}^p]$ and $r$ in NF}
\end{center}
%%%%%%%%%%%%%%%%%%%%%%%%%%%%%%%%%%%%%%%%
%

It follows straight by Definition~\ref{definition:normal_form} that every map in normal form
is a regular morphism. Note that the family of maps in normal form contains all base maps and all
morphisms from $[\mathcal{A}]$. Additionally, it is closed under cotupling $[-,-]$.
Moreover, the following statement holds.

\begin{lem}\label{lemma:normal_closed_under}
  The family $NF(\mathcal{A})$ is closed under the composition $\cdot$,
  finite suprema $\vee$ and saturation $(-)^\ast$.
\end{lem}
\begin{proof}
  The proof is divided into three parts.

  \noindent \textbf{Part 1.} Here, we show that the family $NF(\mathcal{A})$ is closed under
  the composition. Take
  $r_1 = [\perp_{n_1,m_2}, \mathsf{id}_{m_2}]\cdot \alpha^\otimes \cdot \mathsf{in}^{m_1}_{n_1}$
  for $m_1\leq n_1$ and
  $r_2 = [\perp_{n_2,m_3}, \mathsf{id}_{m_3}]\cdot \beta^\otimes \cdot \mathsf{in}^{m_2}_{n_2}$
	for $m_2\leq n_2$,
  where $\alpha: n_1 \rightdcirc n_1+m_2$ and $\beta: n_2 \rightdcirc n_2 + m_3$.
  Consider morphisms $f:n_1\rightdcirc n_1+n_2+m_3$ and $g: n_2\rightdcirc n_1+n_2+m_3$
  defined by
  \[
  f \defeq \alpha  + \perp_{0, n_2-m_2} + \perp_{0,m_3}\text{ and } g \defeq \perp_{0,n_1}+\beta
  \]
  and represented in terms of their string diagrams respectively as follows:
  \begin{center}
  \resizebox{.4\textwidth}{!}
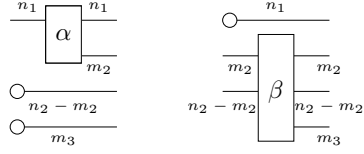
{
  \begin{tikzpicture}
  \pgfmathsetmacro{\przesuniecieX}{0};
  \pgfmathsetmacro{\x}{-0.5};
  \pgfmathsetmacro{\y}{0};
  \pgfmathsetmacro{\shadow}{0.0};

  \draw[fill = gray] (\x+0+\shadow,\y+0+\shadow) rectangle
  (\x+0.5+\shadow,\y+0.8+\shadow);
  \draw[fill = white] (\x+0,\y+0) rectangle (\x+0.5,\y+0.8);

  \draw[-] (\x-0.5,\y+0.6) -- (\x,\y+0.6) node[pos=.5,above] {\tiny $n_1$};
  \draw[-] (\x+0.5,\y+0.6) -- (\x+1,\y+0.6) node[pos=.5,above] {\tiny $n_1$};
  \draw[-] (\x+0.5,\y+0.1) -- (\x+1,\y+0.1) node[pos=.5,below] {\tiny $m_2$};

  \draw[fill = white] (\x+0,\y+0) rectangle (\x+0.5,\y+0.8);
  \draw[-] (\x-0.5,\y-0.4) -- (\x+1,\y-0.4) node[pos=.5,below] {\tiny $n_2-m_2$};
  \draw[-] (\x-0.5,\y-0.9) -- (\x+1,\y-0.9) node[pos=.5,below] {\tiny $m_3$};

  \draw[fill = white] (\x-0.4,\y-0.4) circle (0.1);
  \draw[fill = white] (\x-0.4,\y-0.9) circle (0.1);
  \node (alpha) at (\x+0.25,\y+0.4) {$\alpha$};
  \end{tikzpicture}
  \qquad
%%%%%%%%%%%%%%%%%%%%%%%%%%%%%%%%%%%%%%%%%%%%%%%%%%%%%%%%%%%
  \begin{tikzpicture}
  \pgfmathsetmacro{\przesuniecieX}{0};
  \pgfmathsetmacro{\x}{-0.5};
  \pgfmathsetmacro{\y}{0};
  \pgfmathsetmacro{\shadow}{0.0};

  \draw[-] (\x-0.5,\y+0.6) -- (\x+1,\y+0.6) node[pos=.5,above] {\tiny $n_1$};
  \draw[fill = white] (\x-0.4,\y+0.6) circle (0.1);
  \draw[-] (\x-0.5,\y+0.1) -- (\x,\y+0.1) node[pos=.5,below] {\tiny $m_2$};
  \draw[-] (\x,\y+0.1) -- (\x+1,\y+0.1) node[pos=.5,at end, below] {\tiny $m_2$};
  \draw[-] (\x,\y-0.4) -- (\x+1,\y-0.4) node[pos=.6,at end, below] {\tiny $n_2-m_2$};

  \draw[-] (\x-0.5,\y-0.4) -- (\x,\y-0.4) node[pos=.5,at start, below] {\tiny $n_2 - m_2$};
  \draw[-] (\x,\y-0.9) -- (\x+1,\y-0.9) node[pos=.6,at end, below] {\tiny $m_3$};

  \draw[fill = white] (\x+0,\y-1.1) rectangle (\x+0.5,\y+0.4);
  \node (alpha) at (\x+0.25,\y-0.4) {$\beta$};
  \end{tikzpicture}
  }
  \end{center}
  Let $\gamma = [f,g]$. Then $\gamma : n_1+n_2 \rightdcirc n_1+n_2+m_3$ is in $[\mathcal{A}]$
  and by a careful analysis of the generalized
  star pairing identity it follows that the morphism
  $(\pi^{-1}+\mathsf{id})\cdot k^\otimes$ in~\ref{equation:gspi} is, in our case, given by $\perp_{0,n_1} + \beta^\otimes$ which is diagrammatically
  captured by:
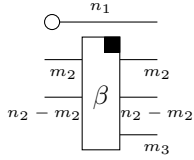
  \begin{center}
  \begin{tikzpicture}
  \pgfmathsetmacro{\przesuniecieX}{0};
  \pgfmathsetmacro{\x}{-0.5};
  \pgfmathsetmacro{\y}{0};
  \pgfmathsetmacro{\shadow}{0.0};

  \draw[-] (\x-0.5,\y+0.6) -- (\x+1,\y+0.6) node[pos=.5,above] {\tiny $n_1$};
  \draw[fill = white] (\x-0.4,\y+0.6) circle (0.1);
  \draw[-] (\x-0.5,\y+0.1) -- (\x,\y+0.1) node[pos=.5,below] {\tiny $m_2$};
  \draw[-] (\x,\y+0.1) -- (\x+1,\y+0.1) node[pos=.5,at end, below] {\tiny $m_2$};
  \draw[-] (\x,\y-0.4) -- (\x+1,\y-0.4) node[pos=.6,at end, below] {\tiny $n_2-m_2$};

  \draw[-] (\x-0.5,\y-0.4) -- (\x,\y-0.4) node[pos=.5,at start, below] {\tiny $n_2 - m_2$};
  \draw[-] (\x,\y-0.9) -- (\x+1,\y-0.9) node[pos=.6,at end, below] {\tiny $m_3$};

  \draw[fill = white] (\x+0,\y-1.1) rectangle (\x+0.5,\y+0.4);
  \draw[fill = black] (\x+0.3,\y+0.2  ) rectangle (\x+0.5,\y+0.4);
  \node (alpha) at (\x+0.25,\y-0.4) {$\beta$};
  \end{tikzpicture}
\end{center}
%Lemma~\ref{lemma:rational_identity} and
Hence, by~\ref{equation:gspi} and~\ref{equation:composition_with_perp} we get:
  \[
    r_2\cdot r_1 = [\perp, \mathsf{id}_{m_3}]\cdot \gamma^\otimes \cdot \mathsf{in}^{m_1}.
  \]

  \noindent \textbf{Part 2.} Here, we show that given two maps
  $r_1 = [\perp_{n_1,m_2}, \mathsf{id}_{m_2}]\cdot \alpha^\otimes \cdot \mathsf{in}^{m_1}_{n_1+m_2}$
  and
  $r_2 = [\perp_{n_2,m_2}, \mathsf{id}_{m_2}]\cdot \beta^\otimes \cdot \mathsf{in}^{m_1}_{n_2+m_2}$
  for $\alpha: n_1\rightdcirc n_1 + m_2$ and $\beta: n_2 \rightdcirc n_2 + m_2$ their join $r_1\vee r_2$
  is in normal form. Let $f: m_1 \rightdcirc m_1 + (n_1+n_2)+m_3$ be defined in terms of supremum of the
  following two morphisms:
  \begin{align*}
   & (\perp_{0,m_1}+\mathsf{id}_{m_1+(n_1-m_1)+n_2+m_2})\cdot (\mathsf{id}_{m_1} + \perp_{0,(n_1-m_1)+n_2+m_2}) = \\
   & ( \perp_{0,m_1}+ \mathsf{id}_{m_1} +\mathsf{id}_{n_1-m_1} + \mathsf{id}_{n_2} + \mathsf{id}_{m_2} )
   \cdot (\mathsf{id}_{m_1}+\perp_{0,n_1-m_1}+\perp_{0,n_2}+\perp_{0,m_2})
\end{align*}
and
\begin{align*}
    & (\perp_{0,m_1}+\mathsf{id}_{m_1+(n_1-m_1)+n_2+m_2})\cdot \sigma \cdot  (\mathsf{id}_{m_1} + \perp_{0,(n_1-m_1)+n_2+m_2}) = \\
    & ( \perp_{0,m_1}+ \mathsf{id}_{m_1} +\mathsf{id}_{n_1-m_1} + \mathsf{id}_{n_2} + \mathsf{id}_{m_2} )
    \cdot \sigma \cdot (\mathsf{id}_{m_1}+\perp_{0,n_1-m_1}+\perp_{0,n_2}+\perp_{0,m_2}),
\end{align*}
where the endomorphism
\[\sigma : m_1 + (n_1-m_1) + m_1 + (n_2-m_1) + m_2 \rightdcirc m_1 + (n_1-m_1) + m_1 + (n_2-m_1) + m_2\]
permutes the first and third
component of the coproduct and is the identity everywhere else.
  Now, let $g:  n_1+n_2 \rightdcirc m_1+(n_1+n_2)+m_2$ be given by:
  \[
  g\defeq \sigma'\cdot (\perp_{0,m_1} + \alpha + \beta),
  \]
  where the endomorphism
  $\sigma':m_1+n_1+m_2+n_2+m_2 \rightdcirc m_1+n_1+m_2+n_2+m_2$ injects the third component of the coproduct into
  the last component and is the identity everywhere else.
  The morphisms $f$ and $g$ are depicted in terms of their string diagrams
  respectively as follows:
  \begin{center}
  \resizebox{.7\textwidth}{!}
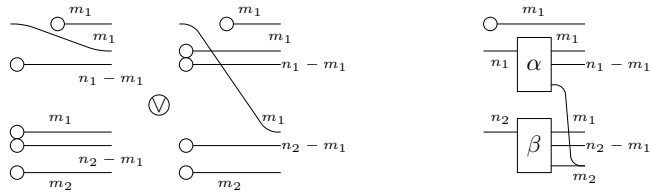
{
  \begin{tikzpicture}
  \pgfmathsetmacro{\przesuniecieX}{0};
  \pgfmathsetmacro{\x}{-7};
  \pgfmathsetmacro{\y}{0};
  \pgfmathsetmacro{\shadow}{0.0};
  \draw[-] (\x+0.1,\y+1) -- (\x+1,\y+1)  node[pos=.5,above] {\tiny $m_1$};
  \draw[fill = white] (\x+0.2,\y+1) circle (0.1);

  \draw[rounded corners=4pt,-]
    (\x-0.5,\y+1) -- (\x-0.3,\y+1) --(\x+0.8,\y+0.6) -- (\x+1,\y+0.6)  node[pos=.5,above] {\tiny $m_1$};
  \draw[-] (\x-0.5,\y+0.4) -- (\x+1,\y+0.4) node[pos=.5, at end, below] {\tiny $n_1-m_1$};
  \draw[fill = white] (\x-0.4,\y+0.4) circle (0.1);

  \pgfmathsetmacro{\y}{-1.2};
   \draw[rounded corners=4pt,-]
    (\x-0.5,\y+0.6) -- (\x+1,\y+0.6)  node[pos=.5,above] {\tiny $m_1$};
  \draw[fill = white] (\x-0.4,\y+0.6) circle (0.1);

  \draw[-] (\x-0.5,\y+0.4) -- (\x+1,\y+0.4) node[pos=.5, at end, below] {\tiny $n_2-m_1$};
  \draw[fill = white] (\x-0.4,\y+0.4) circle (0.1);

  \draw[-] (\x-0.5,\y) -- (\x+1,\y) node[pos=.5,below] {\tiny $m_2$};
  \draw[fill = white] (\x-0.4,\y) circle (0.1);
  \draw[fill = white] (\x+1.7,\y+1) circle (0.15);
  \node (v) at  (\x+1.7,\y+1) {$\vee$};
  \pgfmathsetmacro{\przesuniecieX}{0};
  \pgfmathsetmacro{\x}{-4.5};
  \pgfmathsetmacro{\y}{0};
  \pgfmathsetmacro{\shadow}{0.0};
  \draw[-] (\x+0.1,\y+1) -- (\x+1,\y+1)  node[pos=.5,above] {\tiny $m_1$};
  \draw[fill = white] (\x+0.2,\y+1) circle (0.1);
   \draw[rounded corners=4pt,-]
    (\x-0.5,\y+0.6) -- (\x+1,\y+0.6)  node[pos=.5,at end, above] {\tiny $m_1$};
  \draw[fill = white] (\x-0.4,\y+0.6) circle (0.1);

  \draw[rounded corners=4pt,-]
    (\x-0.5,\y+1) -- (\x-0.3,\y+1) --(\x+0.8,\y-1.2+0.6) -- (\x+1,\y-1.2+0.6)
    node[pos=.5,above] {\tiny $m_1$};
   \draw[-] (\x-0.5,\y+0.4) -- (\x+1,\y+0.4) node at (\x+1.5,\y+0.4) {\tiny $n_1-m_1$};
  \draw[fill = white] (\x-0.4,\y+0.4) circle (0.1);

  \pgfmathsetmacro{\y}{-1.2};

   \draw[-] (\x-0.5,\y+0.4) -- (\x+1,\y+0.4) node at (\x+1.5,\y+0.4) {\tiny $n_2-m_1$};
  \draw[fill = white] (\x-0.4,\y+0.4) circle (0.1);

  \draw[-] (\x-0.5,\y) -- (\x+1,\y) node[pos=.5,below] {\tiny $m_2$};
  \draw[fill = white] (\x-0.4,\y) circle (0.1);

%%%%%%%%%%%%%%%%%%%%%%%%%%%%%%%%%%%%%%%%%%%%%%%%%%%%%%%%%%%%%%%%

  \pgfmathsetmacro{\przesuniecieX}{0};
  \pgfmathsetmacro{\x}{0};
  \pgfmathsetmacro{\y}{0};
  \pgfmathsetmacro{\shadow}{0.0};
  \draw[-] (\x-0.5,\y+1) -- (\x+1,\y+1) node[pos=.5,above] {\tiny $m_1$};
  \draw[fill = white] (\x-0.4,\y+1) circle (0.1);
  \draw[-] (\x-0.5,\y+0.6) -- (\x,\y+0.6) node[pos=.5,below] {\tiny $n_1$};
  \draw[-] (\x+0.5,\y+0.4) -- (\x+1,\y+0.4) node at (\x+1.5,\y+0.4) {\tiny $n_1-m_1$};
  \draw[-] (\x+0.5,\y+0.6) -- (\x+1,\y+0.6) node[pos=.5,above] {\tiny $m_1$};
  \draw[rounded corners=4pt,-]
    (\x+0.5,\y+0.1) --( \x+0.7,\y+0.1) -- (\x+0.8,\y+0.1-1.2) -- (\x+1,\y+0.1-1.2);

  \draw[fill = white] (\x+0,\y+0) rectangle (\x+0.5,\y+0.8);
  \node (alpha) at (\x+0.25,\y+0.4) {$\alpha$};

  \pgfmathsetmacro{\y}{-1.2};

  \draw[-] (\x-0.5,\y+0.6) -- (\x,\y+0.6) node[pos=.5, above] {\tiny $n_2$};
  \draw[-] (\x+0.5,\y+0.6) -- (\x+1,\y+0.6) node[pos=.5,at end, above] {\tiny $m_1$};
  \draw[-] (\x+0.5,\y+0.4) -- (\x+1,\y+0.4) node at (\x+1.5,\y+0.4) {\tiny $n_2-m_1$};

  \draw[-] (\x+0.5,\y+0.1) -- (\x+1,\y+0.1) node[pos=.5,at end, below] {\tiny $m_2$};

  \draw[fill = white] (\x+0,\y+0) rectangle (\x+0.5,\y+0.8);
  \node (alpha) at (\x+0.25,\y+0.4) {$\beta$};

  \end{tikzpicture}
  }
  \end{center}
  Let $\gamma = [f,g]$. Then $\gamma : m_1 +(n_1+n_2)\rightdcirc m_1 + (n_1+n_2) + m_2$
  is in $[\mathcal{A}]$. The morphism $f^\otimes$ is equal to $f\vee (\mathsf{id}_{m_1} + \perp_{0,n_1+n_2+m_2})$ and is
  depicted by:
 \begin{center}
   \resizebox{.5\textwidth}{!}
   {
   \begin{tikzpicture}
   \pgfmathsetmacro{\przesuniecieX}{0};
   \pgfmathsetmacro{\x}{-7};
   \pgfmathsetmacro{\y}{0};
   \pgfmathsetmacro{\shadow}{0.0};
   \draw[-] (\x+0.1,\y+1) -- (\x+1,\y+1)  node[pos=.5,above] {\tiny $m_1$};
   \draw[fill = white] (\x+0.2,\y+1) circle (0.1);

   \draw[rounded corners=4pt,-]
     (\x-0.5,\y+1) -- (\x-0.3,\y+1) --(\x+0.8,\y+0.6) -- (\x+1,\y+0.6)  node[pos=.5,above] {\tiny $m_1$};
   \draw[-] (\x-0.5,\y+0.4) -- (\x+1,\y+0.4) node[pos=.5, at end, below] {\tiny $n_1-m_1$};
   \draw[fill = white] (\x-0.4,\y+0.4) circle (0.1);

   \pgfmathsetmacro{\y}{-1.2};
    \draw[rounded corners=4pt,-]
     (\x-0.5,\y+0.6) -- (\x+1,\y+0.6)  node[pos=.5,above] {\tiny $m_1$};
   \draw[fill = white] (\x-0.4,\y+0.6) circle (0.1);

   \draw[-] (\x-0.5,\y+0.4) -- (\x+1,\y+0.4) node[pos=.5, at end, below] {\tiny $n_2-m_1$};
   \draw[fill = white] (\x-0.4,\y+0.4) circle (0.1);

   \draw[-] (\x-0.5,\y) -- (\x+1,\y) node[pos=.5,below] {\tiny $m_2$};
   \draw[fill = white] (\x-0.4,\y) circle (0.1);
   \draw[fill = white] (\x+1.7,\y+1) circle (0.15);
   \node (v) at  (\x+1.7,\y+1) {$\vee$};
   \pgfmathsetmacro{\przesuniecieX}{0};
   \pgfmathsetmacro{\x}{-4.5};
   \pgfmathsetmacro{\y}{0};
   \pgfmathsetmacro{\shadow}{0.0};
   \draw[-] (\x+0.1,\y+1) -- (\x+1,\y+1)  node[pos=.5,above] {\tiny $m_1$};
   \draw[fill = white] (\x+0.2,\y+1) circle (0.1);
    \draw[rounded corners=4pt,-]
     (\x-0.5,\y+0.6) -- (\x+1,\y+0.6)  node[pos=.5,at end, above] {\tiny $m_1$};
   \draw[fill = white] (\x-0.4,\y+0.6) circle (0.1);

   \draw[rounded corners=4pt,-]
     (\x-0.5,\y+1) -- (\x-0.3,\y+1) --(\x+0.8,\y-1.2+0.6) -- (\x+1,\y-1.2+0.6)
     node[pos=.5,above] {\tiny $m_1$};
    \draw[-] (\x-0.5,\y+0.4) -- (\x+1,\y+0.4) node at (\x+1.5,\y+0.4) {\tiny $n_1-m_1$};
   \draw[fill = white] (\x-0.4,\y+0.4) circle (0.1);

   \pgfmathsetmacro{\y}{-1.2};

    \draw[-] (\x-0.5,\y+0.4) -- (\x+1,\y+0.4) node at (\x+1.5,\y+0.4) {\tiny $n_2-m_1$};
   \draw[fill = white] (\x-0.4,\y+0.4) circle (0.1);

   \draw[-] (\x-0.5,\y) -- (\x+1,\y) node[pos=.5,below] {\tiny $m_2$};
   \draw[fill = white] (\x-0.4,\y) circle (0.1);
   \draw[fill = white] (\x+1.7,\y+1) circle (0.15);
   \node (v) at  (\x+1.7,\y+1) {$\vee$};

   \pgfmathsetmacro{\przesuniecieX}{0};
   \pgfmathsetmacro{\x}{-2};
   \pgfmathsetmacro{\y}{0};
   \pgfmathsetmacro{\shadow}{0.0};
   \draw[-] (\x-0.5,\y+1) -- (\x+1,\y+1)  node[pos=.5,above] {\tiny $m_1$};
    \draw[rounded corners=4pt,-]
     (\x-0.5,\y+0.6) -- (\x+1,\y+0.6)  node[pos=.5,at end, above] {\tiny $m_1$};
   \draw[fill = white] (\x-0.4,\y+0.6) circle (0.1);

   \draw[rounded corners=4pt,-]
     (\x-0.5,\y-0.6) -- (\x+1,\y-1.2+0.6)
     node[pos=.5,above] {\tiny $m_1$};
   \draw[fill = white] (\x-0.4,\y-0.6) circle (0.1);
    \draw[-] (\x-0.5,\y+0.4) -- (\x+1,\y+0.4) node at (\x+1.5,\y+0.4) {\tiny $n_1-m_1$};
   \draw[fill = white] (\x-0.4,\y+0.4) circle (0.1);

   \pgfmathsetmacro{\y}{-1.2};

    \draw[-] (\x-0.5,\y+0.4) -- (\x+1,\y+0.4) node at (\x+1.5,\y+0.4) {\tiny $n_2-m_1$};
   \draw[fill = white] (\x-0.4,\y+0.4) circle (0.1);

   \draw[-] (\x-0.5,\y) -- (\x+1,\y) node[pos=.5,below] {\tiny $m_2$};
   \draw[fill = white] (\x-0.4,\y) circle (0.1);

 \end{tikzpicture}
 }
 \end{center}
  Moreover, $(\pi^{-1}+\mathsf{id})\cdot k^\otimes$ from~\ref{equation:gspi} is, in our case, given by $\sigma'\cdot (\perp_{0,m_1}+\alpha^\otimes +\beta^\otimes)$,
  where $\sigma'$ is as above and is depiced as follows:
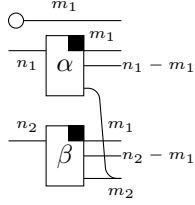
  \begin{center}
  \begin{tikzpicture}
  \pgfmathsetmacro{\przesuniecieX}{0};
  \pgfmathsetmacro{\x}{0};
  \pgfmathsetmacro{\y}{0};
  \pgfmathsetmacro{\shadow}{0.0};
  \draw[-] (\x-0.5,\y+1) -- (\x+1,\y+1) node[pos=.5,above] {\tiny $m_1$};
  \draw[fill = white] (\x-0.4,\y+1) circle (0.1);
  \draw[-] (\x-0.5,\y+0.6) -- (\x,\y+0.6) node[pos=.5,below] {\tiny $n_1$};
  \draw[-] (\x+0.5,\y+0.4) -- (\x+1,\y+0.4) node at (\x+1.5,\y+0.4) {\tiny $n_1-m_1$};
  \draw[-] (\x+0.5,\y+0.6) -- (\x+1,\y+0.6) node[pos=.5,above] {\tiny $m_1$};
  \draw[rounded corners=4pt,-]
    (\x+0.5,\y+0.1) --( \x+0.7,\y+0.1) -- (\x+0.8,\y+0.1-1.2) -- (\x+1,\y+0.1-1.2);

  \draw[fill = white] (\x+0,\y+0) rectangle (\x+0.5,\y+0.8);
  \draw[fill = black] (\x+0.3,\y+0.6) rectangle (\x+0.5,\y+0.8);
  \node (alpha) at (\x+0.25,\y+0.4) {$\alpha$};

  \pgfmathsetmacro{\y}{-1.2};

  \draw[-] (\x-0.5,\y+0.6) -- (\x,\y+0.6) node[pos=.5, above] {\tiny $n_2$};
  \draw[-] (\x+0.5,\y+0.6) -- (\x+1,\y+0.6) node[pos=.5,at end, above] {\tiny $m_1$};
  \draw[-] (\x+0.5,\y+0.4) -- (\x+1,\y+0.4) node at (\x+1.5,\y+0.4) {\tiny $n_2-m_1$};

  \draw[-] (\x+0.5,\y+0.1) -- (\x+1,\y+0.1) node[pos=.5,at end, below] {\tiny $m_2$};

  \draw[fill = white] (\x+0,\y+0) rectangle (\x+0.5,\y+0.8);
  \draw[fill = black] (\x+0.3,\y+0.6) rectangle (\x+0.5,\y+0.8);

  \node (alpha) at (\x+0.25,\y+0.4) {$\beta$};

  \end{tikzpicture}
\end{center}
As before, by a careful analysis of~\ref{equation:gspi}
and~\ref{equation:composition_with_perp} we get:
  \[
  r_1\vee r_2 = [\perp, \mathsf{id}_{m_2}]\cdot \gamma^\otimes \cdot \mathsf{in}^{m_1}.
  \]
  %%%%%%%%%%%%%%%%%%%%%%%%%%%%%%%%%5
  %
  % Part 3 of the proof
  %
  \noindent \textbf{Part 3.} Finally, we show that for a map $r:m\rightdcirc m$ given by its normal
  form
  $r=[ \perp, \mathsf{id}_m]\cdot \alpha^\otimes \cdot \mathsf{in}^m_n$ for $\alpha:n\rightdcirc n+m$
  its saturation
  $r^\ast$ is in $NF(\mathcal{A})$. The proof presented here uses a construction of a morphism
  that will later be used in the proof of Theorem~\ref{theorem:kleene_omega_regular} and
  the lemmas that precede it.

  Let $\alpha':n+m\rightdcirc n+m$ be given by
  $\alpha' = [\alpha,\mathsf{in}^m_{n+m}]$ and
  consider $\gamma:n+m\rightdcirc n+m$ defined by
  \begin{align}
    \label{definition:gamma_map} \gamma \defeq \alpha'\vee \sigma,
  \end{align}
  where $\sigma : m+(n-m)+m \rightdcirc m+(n-m)+m$
  injects the third component of the coproduct into the first one and is the identity everywhere else.
  The map $\gamma$ is depicted by the following diagram:
  
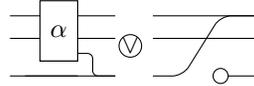
\begin{center}
  \begin{tikzpicture}
  \pgfmathsetmacro{\shadow}{0.00};
  %gwiazdka Kleenego
  %
  \pgfmathsetmacro{\przesuniecieX}{0};
  \pgfmathsetmacro{\x}{\przesuniecieX};
  \pgfmathsetmacro{\y}{-1};
  \pgfmathsetmacro{\shadow}{0};

  %\node (v) at (\x+1.3,\y-0.6) {$r^\ast$};
  %
  %%
  %%
  %%
  \draw[fill = gray] (\x+0+\shadow,\y+0+\shadow) rectangle
  (\x+0.5+\shadow,\y+0.8+\shadow);
  \draw[fill = white] (\x+0,\y+0) rectangle (\x+0.5,\y+0.8);
  \node (alpha) at (\x+0.25,\y+0.4) {$\alpha$};
  \draw[-] (\x-0.4,\y+0.6) -- (\x,\y+0.6);
  \draw[-] (\x-0.4,\y+0.3) -- (\x,\y+0.3);
  \draw[-] (\x+0.5,\y+0.3) -- (\x+1,\y+0.3);
  \draw[rounded corners=4pt,-] (\x-0.4,\y+-0.2) -- (\x+1,\y+-0.2) ;
  \draw[-] (\x+0.5,\y+0.6) -- (\x+1,\y+0.6);

  \draw[rounded corners=4pt,-] (\x+-0.2,\y+-0.2) -- (\x+0.5,\y+-0.2) ;
  \draw[rounded corners=2pt,-] (\x+0.5,\y+0.1) -- (\x+0.7,\y+0.1)
  --(\x+0.7,\y+-0.2)--(\x+1,\y+-0.2);
  %\draw[rounded corners=4pt,-] (\x+0.5,\y+0.2) -- (\x+0.8,\y+0.2) ;
  %
  \draw[fill = white] (\x+1.2,\y+0.2) circle (0.15);
  \node (v) at  (\x+1.2,\y+0.2) {$\vee$};
  \pgfmathsetmacro{\x}{1.9+\przesuniecieX};

  \draw[-] (\x-0.4,\y+0.6) -- (\x+1,\y+0.6);
  \draw[-] (\x-0.4,\y+0.3) -- (\x+1,\y+0.3);
  \draw[rounded corners=4pt,-] (\x-0.4,\y+-0.2) -- (\x,\y+-0.2)--
  (\x+0.5,\y+0.6)-- (\x+1,\y+0.6);

  \draw[rounded corners=4pt,-] (\x+0.5,\y+-0.2) -- (\x+1,\y+-0.2);
  \draw[fill = white] (\x+0.5,\y-0.2) circle (0.1);

  \end{tikzpicture}
  \end{center}

  \noindent Note that $\gamma$ is a $[\mathcal{A}]$-map which satisfies
  $\gamma^\ast = \gamma'^\ast$ for $\gamma'=\alpha'^\ast \vee \sigma$ represented by the diagram:
  
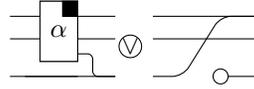
\begin{center}
  \begin{tikzpicture}
  \pgfmathsetmacro{\shadow}{0.00};

  %gwiazdka Kleenego
  %
  \pgfmathsetmacro{\przesuniecieX}{0};
  \pgfmathsetmacro{\x}{\przesuniecieX};
  \pgfmathsetmacro{\y}{-1};
  \pgfmathsetmacro{\shadow}{0};

  %\node (v) at (\x+1.3,\y-0.6) {$r^\ast$};
  %
  %%
  %%
  %%
  \draw[fill = gray] (\x+0+\shadow,\y+0+\shadow) rectangle
  (\x+0.5+\shadow,\y+0.8+\shadow);
  \draw[fill = white] (\x+0,\y+0) rectangle (\x+0.5,\y+0.8);
  \draw[fill = black] (\x+0.3,\y+0.6) rectangle (\x+0.5,\y+0.8);
  \node (alpha) at (\x+0.25,\y+0.4) {$\alpha$};
  \draw[-] (\x-0.4,\y+0.6) -- (\x,\y+0.6);
  \draw[-] (\x-0.4,\y+0.3) -- (\x,\y+0.3);
  \draw[-] (\x+0.5,\y+0.3) -- (\x+1,\y+0.3);
  \draw[rounded corners=4pt,-] (\x-0.4,\y+-0.2) -- (\x+1,\y+-0.2) ;
  \draw[-] (\x+0.5,\y+0.6) -- (\x+1,\y+0.6);

  \draw[rounded corners=4pt,-] (\x+-0.2,\y+-0.2) -- (\x+0.5,\y+-0.2) ;
  \draw[rounded corners=2pt,-] (\x+0.5,\y+0.1) -- (\x+0.7,\y+0.1)
  --(\x+0.7,\y+-0.2)--(\x+1,\y+-0.2);
  %\draw[rounded corners=4pt,-] (\x+0.5,\y+0.2) -- (\x+0.8,\y+0.2) ;
  %
  \draw[fill = white] (\x+1.2,\y+0.2) circle (0.15);
  \node (v) at  (\x+1.2,\y+0.2) {$\vee$};
  \pgfmathsetmacro{\x}{1.9+\przesuniecieX};

  \draw[-] (\x-0.4,\y+0.6) -- (\x+1,\y+0.6);
  \draw[-] (\x-0.4,\y+0.3) -- (\x+1,\y+0.3);
  \draw[rounded corners=4pt,-] (\x-0.4,\y+-0.2) -- (\x,\y+-0.2)--
  (\x+0.5,\y+0.6)-- (\x+1,\y+0.6);

  \draw[rounded corners=4pt,-] (\x+0.5,\y+-0.2) -- (\x+1,\y+-0.2);
  \draw[fill = white] (\x+0.5,\y-0.2) circle (0.1);

  \end{tikzpicture}

  \end{center}

  \noindent We have $[\perp,\mathsf{id}_{m}]\cdot \gamma ' =
  [\perp,\mathsf{id}_{m}]\cdot\alpha'^\ast$ which is depicted by:
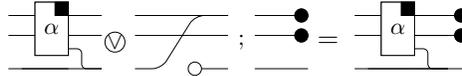
  \begin{center}
  \resizebox{.5\textwidth}{!}{
  \begin{tikzpicture}
  \pgfmathsetmacro{\shadow}{0.00};

  %gwiazdka Kleenego
  %
  \pgfmathsetmacro{\przesuniecieX}{0};
  \pgfmathsetmacro{\x}{\przesuniecieX};
  \pgfmathsetmacro{\y}{-1};
  \pgfmathsetmacro{\shadow}{0};

  %\node (v) at (\x+1.3,\y-0.6) {$r^\ast$};
  %
  %%
  %%
  %%
  \draw[fill = gray] (\x+0+\shadow,\y+0+\shadow) rectangle
  (\x+0.5+\shadow,\y+0.8+\shadow);
  \draw[fill = white] (\x+0,\y+0) rectangle (\x+0.5,\y+0.8);
  \draw[fill = black] (\x+0.3,\y+0.6) rectangle (\x+0.5,\y+0.8);
  \node (alpha) at (\x+0.25,\y+0.4) {$\alpha$};
  \draw[-] (\x-0.4,\y+0.6) -- (\x,\y+0.6);
  \draw[-] (\x-0.4,\y+0.3) -- (\x,\y+0.3);
  \draw[-] (\x+0.5,\y+0.3) -- (\x+1,\y+0.3);
  \draw[rounded corners=4pt,-] (\x-0.4,\y+-0.2) -- (\x+1,\y+-0.2) ;
  \draw[-] (\x+0.5,\y+0.6) -- (\x+1,\y+0.6);

  \draw[rounded corners=4pt,-] (\x+-0.2,\y+-0.2) -- (\x+0.5,\y+-0.2) ;
  \draw[rounded corners=2pt,-] (\x+0.5,\y+0.1) -- (\x+0.7,\y+0.1)
  --(\x+0.7,\y+-0.2)--(\x+1,\y+-0.2);
  %\draw[rounded corners=4pt,-] (\x+0.5,\y+0.2) -- (\x+0.8,\y+0.2) ;
  %
  \draw[fill = white] (\x+1.2,\y+0.2) circle (0.15);
  \node (v) at  (\x+1.2,\y+0.2) {$\vee$};
  %%%%%%%%%%%%%%%%%%%%%%%%%%%%%%%%%%%%%%%%%%%%%%%%%%%%%%%%%%%%%%%%%%
  %%%%%%%%%%%%%%%%%%%%%%%%%%%%%%%%%%%%%%%%%%%%%%%%%%%%%%%%%%%%%%%%%%%
  % druga kolumna
  % %
  % %%
  % %%
  \pgfmathsetmacro{\przesuniecieX}{\przesuniecieX+1.9};
  \pgfmathsetmacro{\x}{\przesuniecieX};

  \draw[-] (\x-0.4,\y+0.6) -- (\x+1,\y+0.6);
  \draw[-] (\x-0.4,\y+0.3) -- (\x+1,\y+0.3);
  \draw[rounded corners=4pt,-] (\x-0.4,\y+-0.2) -- (\x,\y+-0.2)--
  (\x+0.5,\y+0.6)-- (\x+1,\y+0.6);

  \draw[rounded corners=4pt,-] (\x+0.5,\y+-0.2) -- (\x+1,\y+-0.2);
  \draw[fill = white] (\x+0.5,\y-0.2) circle (0.1);

  %\draw[fill = white] (\x+1.2,\y+0.2) circle (0.15);
  %\node (v) at  (\x+1.2,\y+0.2) {$\vee$};
  %\draw[fill = white] (\x+1.2,\y+0.2) circle (0.15);
  \node (v) at  (\x+1.2,\y+0.2) {$;$};

  %%%%%%%%%%%%%%%%%%%%%%%%%%%%%%%%%%%%%%%%%%%%%%%%%%%%%%%%%%%%%%%%%%%%%%%%%%%%%%%%%%%%%%%%%%%%%%%%%%%%%%%%%%%%%%%%%%%%%%%%%
  %%%%%%%%%%%%%%%%%%%%%%%%%%%%%%%%%%%%%%%%%%%%%%%%%%%%%%%%%%%%%%%%%%%%%%%%%%%%%%%%%%%%%%%%%%%%%%%%%%%%%%%%%%%%%%%%%%%%%%%%%%
  % trzecia
  %
  \pgfmathsetmacro{\przesuniecieX}{\przesuniecieX+1.8};
  \pgfmathsetmacro{\x}{\przesuniecieX};

  \draw[-] (\x-0.4,\y+0.6) -- (\x+0.3,\y+0.6);
  \draw[fill = black] (\x+0.3,\y+0.6) circle (0.1);
  \draw[-] (\x-0.4,\y+0.3) -- (\x+0.3,\y+0.3);
  \draw[fill = black] (\x+0.3,\y+0.3) circle (0.1);
  \draw[rounded corners=4pt,-] (\x-0.4,\y+-0.2) -- (\x+0.4,\y+-0.2) ;

  \node (v) at  (\x+0.7,\y+0.2) {\Large $=$};

  %%%%%%%%%%%%%%%%%%%%%%%%%%%%%%%%%%%%%%%%%%%%%%%%%%%%%%%%%%%%%%%%%%%%%%%%%%%%%%%%%%%%%%%%%%%%%%%%%%%%%%%%%%%%%%%%%%%%%%%%%
  %%%%%%%%%%%%%%%%%%%%%%%%%%%%%%%%%%%%%%%%%%%%%%%%%%%%%%%%%%%%%%%%%%%%%%%%%%%%%%%%%%%%%%%%%%%%%%%%%%%%%%%%%%%%%%%%%%%%%%%%%%
  % czwarta
  %
  \pgfmathsetmacro{\przesuniecieX}{\przesuniecieX+1.5};
  \pgfmathsetmacro{\x}{\przesuniecieX};

  \draw[fill = gray] (\x+0+\shadow,\y+0+\shadow) rectangle
  (\x+0.5+\shadow,\y+0.8+\shadow);
  \draw[fill = white] (\x+0,\y+0) rectangle (\x+0.5,\y+0.8);
  \draw[fill = black] (\x+0.3,\y+0.6) rectangle (\x+0.5,\y+0.8);
  \node (alpha) at (\x+0.25,\y+0.4) {$\alpha$};
  \draw[-] (\x-0.4,\y+0.6) -- (\x,\y+0.6);
  \draw[-] (\x-0.4,\y+0.3) -- (\x,\y+0.3);
  \draw[-] (\x+0.5,\y+0.3) -- (\x+1,\y+0.3);
  \draw[rounded corners=4pt,-] (\x-0.4,\y+-0.2) -- (\x+1,\y+-0.2) ;
  \draw[-] (\x+0.5,\y+0.6) -- (\x+1,\y+0.6);

  \draw[rounded corners=4pt,-] (\x+-0.2,\y+-0.2) -- (\x+0.5,\y+-0.2) ;
  \draw[rounded corners=2pt,-] (\x+0.5,\y+0.1) -- (\x+0.7,\y+0.1)
  --(\x+0.7,\y+-0.2)--(\x+1,\y+-0.2);
  %\draw[rounded corners=4pt,-] (\x+0.5,\y+0.2) -- (\x+0.8,\y+0.2) ;
  %
  %
  \pgfmathsetmacro{\przesuniecieX}{\przesuniecieX+0.9};
  \pgfmathsetmacro{\x}{\przesuniecieX};

  \draw[-] (\x-0.4,\y+0.6) -- (\x+0.3,\y+0.6);
  \draw[fill = black] (\x+0.3,\y+0.6) circle (0.1);
  \draw[-] (\x-0.4,\y+0.3) -- (\x+0.3,\y+0.3);
  \draw[fill = black] (\x+0.3,\y+0.3) circle (0.1);
  \draw[rounded corners=4pt,-] (\x-0.4,\y+-0.2) -- (\x+0.4,\y+-0.2) ;

  \end{tikzpicture}

  }
  \end{center}
  Moreover, $[\perp,\mathsf{id}_{m}]\cdot \gamma '^2 =$
  \begin{center}
  \resizebox{.9\textwidth}{!}
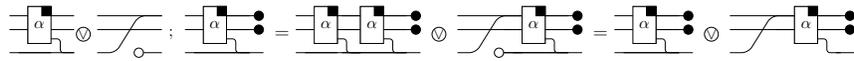
{
  \begin{tikzpicture}
  \pgfmathsetmacro{\shadow}{0.00};

  %gwiazdka Kleenego
  %
  \pgfmathsetmacro{\przesuniecieX}{0};
  \pgfmathsetmacro{\x}{\przesuniecieX};
  \pgfmathsetmacro{\y}{-1};
  \pgfmathsetmacro{\shadow}{0};

  %\node (v) at (\x+1.3,\y-0.6) {$r^\ast$};
  %
  %%
  %%
  %%
  \draw[fill = gray] (\x+0+\shadow,\y+0+\shadow) rectangle
  (\x+0.5+\shadow,\y+0.8+\shadow);
  \draw[fill = white] (\x+0,\y+0) rectangle (\x+0.5,\y+0.8);
  \draw[fill = black] (\x+0.3,\y+0.6) rectangle (\x+0.5,\y+0.8);
  \node (alpha) at (\x+0.25,\y+0.4) {$\alpha$};
  \draw[-] (\x-0.4,\y+0.6) -- (\x,\y+0.6);
  \draw[-] (\x-0.4,\y+0.3) -- (\x,\y+0.3);
  \draw[-] (\x+0.5,\y+0.3) -- (\x+1,\y+0.3);
  \draw[rounded corners=4pt,-] (\x-0.4,\y+-0.2) -- (\x+1,\y+-0.2) ;
  \draw[-] (\x+0.5,\y+0.6) -- (\x+1,\y+0.6);

  \draw[rounded corners=4pt,-] (\x+-0.2,\y+-0.2) -- (\x+0.5,\y+-0.2) ;
  \draw[rounded corners=2pt,-] (\x+0.5,\y+0.1) -- (\x+0.7,\y+0.1)
  --(\x+0.7,\y+-0.2)--(\x+1,\y+-0.2);
  %\draw[rounded corners=4pt,-] (\x+0.5,\y+0.2) -- (\x+0.8,\y+0.2) ;
  %
  \draw[fill = white] (\x+1.2,\y+0.2) circle (0.15);
  \node (v) at  (\x+1.2,\y+0.2) {$\vee$};
  %%%%%%%%%%%%%%%%%%%%%%%%%%%%%%%%%%%%%%%%%%%%%%%%%%%%%%%%%%%%%%%%%%
  %%%%%%%%%%%%%%%%%%%%%%%%%%%%%%%%%%%%%%%%%%%%%%%%%%%%%%%%%%%%%%%%%%%
  % druga kolumna
  % %
  % %%
  % %%
  \pgfmathsetmacro{\przesuniecieX}{\przesuniecieX+1.9};
  \pgfmathsetmacro{\x}{\przesuniecieX};

  \draw[-] (\x-0.4,\y+0.6) -- (\x+1,\y+0.6);
  \draw[-] (\x-0.4,\y+0.3) -- (\x+1,\y+0.3);
  \draw[rounded corners=4pt,-] (\x-0.4,\y+-0.2) -- (\x,\y+-0.2)--
  (\x+0.5,\y+0.6)-- (\x+1,\y+0.6);

  \draw[rounded corners=4pt,-] (\x+0.5,\y+-0.2) -- (\x+1,\y+-0.2);
  \draw[fill = white] (\x+0.5,\y-0.2) circle (0.1);

  %\draw[fill = white] (\x+1.2,\y+0.2) circle (0.15);
  %\node (v) at  (\x+1.2,\y+0.2) {$\vee$};
  %\draw[fill = white] (\x+1.2,\y+0.2) circle (0.15);
  \node (v) at  (\x+1.2,\y+0.2) {$;$};

  %%%%%%%%%%%%%%%%%%%%%%%%%%%%%%%%%%%%%%%%%%%%%%%%%%%%%%%%%%%%%%%%%%%%%%%%%%%%%%%%%%%%%%%%%%%%%%%%%%%%%%%%%%%%%%%%%%%%%%%%%
  %%%%%%%%%%%%%%%%%%%%%%%%%%%%%%%%%%%%%%%%%%%%%%%%%%%%%%%%%%%%%%%%%%%%%%%%%%%%%%%%%%%%%%%%%%%%%%%%%%%%%%%%%%%%%%%%%%%%%%%%%%
  % trzecia
  %
  %
  \pgfmathsetmacro{\przesuniecieX}{\przesuniecieX+1.9};
  \pgfmathsetmacro{\x}{\przesuniecieX};

  \draw[fill = gray] (\x+0+\shadow,\y+0+\shadow) rectangle
  (\x+0.5+\shadow,\y+0.8+\shadow);
  \draw[fill = white] (\x+0,\y+0) rectangle (\x+0.5,\y+0.8);
  \draw[fill = black] (\x+0.3,\y+0.6) rectangle (\x+0.5,\y+0.8);
  \node (alpha) at (\x+0.25,\y+0.4) {$\alpha$};
  \draw[-] (\x-0.4,\y+0.6) -- (\x,\y+0.6);
  \draw[-] (\x-0.4,\y+0.3) -- (\x,\y+0.3);
  \draw[-] (\x+0.5,\y+0.3) -- (\x+1,\y+0.3);
  \draw[rounded corners=4pt,-] (\x-0.4,\y+-0.2) -- (\x+1,\y+-0.2) ;
  \draw[-] (\x+0.5,\y+0.6) -- (\x+1,\y+0.6);

  \draw[rounded corners=4pt,-] (\x+-0.2,\y+-0.2) -- (\x+0.5,\y+-0.2) ;
  \draw[rounded corners=2pt,-] (\x+0.5,\y+0.1) -- (\x+0.7,\y+0.1)
  --(\x+0.7,\y+-0.2)--(\x+1,\y+-0.2);
  %\draw[rounded corners=4pt,-] (\x+0.5,\y+0.2) -- (\x+0.8,\y+0.2) ;
  %
  %
  \pgfmathsetmacro{\przesuniecieX}{\przesuniecieX+0.9};
  \pgfmathsetmacro{\x}{\przesuniecieX};

  \draw[-] (\x-0.4,\y+0.6) -- (\x+0.3,\y+0.6);
  \draw[fill = black] (\x+0.3,\y+0.6) circle (0.1);
  \draw[-] (\x-0.4,\y+0.3) -- (\x+0.3,\y+0.3);
  \draw[fill = black] (\x+0.3,\y+0.3) circle (0.1);
  \draw[rounded corners=4pt,-] (\x-0.4,\y+-0.2) -- (\x+0.4,\y+-0.2) ;

  \node (v) at  (\x+0.8,\y+0.2) {\Large $=$};

  %%%%%%%%%%%%%%%%%%%%%%%%
  %%%%%%%%%%%%%%%%%%%%%%%%%
  %%%%%%%%%%%%%%%%%%%%%%%%%%
  % czwarta
  % %
  % %%
  \pgfmathsetmacro{\przesuniecieX}{\przesuniecieX+1.5};
  \pgfmathsetmacro{\x}{\przesuniecieX};

  \draw[fill = gray] (\x+0+\shadow,\y+0+\shadow) rectangle
  (\x+0.5+\shadow,\y+0.8+\shadow);
  \draw[fill = white] (\x+0,\y+0) rectangle (\x+0.5,\y+0.8);
  \draw[fill = black] (\x+0.3,\y+0.6) rectangle (\x+0.5,\y+0.8);
  \node (alpha) at (\x+0.25,\y+0.4) {$\alpha$};
  \draw[-] (\x-0.4,\y+0.6) -- (\x,\y+0.6);
  \draw[-] (\x-0.4,\y+0.3) -- (\x,\y+0.3);
  \draw[-] (\x+0.5,\y+0.3) -- (\x+1,\y+0.3);
  \draw[rounded corners=4pt,-] (\x-0.4,\y+-0.2) -- (\x+1,\y+-0.2) ;
  \draw[-] (\x+0.5,\y+0.6) -- (\x+1,\y+0.6);

  \draw[rounded corners=4pt,-] (\x+-0.2,\y+-0.2) -- (\x+0.5,\y+-0.2) ;
  \draw[rounded corners=2pt,-] (\x+0.5,\y+0.1) -- (\x+0.7,\y+0.1)
  --(\x+0.7,\y+-0.2)--(\x+1,\y+-0.2);
  %\draw[rounded corners=4pt,-] (\x+0.5,\y+0.2) -- (\x+0.8,\y+0.2) ;
  %
  \pgfmathsetmacro{\przesuniecieX}{\przesuniecieX+1};
  \pgfmathsetmacro{\x}{\przesuniecieX};

  \draw[fill = gray] (\x+0+\shadow,\y+0+\shadow) rectangle
  (\x+0.5+\shadow,\y+0.8+\shadow);
  \draw[fill = white] (\x+0,\y+0) rectangle (\x+0.5,\y+0.8);
  \draw[fill = black] (\x+0.3,\y+0.6) rectangle (\x+0.5,\y+0.8);
  \node (alpha) at (\x+0.25,\y+0.4) {$\alpha$};
  \draw[-] (\x-0.4,\y+0.6) -- (\x,\y+0.6);
  \draw[-] (\x-0.4,\y+0.3) -- (\x,\y+0.3);
  \draw[-] (\x+0.5,\y+0.3) -- (\x+1,\y+0.3);
  \draw[rounded corners=4pt,-] (\x-0.4,\y+-0.2) -- (\x+1,\y+-0.2) ;
  \draw[-] (\x+0.5,\y+0.6) -- (\x+1,\y+0.6);

  \draw[rounded corners=4pt,-] (\x+-0.2,\y+-0.2) -- (\x+0.5,\y+-0.2) ;
  \draw[rounded corners=2pt,-] (\x+0.5,\y+0.1) -- (\x+0.7,\y+0.1)
  --(\x+0.7,\y+-0.2)--(\x+1,\y+-0.2);
  %\draw[rounded corners=4pt,-] (\x+0.5,\y+0.2) -- (\x+0.8,\y+0.2) ;
  %
  %
  \pgfmathsetmacro{\przesuniecieX}{\przesuniecieX+0.9};
  \pgfmathsetmacro{\x}{\przesuniecieX};

  \draw[-] (\x-0.4,\y+0.6) -- (\x+0.3,\y+0.6);
  \draw[fill = black] (\x+0.3,\y+0.6) circle (0.1);
  \draw[-] (\x-0.4,\y+0.3) -- (\x+0.3,\y+0.3);
  \draw[fill = black] (\x+0.3,\y+0.3) circle (0.1);
  \draw[rounded corners=4pt,-] (\x-0.4,\y+-0.2) -- (\x+0.4,\y+-0.2) ;

  \draw[fill = white] (\x+0.8,\y+0.2) circle (0.15);
  \node (v) at  (\x+0.8,\y+0.2) {$\vee$};

  %%%%%%%%%%%%%%%%%%%%%%%%%%%%%%%%%%%%%%%%%%%%%%%%%%%%%%%%%%%%%%%%%%%
  %%%%%%%%%%%%%%%%%%%%%%%%%%%%%%%%%%%%%%%%%%%%%%%%%%%%%%%%%%%%%%%%%%%%
  %%%%%%%%%%%%%%%%%%%%%%%%%%%%%%%%%%%%%%%%%%%%%%%%%%%%%%%%%%%%%%%%%%%%%
  % piąta
  % %
  % %%
  % %%
  \pgfmathsetmacro{\przesuniecieX}{\przesuniecieX+1.6};
  \pgfmathsetmacro{\x}{\przesuniecieX};

  \draw[-] (\x-0.4,\y+0.6) -- (\x+1,\y+0.6);
  \draw[-] (\x-0.4,\y+0.3) -- (\x+1,\y+0.3);
  \draw[rounded corners=4pt,-] (\x-0.4,\y+-0.2) -- (\x,\y+-0.2)--
  (\x+0.5,\y+0.6)-- (\x+1,\y+0.6);

  \draw[rounded corners=4pt,-] (\x+0.5,\y+-0.2) -- (\x+1,\y+-0.2);
  \draw[fill = white] (\x+0.5,\y-0.2) circle (0.1);

  \pgfmathsetmacro{\przesuniecieX}{\przesuniecieX+1};
  \pgfmathsetmacro{\x}{\przesuniecieX};

  \draw[fill = gray] (\x+0+\shadow,\y+0+\shadow) rectangle
  (\x+0.5+\shadow,\y+0.8+\shadow);
  \draw[fill = white] (\x+0,\y+0) rectangle (\x+0.5,\y+0.8);
  \draw[fill = black] (\x+0.3,\y+0.6) rectangle (\x+0.5,\y+0.8);
  \node (alpha) at (\x+0.25,\y+0.4) {$\alpha$};
  \draw[-] (\x-0.4,\y+0.6) -- (\x,\y+0.6);
  \draw[-] (\x-0.4,\y+0.3) -- (\x,\y+0.3);
  \draw[-] (\x+0.5,\y+0.3) -- (\x+1,\y+0.3);
  \draw[rounded corners=4pt,-] (\x-0.4,\y+-0.2) -- (\x+1,\y+-0.2) ;
  \draw[-] (\x+0.5,\y+0.6) -- (\x+1,\y+0.6);

  \draw[rounded corners=4pt,-] (\x+-0.2,\y+-0.2) -- (\x+0.5,\y+-0.2) ;
  \draw[rounded corners=2pt,-] (\x+0.5,\y+0.1) -- (\x+0.7,\y+0.1)
  --(\x+0.7,\y+-0.2)--(\x+1,\y+-0.2);
  %\draw[rounded corners=4pt,-] (\x+0.5,\y+0.2) -- (\x+0.8,\y+0.2) ;
  %
  %
  \pgfmathsetmacro{\przesuniecieX}{\przesuniecieX+0.9};
  \pgfmathsetmacro{\x}{\przesuniecieX};

  \draw[-] (\x-0.4,\y+0.6) -- (\x+0.3,\y+0.6);
  \draw[fill = black] (\x+0.3,\y+0.6) circle (0.1);
  \draw[-] (\x-0.4,\y+0.3) -- (\x+0.3,\y+0.3);
  \draw[fill = black] (\x+0.3,\y+0.3) circle (0.1);
  \draw[rounded corners=4pt,-] (\x-0.4,\y+-0.2) -- (\x+0.4,\y+-0.2) ;

  %\draw[fill = white] (\x+0.8,\y+0.2) circle (0.15);
  \node (v) at  (\x+0.8,\y+0.2) {\Large $=$};

  %%%%%%%%%%%%%%%%%%%%%%%%%%%%%%%%%%%%%%%%%%%%%%%%%%%%%%%%%%%%%%%%%%%%%%%%%%%%%%%%%%%%%%%%%%
  %%%%%%%%%%%%%%%%%%%%%%%%%%%%%%%%%%%%%%%%%%%%%%%%%%%%%%%%%%%%%%%%%%%%%%%%%%%%%%%%%%%%%%%%%%%
  % szósta
  % %
  % %%
  % %%%
  % %%%
  \pgfmathsetmacro{\przesuniecieX}{\przesuniecieX+1.5};
  \pgfmathsetmacro{\x}{\przesuniecieX};

  \draw[fill = gray] (\x+0+\shadow,\y+0+\shadow) rectangle
  (\x+0.5+\shadow,\y+0.8+\shadow);
  \draw[fill = white] (\x+0,\y+0) rectangle (\x+0.5,\y+0.8);
  \draw[fill = black] (\x+0.3,\y+0.6) rectangle (\x+0.5,\y+0.8);
  \node (alpha) at (\x+0.25,\y+0.4) {$\alpha$};
  \draw[-] (\x-0.4,\y+0.6) -- (\x,\y+0.6);
  \draw[-] (\x-0.4,\y+0.3) -- (\x,\y+0.3);
  \draw[-] (\x+0.5,\y+0.3) -- (\x+1,\y+0.3);
  \draw[rounded corners=4pt,-] (\x-0.4,\y+-0.2) -- (\x+1,\y+-0.2) ;
  \draw[-] (\x+0.5,\y+0.6) -- (\x+1,\y+0.6);

  \draw[rounded corners=4pt,-] (\x+-0.2,\y+-0.2) -- (\x+0.5,\y+-0.2) ;
  \draw[rounded corners=2pt,-] (\x+0.5,\y+0.1) -- (\x+0.7,\y+0.1)
  --(\x+0.7,\y+-0.2)--(\x+1,\y+-0.2);
  %\draw[rounded corners=4pt,-] (\x+0.5,\y+0.2) -- (\x+0.8,\y+0.2) ;
  %
  %
  \pgfmathsetmacro{\przesuniecieX}{\przesuniecieX+0.9};
  \pgfmathsetmacro{\x}{\przesuniecieX};

  \draw[-] (\x-0.4,\y+0.6) -- (\x+0.3,\y+0.6);
  \draw[fill = black] (\x+0.3,\y+0.6) circle (0.1);
  \draw[-] (\x-0.4,\y+0.3) -- (\x+0.3,\y+0.3);
  \draw[fill = black] (\x+0.3,\y+0.3) circle (0.1);
  \draw[rounded corners=4pt,-] (\x-0.4,\y+-0.2) -- (\x+0.4,\y+-0.2) ;

  \draw[fill = white] (\x+0.8,\y+0.2) circle (0.15);
  \node (v) at  (\x+0.8,\y+0.2) {$\vee$};

  %%%%%%%%%%%%%%%%%%%%%%%%%%%%%%%%%%%%%%%%%%%%%%%%%%%%%%%%%%%%%%%%%%%%%%%%%%%%%%%%%%%%%%%%%%
  %%%%%%%%%%%%%%%%%%%%%%%%%%%%%%%%%%%%%%%%%%%%%%%%%%%%%%%%%%%%%%%%%%%%%%%%%%%%%%%%%%%%%%%%%%%
  % siódma
  % %
  % %%
  % %%%
  % %%%
  % %%%
  % %%%
  \pgfmathsetmacro{\przesuniecieX}{\przesuniecieX+1.6};
  \pgfmathsetmacro{\x}{\przesuniecieX};

  \draw[-] (\x-0.4,\y+0.6) -- (\x+1,\y+0.6);
  \draw[-] (\x-0.4,\y+0.3) -- (\x+1,\y+0.3);
  \draw[rounded corners=4pt,-] (\x-0.4,\y+-0.2) -- (\x,\y+-0.2)--
  (\x+0.5,\y+0.6)-- (\x+1,\y+0.6);

  %\draw[rounded corners=4pt,-] (\x+0.5,\y+-0.2) -- (\x+1,\y+-0.2);
  %\draw[fill = white] (\x+0.5,\y-0.2) circle (0.1);
  %
  \pgfmathsetmacro{\przesuniecieX}{\przesuniecieX+1};
  \pgfmathsetmacro{\x}{\przesuniecieX};

  \draw[fill = gray] (\x+0+\shadow,\y+0+\shadow) rectangle
  (\x+0.5+\shadow,\y+0.8+\shadow);
  \draw[fill = white] (\x+0,\y+0) rectangle (\x+0.5,\y+0.8);
  \draw[fill = black] (\x+0.3,\y+0.6) rectangle (\x+0.5,\y+0.8);
  \node (alpha) at (\x+0.25,\y+0.4) {$\alpha$};
  \draw[-] (\x-0.4,\y+0.6) -- (\x,\y+0.6);
  \draw[-] (\x-0.4,\y+0.3) -- (\x,\y+0.3);
  \draw[-] (\x+0.5,\y+0.3) -- (\x+1,\y+0.3);
  %\draw[rounded corners=4pt,-] (\x-0.4,\y+-0.2) -- (\x+1,\y+-0.2) ;
  \draw[-] (\x+0.5,\y+0.6) -- (\x+1,\y+0.6);

  %\draw[rounded corners=4pt,-] (\x+-0.2,\y+-0.2) -- (\x+0.5,\y+-0.2) ;
  \draw[rounded corners=2pt,-] (\x+0.5,\y+0.1) -- (\x+0.7,\y+0.1)
  --(\x+0.7,\y+-0.2)--(\x+1,\y+-0.2);
  %\draw[rounded corners=4pt,-] (\x+0.5,\y+0.2) -- (\x+0.8,\y+0.2) ;
  %
  %
  \pgfmathsetmacro{\przesuniecieX}{\przesuniecieX+0.9};
  \pgfmathsetmacro{\x}{\przesuniecieX};

  \draw[-] (\x-0.4,\y+0.6) -- (\x+0.3,\y+0.6);
  \draw[fill = black] (\x+0.3,\y+0.6) circle (0.1);
  \draw[-] (\x-0.4,\y+0.3) -- (\x+0.3,\y+0.3);
  \draw[fill = black] (\x+0.3,\y+0.3) circle (0.1);
  \draw[rounded corners=4pt,-] (\x-0,\y+-0.2) -- (\x+0.4,\y+-0.2) ;

  \end{tikzpicture}

  }
  \end{center}
  \noindent Hence, by right distributivity w.r.t.\ the base morphisms we have
  $[\perp,\mathsf{id}_{m}]\cdot \gamma '^2\cdot \mathsf{in}^{m}_{n+m} =
  \mathsf{id}_{m} \vee r$ and $[\perp,\mathsf{id}_{m}]\cdot \gamma '^2\cdot
  \mathsf{in}^{n} = [\perp,\mathsf{id}_{m}]\cdot \gamma '\cdot \mathsf{in}^{n}$.
  Since $[\perp,\mathsf{id}_{m}]\cdot \gamma '\cdot \mathsf{in}^{m}_{n+m}
  =\mathsf{id}_{m}$ we conclude that:
  \[
  [\perp,\mathsf{id}_{m}]\cdot \gamma '^2\leq (\mathsf{id}\vee r)\cdot
  [\perp,\mathsf{id}_{m}]\cdot \gamma'.
  \]

  %\begin{align*}
  %&[0,\mathsf{id}_{m_2}]\cdot \gamma '^2 =
  %[0,\mathsf{id}_{m_2}]\cdot[\alpha,\mathsf{in}_{m'+m_2}^{m_2}]^\ast \cdot (
  %[\alpha,\mathsf{in}_{m'+m_2}^{m_2}]^\ast\vee  \mathsf{t} ) =\\
  %&[0,\mathsf{id}_{m_2}]\cdot[\alpha,\mathsf{in}_{m'+m_2}^{m_2}]^\ast \vee
  %[0,\mathsf{id}_{m_2}]\cdot[\alpha,\mathsf{in}_{m'+m_2}^{m_2}]^\ast\cdot
  %\mathsf{t} = \\
  %&[[0,\mathsf{id}_{m_2}]\cdot[\alpha,\mathsf{in}_{m'+m_2}^{m_2}]^\ast\cdot
  %\mathsf{in}^{m'}_{m'+m_2},\mathsf{id}_{m_2} \vee r] =\\
  %& [\mathsf{id}_{m_2},\mathsf{id}_{m_2}\vee r]\cdot
  %([0,\mathsf{id}_{m_2}]\cdot[\alpha,\mathsf{in}_{m'+m_2}^{m_2}]^\ast\cdot
  %\mathsf{in}^{m'}_{m'+m_2}+\mathsf{id}_{m_2})=\\
  %& [\mathsf{id}_{m_2},\mathsf{id}_{m_2}\vee r]\cdot
  %([0,\mathsf{id}_{m_2}]\cdot[\alpha,\mathsf{in}_{m'+m_2}^{m_2}]^\ast\cdot
  %\mathsf{in}^{m'}_{m'+m_2}+[0,\mathsf{id}_{m_2}]\cdot[\alpha,\mathsf{in}_{m'+m_2
  %}^{m_2}]^\ast\cdot \mathsf{in}^{m_2}_{m'+m_2})\leq \\
  %& [\mathsf{id}_{m_2}\vee r,\mathsf{id}_{m_2}\vee r]\cdot
  %([0,\mathsf{id}_{m_2}]\cdot[\alpha,\mathsf{in}_{m'+m_2}^{m_2}]^\ast\cdot
  %\mathsf{in}^{m'}_{m'+m_2}+[0,\mathsf{id}_{m_2}]\cdot[\alpha,\mathsf{in}_{m'+m_2
  %}^{m_2}]^\ast\cdot \mathsf{in}^{m_2}_{m'+m_2})=\\
  %& (\mathsf{id}_{m_2}\vee r)\cdot
  %[0,\mathsf{id}_{m_2}]\cdot[\alpha,\mathsf{in}_{m'+m_2}^{m_2}]^\ast =
  %(\mathsf{id}_{m_2}\vee r)\cdot  [0,\mathsf{id}_{m_2}]\cdot\gamma'.
  %\end{align*}
  %
  \noindent Moreover,
  \[
  [\perp,\mathsf{id}_{m}]\cdot \gamma '^2  = [r',\mathsf{id}\vee r] \geq
  [\perp,\mathsf{id}_{m} \vee r] = (\mathsf{id}\vee r)\cdot
  [\perp,\mathsf{id}_{m}],
  \]
  for $r' \defeq [\perp,\mathsf{id}]\cdot\alpha^\ast$.
  Note that $r'\cdot \mathsf{in}^{m}_{n} = r$.
  Hence, to summarize:
  \[
  (\mathsf{id}\vee r)\cdot [\perp,\mathsf{id}]=[\perp,\mathsf{id}\vee r] \leq
  [r',\mathsf{id}\vee r]  \leq [\perp,\mathsf{id}]\cdot \gamma '^2 \leq
  (\mathsf{id}\vee r)\cdot [\perp,\mathsf{id}]\cdot \gamma'.
  \]
  If we let $f = [\perp,\mathsf{id}]$ and $g= f\cdot \gamma'$ then the above inequalities
  are rephrased as follows:
  \[
  (\mathsf{id}\vee r) \cdot f \leq f\cdot (\gamma')^2 \text{ and } g\cdot \gamma'
  \leq (\mathsf{id} \vee r) \cdot g.
  \]
  By~\ref{prop:0} in Lemma~\ref{theorem:least_fixpoint},~\ref{condition:1}
  in Assumption~\ref{assumptions:kleisli_cat} and  $\mathsf{id}\leq \gamma'$ we get:
  \[
  r^\ast \cdot f \leq f\cdot \gamma'^\ast = g\cdot \gamma'^\ast \leq r^\ast \cdot g.
  \]
  Since $f\cdot \mathsf{in}^m_{n+m} = g\cdot \mathsf{in}^m_{n+m} =\mathsf{id}_m$ and
	$\gamma^\ast = \gamma'^\ast$ we get:
  $
  r^\ast = [\perp,\mathsf{id} ]\cdot  \gamma^\ast \cdot \mathsf{in}^m_{n+m}.
  $
  Similarily, we show
  $
    r^+ = [\perp,\mathsf{id}] \cdot
    \gamma^\ast \cdot \mathsf{in}^n_{n+m} \cdot \mathsf{in}^m_n.
  $
  This proves that $r^+,r^\ast \in NF(\mathcal{A})$.
\end{proof}

We are now ready to present the following proof.

\begin{proof}(Theorem~\ref{theorem:kleene_regular})
  Note that all maps from $NF(\mathcal{A})$ are regular and
	all regular maps are in $\mathfrak{Rat}(\mathcal{A})$.
  By Lemma~\ref{lemma:normal_closed_under}
  morphisms from $NF(\mathcal{A})$ form a theory closed under
	finite suprema and saturation. Hence,
  $NF(\mathcal{A}) = \mathfrak{Reg}(\mathcal{A}) = \mathfrak{Rat}(\mathcal{A})$.
	This completes the proof.
\end{proof}

%Hence, regular tree behaviours $r_n$ on $n$-variables can be defined by the
%following grammar:
%\begin{align*}
%r_n= TF_\varepsilon n \mid  r_n \vee r_n \mid  [r_n,\ldots, r_n]\cdot r_m \mid
%r^{\ast,i}_n
%\end{align*}
%

Before we proceed with the proof of Theorem~\ref{theorem:kleene_omega_regular}
we require one extra statement. Let us define:
\begin{align*}
\omega\mathfrak{Rat}(\mathcal{A})(n) \defeq & \{ [r_1,\ldots, r_{m}]^\omega  \cdot r
\mid r\in \mathfrak{Reg}(n,m), r_i \in \mathfrak{Reg}(1,m)   \text{ for } m
<\omega   \},\\
\omega\mathfrak{Reg}(\mathcal{A})(n) \defeq & \{
||\alpha,\mathfrak{F}||_{\omega}\cdot \mathsf{in}^n_m:n\rightdcirc 0 \mid
(\alpha,\mathfrak{F}) \text{ is } \mathcal{A}\text{-aut.} \text{ with } \alpha:m\rightdcirc m\}.
\end{align*}
and  note that $\omega\mathfrak{Rat}(\mathcal{A}) = \omega\mathfrak{Rat}(\mathcal{A})(1)$ and
$\omega\mathfrak{Reg}(\mathcal{A})= \omega\mathfrak{Reg}(\mathcal{A})(1)$. Additionally,
the following holds.
\begin{lem}
For any $r\in \mathfrak{Reg}(m,m)$ we have:
\[r^\omega \in \omega\mathfrak{Reg}(\mathcal{A})(m).\]
\end{lem}
\begin{proof}
 % For sake of clarity of notation let us define  $m_1\defeq m$ and $m_2\defeq m$. We introduce
 % $m_1$ and $m_2$ to denote the same number $m$ for the same reason as in the
 % proof of Theorem~\ref{theorem:kleene_regular},
 % i.e.\ to distinguish between the domain and codomain of $r$. Hence,
 % $r:m_1\rightdcirc m_2$. Since, as we have seen in the proof of Theorem~\ref{theorem:kleene_regular},
All regular maps can be given in their normal form. Hence, we have
\[
r=[\perp,\mathsf{id}_{m}]\cdot \alpha^\otimes \cdot \mathsf{in}^{m}_n\]
\noindent for $\alpha:n\to n+m$, where
$m\leq n$. We depict $r$ by \resizebox{.08\textwidth}{!}{
\begin{tikzpicture}
\pgfmathsetmacro{\przesuniecieX}{0};

\pgfmathsetmacro{\shadow}{0.0};
\pgfmathsetmacro{\x}{\przesuniecieX};
\pgfmathsetmacro{\y}{0};

\draw[fill = gray] (\x+0+\shadow,\y+0+\shadow) rectangle
(\x+0.5+\shadow,\y+0.8+\shadow);
\draw[fill = white] (\x+0,\y+0) rectangle (\x+0.5,\y+0.8);
\draw[fill = black] (\x+0.3,\y+0.6) rectangle (\x+0.5,\y+0.8);
\draw[-] (\x-0.5,\y+0.6) -- (\x,\y+0.6);

\draw[-] (\x-0.4,\y+0.3) -- (\x,\y+0.3);

\draw[-] (\x+0.5,\y+0.3) -- (\x+1,\y+0.3);
\draw[fill = black] (\x+1,\y+0.3) circle (0.1);
\draw[rounded corners=4pt,-] (\x-0.4,\y+-0.2) -- (\x+1,\y+-0.2) ;
\draw[-] (\x+0.5,\y+0.6) -- (\x+1,\y+0.6);
\draw[fill = black] (\x+1,\y+0.6) circle (0.1);
\node (alpha) at (\x+0.25,\y+0.4) {$\alpha$};
\draw[rounded corners=4pt,-] (\x+-0.2,\y+-0.2) -- (\x+0.5,\y+-0.2) ;
\draw[rounded corners=2pt,-] (\x+0.5,\y+0.1) -- (\x+0.7,\y+0.1)
--(\x+0.7,\y+-0.2)--(\x+1.2,\y+-0.2);
%\draw[rounded corners=2pt,-] (\x+1,\y-0.2) -- (\x+1.2,\y-0.2)
%--(\x+1.2,\y-0.7)--(\x+1.4,\y-0.7);
%\draw[rounded corners=4pt,-] (\x+0.5,\y+0.2) -- (\x+0.8,\y+0.2) ;
%
\draw[fill = white] (\x-0.3,\y+0.3) circle (0.1);
\draw[fill = white] (\x-0.3,\y-0.2) circle (0.1);
%   Our aim is to find an automaton with the specified set of initial states
% associated with the morphism $r_2\cdot r_1$.
\end{tikzpicture}
}. Consider the morphism $\gamma\defeq [\alpha,\mathsf{in}]\vee \sigma = \alpha'\vee \sigma$
defined as in~\ref{definition:gamma_map} depicted by the following diagram:
\begin{center}
\begin{tikzpicture}
\pgfmathsetmacro{\shadow}{0.00};

%gwiazdka Kleenego
%
\pgfmathsetmacro{\przesuniecieX}{0};
\pgfmathsetmacro{\x}{\przesuniecieX};
\pgfmathsetmacro{\y}{-1};
\pgfmathsetmacro{\shadow}{0};

%\node (v) at (\x+1.3,\y-0.6) {$r^\ast$};
%
%%
%%
%%
\draw[fill = gray] (\x+0+\shadow,\y+0+\shadow) rectangle
(\x+0.5+\shadow,\y+0.8+\shadow);
\draw[fill = white] (\x+0,\y+0) rectangle (\x+0.5,\y+0.8);
\node (alpha) at (\x+0.25,\y+0.4) {$\alpha$};
\draw[-] (\x-0.4,\y+0.6) -- (\x,\y+0.6);
\draw[-] (\x-0.4,\y+0.3) -- (\x,\y+0.3);
\draw[-] (\x+0.5,\y+0.3) -- (\x+1,\y+0.3);
\draw[rounded corners=4pt,-] (\x-0.4,\y+-0.2) -- (\x+1,\y+-0.2) ;
\draw[-] (\x+0.5,\y+0.6) -- (\x+1,\y+0.6);

\draw[rounded corners=4pt,-] (\x+-0.2,\y+-0.2) -- (\x+0.5,\y+-0.2) ;
\draw[rounded corners=2pt,-] (\x+0.5,\y+0.1) -- (\x+0.7,\y+0.1)
--(\x+0.7,\y+-0.2)--(\x+1,\y+-0.2);
%\draw[rounded corners=4pt,-] (\x+0.5,\y+0.2) -- (\x+0.8,\y+0.2) ;
%
\draw[fill = white] (\x+1.2,\y+0.2) circle (0.15);
\node (v) at  (\x+1.2,\y+0.2) {$\vee$};
\pgfmathsetmacro{\x}{1.9+\przesuniecieX};

\draw[-] (\x-0.4,\y+0.6) -- (\x+1,\y+0.6);
\draw[-] (\x-0.4,\y+0.3) -- (\x+1,\y+0.3);
\draw[rounded corners=4pt,-] (\x-0.4,\y+-0.2) -- (\x,\y+-0.2)--
(\x+0.5,\y+0.6)-- (\x+1,\y+0.6);

\draw[rounded corners=4pt,-] (\x+0.5,\y+-0.2) -- (\x+1,\y+-0.2);
\draw[fill = white] (\x+0.5,\y-0.2) circle (0.1);

\end{tikzpicture}
\end{center}
By the properties listed in \textbf{Part 3.} of the proof of Lemma~\ref{lemma:normal_closed_under} we have:
\[
r^+ = [\perp,\mathsf{id}] \cdot \gamma^\ast \cdot
      \mathsf{in}^{n}_{n+m}\cdot \mathsf{in}^{m}_{n}.
\]
Moreover, let $\xi$ be defined by $\xi \defeq = \sigma \cdot \alpha' = \sigma \cdot [\alpha,\mathsf{in}]$ and depicted in the following diagram:
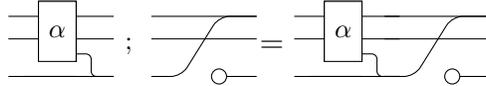
\begin{center}
\begin{tikzpicture}
\pgfmathsetmacro{\shadow}{0.00};

%gwiazdka Kleenego
%
\pgfmathsetmacro{\przesuniecieX}{0};
\pgfmathsetmacro{\x}{\przesuniecieX};
\pgfmathsetmacro{\y}{-1};
\pgfmathsetmacro{\shadow}{0};

%\node (v) at (\x+1.3,\y-0.6) {$r^\ast$};
%
%%
%%
%%
\draw[fill = gray] (\x+0+\shadow,\y+0+\shadow) rectangle
(\x+0.5+\shadow,\y+0.8+\shadow);
\draw[fill = white] (\x+0,\y+0) rectangle (\x+0.5,\y+0.8);
\node (alpha) at (\x+0.25,\y+0.4) {$\alpha$};
\draw[-] (\x-0.4,\y+0.6) -- (\x,\y+0.6);
\draw[-] (\x-0.4,\y+0.3) -- (\x,\y+0.3);
\draw[-] (\x+0.5,\y+0.3) -- (\x+1,\y+0.3);
\draw[rounded corners=4pt,-] (\x-0.4,\y+-0.2) -- (\x+1,\y+-0.2) ;
\draw[-] (\x+0.5,\y+0.6) -- (\x+1,\y+0.6);

\draw[rounded corners=4pt,-] (\x+-0.2,\y+-0.2) -- (\x+0.5,\y+-0.2) ;
\draw[rounded corners=2pt,-] (\x+0.5,\y+0.1) -- (\x+0.7,\y+0.1)
--(\x+0.7,\y+-0.2)--(\x+1,\y+-0.2);
%\draw[rounded corners=4pt,-] (\x+0.5,\y+0.2) -- (\x+0.8,\y+0.2) ;
%
%\draw[fill = white] (\x+1.2,\y+0.2) circle (0.15);
\node (v) at  (\x+1.2,\y+0.2) {$;$};

\pgfmathsetmacro{\przesuniecieX}{\przesuniecieX+1.9};
\pgfmathsetmacro{\x}{\przesuniecieX};

\draw[-] (\x-0.4,\y+0.6) -- (\x+1,\y+0.6);
\draw[-] (\x-0.4,\y+0.3) -- (\x+1,\y+0.3);
\draw[rounded corners=4pt,-] (\x-0.4,\y+-0.2) -- (\x,\y+-0.2)--
(\x+0.5,\y+0.6)-- (\x+1,\y+0.6);

\draw[rounded corners=4pt,-] (\x+0.5,\y+-0.2) -- (\x+1,\y+-0.2);
\draw[fill = white] (\x+0.5,\y-0.2) circle (0.1);
\node (v) at  (\x+1.2,\y+0.2) {\Large $=$};
%%%%%%%%%%%%%%%%%%%%%%%%%%%%%%%%%%%%
%%%%%%%%%%%%%%%%%%%%%%%%%%%%%%%%%%%%%
%%%%%%%%%%%%%%%%%%%%%%%%%%%%%%%%%%%%%%
%%%%%%%%%%%%%%%%%%%%%%%%%%%%%%%%%%%%%%%
%%%%%%%%%%%%%%%%%%%%%%%%%%%%%%%%%%%%%%%%
\pgfmathsetmacro{\przesuniecieX}{\przesuniecieX+1.9};
\pgfmathsetmacro{\x}{\przesuniecieX};
\draw[fill = gray] (\x+0+\shadow,\y+0+\shadow) rectangle
(\x+0.5+\shadow,\y+0.8+\shadow);
\draw[fill = white] (\x+0,\y+0) rectangle (\x+0.5,\y+0.8);
\node (alpha) at (\x+0.25,\y+0.4) {$\alpha$};
\draw[-] (\x-0.4,\y+0.6) -- (\x,\y+0.6);
\draw[-] (\x-0.4,\y+0.3) -- (\x,\y+0.3);
\draw[-] (\x+0.5,\y+0.3) -- (\x+1,\y+0.3);
\draw[rounded corners=4pt,-] (\x-0.4,\y+-0.2) -- (\x+1,\y+-0.2) ;
\draw[-] (\x+0.5,\y+0.6) -- (\x+1,\y+0.6);

\draw[rounded corners=4pt,-] (\x+-0.2,\y+-0.2) -- (\x+0.5,\y+-0.2) ;
\draw[rounded corners=2pt,-] (\x+0.5,\y+0.1) -- (\x+0.7,\y+0.1)
--(\x+0.7,\y+-0.2)--(\x+1,\y+-0.2);
%\draw[rounded corners=4pt,-] (\x+0.5,\y+0.2) -- (\x+0.8,\y+0.2) ;
%
%\draw[fill = white] (\x+1.2,\y+0.2) circle (0.15);
\pgfmathsetmacro{\przesuniecieX}{\przesuniecieX+1.2};
\pgfmathsetmacro{\x}{\przesuniecieX};

\draw[-] (\x-0.4,\y+0.6) -- (\x+1,\y+0.6);
\draw[-] (\x-0.4,\y+0.3) -- (\x+1,\y+0.3);
\draw[rounded corners=4pt,-] (\x-0.4,\y+-0.2) -- (\x,\y+-0.2)--
(\x+0.5,\y+0.6)-- (\x+1,\y+0.6);

\draw[rounded corners=4pt,-] (\x+0.5,\y+-0.2) -- (\x+1,\y+-0.2);
\draw[fill = white] (\x+0.5,\y-0.2) circle (0.1);
\end{tikzpicture}
\end{center}

\noindent The map $\xi$ is a $[\mathcal{A}]$-map which satisfies $\xi^\ast =
\gamma^\ast$. Additionally,
\begin{align*}
& r^+ = [\perp,\mathsf{id}] \cdot \xi^\ast \cdot
\mathsf{in}^{n}_{n+m}\cdot \mathsf{in}^{m}_n= \\
& [\perp,\mathsf{id}_{m}] \cdot \xi^\ast \cdot \xi\cdot
\mathsf{in}^{m}_{n+m}= \\
& [\perp,\mathsf{id}_{m}] \cdot \xi^+\cdot
\mathsf{in}^{m}_{n+m}.
\end{align*}
Since
$(\perp+\mathsf{id})=\mathsf{in}^{m}_{n+m}\cdot [\perp,\mathsf{id}]$ we get that
\[
\mathsf{in}^{m}_{n+m} \cdot r^+= \mathsf{in}^{m}\cdot [\perp,\mathsf{id}_{m}] \cdot
\xi^+\cdot \mathsf{in}^{m}_{n+m}= (\perp+\mathsf{id}_{m}) \cdot \xi^+\cdot
\mathsf{in}^{m}_{n+m}.
\]
From the above we get:
\[
r^\omega \stackrel{\diamond}{=} (r^+)^\omega \stackrel{\dagger}{=}
((\perp+\mathsf{id}_{m})\cdot \xi^+)^\omega\cdot
\mathsf{in}^{m}_{n+m}.
\]
The identity $(\diamond)$ follows by Lemma~\ref{theorem:least_fixpoint}.
The identity $(\dagger)$ follows from a more general property:
given two coalgebras $\alpha:X\rightdcirc X=X\to TX$ and $\beta:Y\rightdcirc Y=Y\to TY$
and a $\mathsf{Set}$-map $j:X\to Y$ which is a coalgebra homomomorphism (or, equivalently,
 $j^\sharp \cdot \alpha = \beta \cdot j^\sharp$ in $\mathcal{K}l(T)$)
 we have: $\alpha^\omega = \beta^\omega \cdot j^\sharp = \beta^\omega \circ j$.
 This property known as
  \emph{uniformity} of $(-)^\omega$ \emph{w.r.t.\ the base maps} (see e.g.~\cite{Simpson:2000:CAC:788022.788996}).
  In our setting, the fixpoint operator $(-)^\omega$ is uniform w.r.t.\ the base maps
  since the order of $\mathcal{K}l(T)$ is pointwise induced
   and since $\alpha^\omega =\bigwedge_{\kappa \in \mathsf{Ord}}(x\mapsto x\cdot \alpha)^\kappa(\top)$
   (see Remark~\ref{remark:right_distributivity}).

This completes the proof of the lemma.
\end{proof}

We are now ready to proceed with the proof of Theorem~\ref{theorem:kleene_omega_regular}.

\begin{proof}(Theorem~\ref{theorem:kleene_omega_regular})
 We have
\[\omega\mathfrak{Rat}(\mathcal{A}) \supseteq \omega\mathfrak{Reg}(\mathcal{A})\] as it is
enough to note that
\[||\alpha,\mathfrak{F}||_{\omega}\cdot i_n = (\mathfrak{f_F}\cdot
\alpha^+)^\omega \cdot i_n = (\mathfrak{f_F}\cdot \alpha^+)^\omega \cdot
\mathfrak{f_F}\cdot  \alpha^+ \cdot i_n\] and take $r= \mathfrak{f_F} \cdot
\alpha^+\cdot  i_n$, $r_k =  \mathfrak{f_F} \cdot  \alpha^+ \cdot k_n$.

Conversely, let $r, r_i \in \mathfrak{Reg}(1,m)$. Put $s=[r_1,\ldots,r_m]$ and
consider any regular morphism $s'\in \mathfrak{Reg}(m,m)$ such that $r = s'\cdot
1_m$. For sake of clarity of notation let $m_1\defeq m$ and $m_2\defeq m$.
Consider the map $\gamma\defeq \sigma \cdot (s'+s)$, where $\sigma : m+m\rightdcirc m+m$ injects the first component
of the coproduct into the second one and is the identity everywhere else. The morphism $\gamma$ is depicted below:
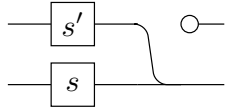
\begin{center}
\resizebox{.2\textwidth}{!}{
\begin{tikzpicture}
\pgfmathsetmacro{\przesuniecieX}{0};

\pgfmathsetmacro{\shadow}{0.0};
\pgfmathsetmacro{\x}{\przesuniecieX};
\pgfmathsetmacro{\y}{0};

\draw[-] (\x-0.5,\y+0.25) -- (\x+1,\y+0.25);
\draw[fill = white] (\x+0,\y+0) rectangle (\x+0.5,\y+0.5);
\node (alpha) at (\x+0.25,\y+0.25) {$s'$};

\pgfmathsetmacro{\y}{-0.7};

\draw[-] (\x-0.5,\y+0.25) -- (\x+1,\y+0.25);
\draw[fill = white] (\x+0,\y+0) rectangle (\x+0.5,\y+0.5);
\node (alpha) at (\x+0.25,\y+0.25) {$s$};

%%%%%%%%%%%%%%%%%%%%%%%%%%%%%%%%%%
%%%%%%%%%%%%%%%%%%%%%%%%%%%%%%%%%%%
%%%%%%%%%%%%%%%%%%%%%%%%%%%%%%%%%%%%
%%%%%%%%%%%%%%%%%%%%%%%%%%%%%%%%%%%%%
%%%%%%%%%%%%%%%%%%%%%%%%%%%%%%%%%%%%%%
\pgfmathsetmacro{\przesuniecieX}{\przesuniecieX+1};
\pgfmathsetmacro{\x}{\przesuniecieX};
\pgfmathsetmacro{\y}{0};

\draw[-] [rounded corners=4pt,-] (\x+0.5,\y+0.25) -- (\x+1,\y+.25);
\draw[fill = white] (\x+0.6,\y+0.25) circle (0.1);
\draw[-] [rounded corners=4pt,-] (\x,\y+0.25)--(\x+0.1,\y+0.25)--(\x+0.2,\y-0.45)-- (\x+0.5,\y-0.45);

\pgfmathsetmacro{\y}{-0.7};
\draw[-] (\x,\y+0.25) -- (\x+1,\y+0.25);
\end{tikzpicture}
}
\end{center}
\noindent Note that this map is a regular morphism, so the map $ \gamma ^\omega
\cdot 1_{m_1+m_2}$ is $\omega$-regular. Moreover, we get:
\begin{align*}
&\gamma ^\omega \cdot 1_{m_1+m_2} = \gamma ^\omega \cdot  \gamma \cdot
1_{m_1+m_2}  =    \gamma^\omega \cdot \mathsf{in}^{m_2}\cdot  s' \cdot 1_{m_1} =
[r_1,\ldots,r_m]^\omega \cdot r.
\end{align*}
This completes the proof.
\end{proof}

%%%% section %%%%%%%%%%%%%%%%%%%%%%%%%%%%%%%%%%%%%%%%%%%%%%%%%%%%
\section{Probabilistic automata}%
\label{section:probabilistic_systems}
%%%%%%%%%%%%%%%%%%%%%%%%%%%%%%%%%%%%%%%%%%%%%%%%%%%%%%%%%%%%%%%%

The main purpose of this section is to put probabilistic systems~\cite{baier97:cav,sokolova09:sacs,sokolova11,
urabe_et_al:LIPIcs:2016:6186,baier:2005:RRL:1078035.1079689,baier:2012:PO:2108242.2108243}
into the framework of Section~\ref{section:automata}. Here, we focus our attention
on probabilistic automata which are akin to
fully probabilistic systems from
Example~\ref{example:fully_prob}\cite{baier97:cav,brengos2015:jlamp,baier00,sokolova11}.

\subsection{Preliminaries}
A \emph{probabilistic
automaton} is a tuple \[(X,\Sigma,P:X\times \Sigma \times X\to [0,1],\mathfrak{F}),\]
where $X$ is a set of \emph{states}, $\Sigma$ an \emph{alphabet},
$P$ is a \emph{probability transition function},
i.e.\ a function such that
for any $x\in X$ we have $\sum_{(a,y)\in \Sigma\times X} P(x,a,y) = 1$,
and $\mathfrak{F}\subseteq X$ the set of \emph{accepting states}.

 For $C\subseteq X$ we define
$P(x,a,C) \defeq \sum_{y\in C} P(x,a,y)$. An \emph{execution fragment} is a finite sequence
$\mathfrak{s} = x_0\stackrel{a_0}{\to} x_1 \stackrel{a_1}{\to}x_2 \ldots x_{n-1}\stackrel{a_{n-1}}{\to} x_n$ such
that $P(x_i,a_i,x_{i+1}) > 0$. We define $first (\mathfrak{s} ) = x_0$,
 $last (\mathfrak{s}) = x_n$, $length(\mathfrak{s}) = n$, $trace(\mathfrak{s}) = a_0\ldots a_{n-1}$ and
$P(\mathfrak{s}) = \prod_{i=0,\ldots,n-1} P(x_i,a_i,x_{i+1})$. An \emph{execution} is an infinite sequence
$\mathfrak{p} = x_0\stackrel{a_0}{\to} x_1\stackrel{a_1}{\to} x_2 \stackrel{a_2}{\to}\dots$
with $P(x_i,a_i,x_{i+1})>0$.

Let $first(\mathfrak{p}) \defeq  x_0$, $trace(\mathfrak{p}) \defeq a_0
a_1\ldots$,
$\mathfrak{p}^{(n)} \defeq x_0\stackrel{a_0}{\to} \ldots \stackrel{a_{n-1}}{\to}x_n$ and $\mathfrak{p}_n \defeq x_n$.
For an execution fragment
$\mathfrak{s}$ of length $n$ let $\mathfrak{s}\uparrow$ denote the set of all executions
$\mathfrak{p}$ such that $\mathfrak{p}^{(n)} = \mathfrak{s}$.

Let $Exec(x)$ denote the set of all executions $\mathfrak{p}$ such that $first (\mathfrak{p})=x$.
Let $\Sigma(x)$ be the smallest sigma field on $Exec(x)$ which contains all sets $\mathfrak{s}\uparrow$ for any
execution fragment $\mathfrak{s}$ with $first(\mathfrak{s}) = x$. Finally, let $\mathcal{Q}_x$ denote the unique
probability measure on $\Sigma(x)$ such that
$\mathcal{Q}_x(\mathfrak{s}\uparrow) = P(\mathfrak{s})$ for any execution
fragment $\mathfrak{s}$ with $first(\mathfrak{s}) = x$. We will often drop the subscript
and write $\mathcal{Q}$ instead of $\mathcal{Q}_x$ if the measure can be deduced from the context.

For $\Lambda \subseteq \Sigma^\ast$ and $C\subseteq X$ define $Exec(\Lambda,C)$ to be the set of all
 executions $\mathfrak{p}=x_0\stackrel{a_0}{\to}x_1\ldots $ for which there is $n$ with $trace(\mathfrak{p}^{(n)})
 \in \Lambda$ and $x_n\in C$ and consider $Exec(x,\Lambda, C) \defeq Exec(\Lambda,C)\cap Exec(x)$.
 As stated in~\cite{baier97:cav} the set $Exec(x,\Lambda,C)$ is $\Sigma(x)$-measurable. Additionally,
 put $Exec(\Lambda) \defeq Exec(\Lambda,X)$ and $Exec(x,\Lambda) \defeq Exec(x,\Lambda,X)$.

An execution $\mathfrak{p}=x=x_0\stackrel{a_0}{\to}x_1\stackrel{a_1}{\to}\ldots $
starting at $x$ is called \emph{$C$-accepting} provided that it visists $C$ infinitely often, i.e.
the set $\{i<\omega \mid x_i \in C\}$ is infinite. Let $AccExec(C)$ denote all $C$-accepting executions and let
$AccExec(x,C) = AccExec(C) \cap Exec(x)$. The set $AccExec(x,C)$ is $\Sigma(x)$-measurable as

\begin{align*}
  & AccExec(x,C) = \bigcap_{n\geq 0}\bigcup_{k\geq n}
    \{\mathfrak{p}\in Exec(x) \mid \mathfrak{p}_k \in C \}=\\
  & \bigcap_{n\geq 0}\bigcup_{k\geq n}
    Exec(x,\{ \sigma\in \Sigma^\ast \mid \text{ length of }\sigma = k\} ,C)=\\
  & \bigcap_{n\geq 0} Exec(x, \{\sigma\in \Sigma^\ast \mid \text{ length of }\sigma \geq n\},C).
\end{align*}

\noindent A curious reader is referred to e.g.~\cite{baier00,baier97:cav} for more details on fully probabilistic systems
and probability measures they induce.
\begin{exa}\label{example:probabilistic_auto}
  Consider the automaton $(\{s_0,s_1\},\Sigma = \{0,1\},P, \mathfrak{F}=\{s_1\})$, where
	$P$ is the probability transition function given by the diagram below:
	\begin{center}
    \resizebox{0.22\textwidth}{!}
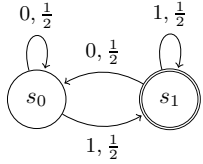
{
    \begin{tikzpicture}[shorten >=1pt,node distance=2cm,on grid]
      \node[state]   (q_0)                {$s_0$};
      \node[state,accepting] (q_2) [right=of q_0] {$s_1$};
      \path[->] (q_0) edge[bend right] node [below] {\small $1,\frac12$}
							  (q_2) edge[loop above] node         {\small $0,\frac12$} ()
                (q_2) edge[bend right] node [above] {\small $0,\frac12$} (q_0);
      \path[->] (q_2) edge[loop above] node         {\small $1,\frac12$} ();
    \end{tikzpicture}
    }
	\end{center}
	Let $\Lambda \subseteq \Sigma^\ast$. Then
  $Exec(s_0,\Lambda, \mathfrak{F})$ consists of all executions
	$\mathfrak{p}$ such that $\mathfrak{p} =
  x_0 \stackrel{a_0}{\to} x_1\stackrel{a_1}{\to} x_2 \stackrel{a_2}{\to} \ldots$
	for which $x_0 = s_0$,
  \begin{align}
    (x_i,a_i,x_{i+1}) \text{ is from }
    \{(s_0,0,s_0), (s_0,1,s_1), (s_1,0,s_0), (s_1,1,s_1)\}, \label{condition:exec}
  \end{align}
  and such that there is $n$ with $a_0\ldots a_n\in \Lambda$ and $a_n=1$.
  Similarily, $Exec(s_1,\Lambda,\mathfrak{F})$
	contains all executions $\mathfrak{p}$ such that
  $first(\mathfrak{p}) = s_1$, satisfying~\ref{condition:exec}
  and whose $trace(\mathfrak{p}^{(n)})\in \Lambda$ with its last
	letter equal to $1$ for some $n\in\mathbb{N}$. In particular this means that
  $\mathcal{Q}(Exec(s_i,\{w0\},\mathfrak{F})) = 0$ and
  $\mathcal{Q}(Exec(s_i,\{w1\},\mathfrak{F})) = \left(\frac12 \right)^{n+1}$
  for any $w\in \Sigma^\ast$ whose length equals to $n$.

  Now, $AccExec(s_0,\mathfrak{F})$ is the set of all executions
	$\mathfrak{p} = s_0 = x_0 \stackrel{a_0}{\to} x_1\stackrel{a_1}{\to} x_2
  \stackrel{a_2}{\to} \ldots$
  such that it satisfies~\ref{condition:exec}
  and for any $n\in \mathbb{N}$
  there is $i_n\geq n$ such that $(x_{i_n},a_{i_n}, x_{i_n+1})$ equals to
  $(s_i,1,s_1)$.
	Similarily, $AccExec(s_1, \mathfrak{F})$ is the set of all
  executions whose first state is $s_1$ which
	satisfy the same argument as above. It is easy to see that
  $\mathcal{Q}(AccExec(s_0,\mathfrak{F}))=\mathcal{Q}(AccExec(s_1,\mathfrak{F}))=1$%
\footnote{This follows by the fact that any execution
      $\mathfrak{p}=x_0\stackrel{a_0}{\to}x_1\stackrel{a_1}{\to}\ldots$ satisfying~\ref{condition:exec} that is not a member of $AccExec(s_i, \mathfrak{F})$
      satisfies the following condition: there is a natural number
      $n$ such that for all $i\geq n$ we have $x_i = s_0$ and $a_i = 0$.
      Hence, the probability of the set of all such executions equals $0$.
  }.

\end{exa}
\noindent The remaining part of this section will focus on finding a suitable setting in
which we can model probabilistic automata and their (in)finite behaviour using
the framework presented in the previous section.

\begin{rem}\label{remark:probabilistic_automata_coalgebraically}
   Any probabilistic automaton can be modelled coalgebraically as a pair
   $(\alpha: X \to \mathcal{D}(\Sigma \times X),\mathfrak{F}\subseteq X)$,
	 where $\mathcal{D}$ is the subdistribution monad from Example~\ref{example:distr_monad}
	 and $\alpha$ is given by (see e.g.~\cite{sokolova09:sacs,bonchi2015killing}):
   \[
     \alpha(x)(a,y) \defeq P(x,a,y).
   \]
   The Kleisli category associated with the monad
   $\mathcal{D}(\Sigma^\ast \times\mathcal{I}d+\Sigma^\omega )$\footnote{By applying the construction from
   Section~\ref{section:monads} to $T=\mathcal{D}$ and
	 $F=\Sigma\times \mathcal{I}d$ we obtain a monadic structure
   on the aforementioned functor.}
	 is order enriched with the hom-set ordering given for
   $f,g:X\to \mathcal{D}(\Sigma^\ast\times Y +\Sigma^\omega) $ by:
   \[
     f\leq g \iff f(x)(y)\leq g(x)(y).
   \]
   Although the Kleisli category for this monad is $\cpo$-enriched,
   its hom-posets do \emph{not} admit arbitrary finite suprema~\cite{hasuo07:trace,brengos2015:jlamp}. In other words, the setting
   is incompatible with the setting from the previous section. Hence, the remaining
   part of this section is focused on solving this issue based on ideas given in~\cite{gp:icalp2014,brengos2015:jlamp}.
\end{rem}

\subsection{Choosing the right monad}
Here, we introduce a monad which is a suitable replacement for $\mathcal{D}$, i.e.
it satisfies the desired properties to make it
suitable for modeling (in)finite behaviours of probabilistic automata.
Inspired by~\cite{gp:icalp2014} we consider the \emph{continuous continuation
  monad} parametrized by the set $[0,1]$ whose functorial part is
defined for any set $X$ by:
\[
  \mathfrak{D} X \defeq (X\to [0,1])\to_\omega [0,1],
\]
where $P\to_\omega Q$ denotes the set of functions between two
\cpo's $P$ and $Q$ which preserve suprema of $\omega$-chains.
The identity maps and the composition in the Kliesli category for
$\mathfrak{D}$ are as follows:
\begin{align*}
  & \mathsf{id}:X\to (X\to [0,1])\to_\omega [0,1]; x\mapsto \Delta_x,
                                 \text{ where } \Delta_x(d) \defeq d (x). \\
  & g\cdot f: X\to (Z\to [0,1])\to_\omega [0,1],
    \text{ where } (g\cdot f)(x)( d ) \defeq f(x)(y\mapsto g(y)(d))
\end{align*}
with  $f:X\to (Y\to [0,1])\to_\omega [0,1]$ and
$g:Y\to (Z\to [0,1])\to_\omega [0,1]$. Note that any
$d:X\to [0,1]\in \mathcal{D}X$ can be assigned a function
$\nu_X(d):(X\to [0,1])\to_\omega [0,1]$
which maps any $d':X\to [0,1]$ onto
\begin{align}
  \nu_X(d)(d')=\sum_{x\in X} d'(x)\cdot d(x).\label{formula_natural_transformation_probabilistic}
\end{align}
It is not hard to see that this turns the family $\{\nu_X:\mathcal{D}X\to \mathfrak{D} X\}$
into a natural transformation $\nu:\mathcal{D}\implies \mathfrak{D}$
which is a monad morphism between the monads $\mathcal{D}$ and $\mathfrak{D}$.
Additionally, it is easy to see that there is a natural ordering of arrows in
$\mathcal{K}l(\mathfrak{D})$ which share the same domain and codomain.
Indeed, for $f,g:X\to \mathfrak{D} Y$ we have:
\[
f\leq g \iff f(x)(d)\leq g(x)(d) \text{ for any }x\in X \text{ and } d:Y\to [0,1].
\]
This turns the Kleisli category for the monad  $\mathfrak{D}$ into an
order enriched category. Moreover, the following theorem is true.

\begin{lem} The order enrichment of $\mathcal{K}l(\mathfrak{D})$ is pointwise induced,
every hom-set of $\mathcal{K}l(\mathfrak{D})$ is a complete lattice and
the category is $\cpo$-enriched and left distributive.
\end{lem}
\begin{proof}
  The fact that the partial order is a complete lattice order is a direct corollary from the definition
  of $\mathfrak{D}$.
  We will now prove that Kleisli category for $\mathfrak{D}$ is $\cpo$-enriched.
  Take any ascending chain $\{f_i:X\to \mathfrak{D}Y\}_{i<\omega}$ of morphisms and note that
  \begin{align}
    & \label{eq:left_dist_1} [g\cdot (\bigvee_i f_i)](x)(d) {=} (\bigvee_i f_i)(x)(y\mapsto g(y)(d)) =  \\
    & \label{eq:left_dist_2} \bigvee_i f_i (x)(y\mapsto g(y)(d)) = \bigvee_i g\cdot f_i(x)(d) \text{ and }\\
    &  [(\bigvee_i f_i)\cdot h](z)(d) = h(z)(x\mapsto \bigvee_i f_i(x)(d)) \stackrel{\dagger}{=}\\
    &  \bigvee_i h(z)(x\mapsto f_i(x)(d)) = \bigvee_i (h\cdot f_i)(z)(d),
  \end{align}
  where the equality marked with ($\dagger$) follows from the fact that $\mathfrak{D}X$ consists
  of functions $(X\to [0,1])\to_\omega [0,1]$ that preserve $\omega$-chains.
  Note that the identities~\ref{eq:left_dist_1} and~\ref{eq:left_dist_2} hold more generally for an
  arbitrary family of morphisms $\{f_i\}_{i\in I}:X\to \mathfrak{D} Y $ and $g:Y\to
  \mathfrak{D} Z$. This shows left distributivity of the Kliesli category and ends the proof.
\end{proof}
From the above it follows that $\mathcal{K}l(\mathfrak{D})$ satisfies~\ref{condition:0} -~\ref{condition:3}
from Section~\ref{section:automata}.

\subsubsection{ The monad for probabilistic automata and their behaviours}

By instantiating the construction from Section~\ref{section:monads}
for $T=\mathfrak{D}$ and $F=\Sigma\times \mathcal{I}d$ we obtain
the monad $\mathfrak{D}(\Sigma^\ast \times \mathcal{I}d + \Sigma^\omega)$.
We explicitly spell out the formula for the identity maps and the composition in
$\mathcal{K}l(\mathfrak{D}(\Sigma^\ast\times \mathcal{I}d+\Sigma^\omega))$ as it
will be used throughout the remaining part of this section.
For any set $X$ the identity map $\mathsf{id}$ is given by:
\begin{align*}
    & \mathsf{id}:X\to (\Sigma^\ast \times X +\Sigma^\omega\to [0,1])\to_\omega [0,1], \\
    & \mathsf{id}(x)(d) = d(\varepsilon,x).
\end{align*}
Moreover, for $f:X\to \mathfrak{D}(\Sigma^\ast \times Y+\Sigma^\omega)$ and
 $g:Y\to \mathfrak{D}(\Sigma^\ast\times Z+\Sigma^\omega)$ the map
 $g\cdot f: X\to \mathfrak{D}(\Sigma^\ast \times Z+\Sigma^\omega)$ is:
\begin{align*}
    & g\cdot f :X\to (\Sigma^\ast \times Z+\Sigma^\omega\to [0,1])\to_\omega [0,1], \\
    & (g\cdot f) (x)(d) = f(x)((\sigma,y) \mapsto g(y)(d_{|\sigma}) \text{ and }  v\mapsto d(v) ),
\end{align*}
where for $d:\Sigma^\ast\times Z+\Sigma^\omega \to [0,1]$ and $\sigma\in \Sigma^\ast$
the map $d_{|\sigma}:\Sigma^\ast\times Z+\Sigma^\omega\to [0,1]$ is given by
$d_{|\sigma}(\tau,z) = d(\sigma\tau,z)$ for $\tau\in \Sigma^\ast$, $z\in Z$ and
$d_{|\sigma}(v) = d(\sigma v)$ for $v\in \Sigma^\omega$.

The following statement is a direct consequence of the definition of the monad
$\mathfrak{D}(\Sigma^\ast\times \mathcal{I}d+\Sigma^\omega)$ and the properties
of $\mathfrak{D}$ from the previous subsection.

\begin{thm}\label{theorem:probabilistic_fit_the_framework}
  $\mathcal{K}l(\mathfrak{D}(\Sigma^\ast \times \mathcal{I}d +
  \Sigma^\omega))$ satisfies~\ref{condition:0} -~\ref{condition:3} from Section~\ref{section:automata}.
\end{thm}

\subsection{(In)finite behaviour}

Let $(\alpha,\mathfrak{F})$ be a pair as in Remark~\ref{remark:probabilistic_automata_coalgebraically}
and consider
\[
(\widehat{\alpha}:X\stackrel{\alpha}{\to}\mathcal{D}(\Sigma\times X)
\stackrel{\nu_{\Sigma\times X}}{\to} \mathfrak{D}
(\Sigma \times X)
\hookrightarrow \mathfrak{D}(\Sigma^\ast \times X+\Sigma^\omega),\mathfrak{F}),
\]
where $\nu$ is given
by~\ref{formula_natural_transformation_probabilistic}. We see that the map
$\widehat{\alpha}$ can be viewed as an endomorphism in the Kleisli category for the monad
$\mathfrak{D} (\Sigma^\ast \times \mathcal{I}d +\Sigma^\omega)$.

The rest of this section is devoted to presenting some properties of
 $||\widehat{\alpha},\mathfrak{F}||$ and $||\widehat{\alpha},\mathfrak{F}||_\omega$ for the pair
$(\widehat{\alpha},\mathfrak{F})$. We show that the values $||\widehat{\alpha},\mathfrak{F}||(x)$
and $||\widehat{\alpha},\mathfrak{F}||_\omega(x)$ of behaviour functions
encode probabilities of certain events from $\Sigma(x)$. The main results are summarized in
Theorem~\ref{theorem:saturation_formula_probabilistic} and~\ref{theorem:infinite_beh_probabilistic}.

\subsubsection{Finite behaviour} At first let us focus on describing the finite behaviour map
$||\widehat{\alpha},\mathfrak{F}||$. Before we do that (see Theorem~\ref{theorem:saturation_formula_probabilistic} for details)
we need to present some intermediate results first.
The following lemma is a direct consequence of the definition of
$\widehat{\alpha}^\ast,\widehat{\alpha}^+$
and the composition in $\mathcal{K}l(\mathfrak{D}(\Sigma^\ast\times \mathcal{I}d+\Sigma^\omega))$.
\begin{lem}\label{lemma:saturation_for_probabilistic}
  We have:
  \begin{align*}
    & \widehat{\alpha}^+,\widehat{\alpha}^\ast:X\to (\Sigma^\ast \times X +\Sigma^\omega\to [0,1])
      \to_\omega [0,1]; \\
    & \widehat{\alpha}^+(x)(d) =
      \sum_{(a,y)\in \Sigma \times X} \widehat{\alpha}^\ast(y)(d_{|a})\cdot \alpha(x)(a,y)\\
    & \widehat{\alpha}^\ast (x)(d) =
      \max \{d(\varepsilon,x) \ , \   \sum_{(a,y)\in \Sigma \times X }
      \widehat{\alpha}^\ast (y)(d_{|a}) \cdot \alpha(x)(a,y) \}.
     \end{align*}
\end{lem}

Now, consider a subset $\mathfrak{F}\subseteq X$ and note that, in the setting of this
subsection, the map $\mathfrak{f_{F}}$ is explicitly given by:

\begin{align*}
  & \mathfrak{f}_{\mathfrak{F}}:X\to (\Sigma^\ast\times X+\Sigma^\omega\to
    [0,1])\to_\omega [0,1], \\
  & \mathfrak{f}_{\mathfrak{F}}(x)(d) =\left \{
    \begin{array}{cc} d(\varepsilon,x) & \text{ for }x\in \mathfrak{F},\\
                               0       & \text{ otherwise.}
    \end{array} \right.
\end{align*}
Hence, the map $!\cdot \mathfrak{f_{{F}}}:X\to (\Sigma^\ast\times
1+\Sigma^\omega\to [0,1])\to_\omega [0,1]$ is given by:

\begin{align*}
  & (!\cdot \mathfrak{f}_{\mathfrak{F}})(x)(d) =\left \{
    \begin{array}{cc} d(\varepsilon,1) & \text{ for }x\in \mathfrak{F},\\
                               0       & \text{ otherwise.}
    \end{array} \right.
\end{align*}
Therefore, the finite behaviour of $(\widehat{\alpha},\mathfrak{F})$ is:
\begin{align*}
    & ||\widehat{\alpha},\mathfrak{F}||:X\to
      (\Sigma^\ast\times 1 +\Sigma^\omega\to [0,1])\to_\omega [0,1] \\
    & ||\widehat{\alpha},\mathfrak{F} ||(x)(d) = (!\cdot \mathfrak{f_F}\cdot \widehat{\alpha}^*)(x)(d) =  \\
    & \widehat{\alpha}^\ast (x) \left(
      (\sigma, y)\mapsto \left \{ \begin{array}{cc}
                                    d(\sigma,1) & y\in \mathfrak{F}, \\
                                              0 & \text{otherwise}
                                  \end{array}
                          \right. \text{ and } v\mapsto 0
                                \right).
\end{align*}
For $\Lambda \subseteq \Sigma^\ast$ and $\sigma \in \Sigma$
put  $\Lambda/\sigma \defeq \{ \tau \mid \sigma \tau \in \Lambda\}$ and consider a mapping
$\chi_{\Lambda}:\Sigma^\ast \times 1+\Sigma^\omega \to [0,1]$ given by:
\begin{align}
  \chi_\Lambda (x) = \left \{
    \begin{array}{cc} 1 & x=(\sigma,1) \text{ and }\sigma\in \Lambda \\ 0
    & \text{ otherwise}.
    \end{array} \right.\label{equation_chi}
\end{align}
If we define a function
$\chi_{\mathfrak{F},\Lambda}:\Sigma^\ast\times X+\Sigma^\omega \to [0,1]$
by:
\begin{align}
  \chi_{\mathfrak{F},\Lambda}(\tau,y) = \left \{ \begin{array}{cc}
                                              1 & y\in \mathfrak{F}\text{ and
                                                  }\tau \in \Lambda, \\
                                              0 & \text{otherwise}
                                  \end{array}
                          \right. \text{ and } \chi_{\mathfrak{F},\Lambda}(v)= 0.%
                          \label{equation:chi_extended}
\end{align}
then $||\widehat{\alpha},\mathfrak{F}||(x)(\chi_{\Lambda}) =
      \widehat{\alpha}^\ast(x)(\chi_{\mathfrak{F},\Lambda})$ and:
\begin{align}
  & ||\widehat{\alpha},\mathfrak{F}||(x)(\chi_{\Lambda}) =
  & \left \{ \begin{array}{cc}
               1 &  \varepsilon\in \Lambda \text{ and }x\in \mathfrak{F} \\
               \sum_{(a,y)\in \Sigma \times X }
               \widehat{\alpha}^\ast (y) (\chi_{\mathfrak{F},\Lambda/a} ) \cdot \alpha(x)(a,y)  &
                                                                           \text{ otherwise.}
             \end{array}
           \right.\label{equation:behaviour}
\end{align}
A careful analysis of the formulae from~\cite{baier97:cav,brengos2015:jlamp} describing the value
$\mathcal{Q}(Exec(x,\Lambda,C))$ and the above observations lead us to the statement below.
It turns out that for any state $x$ the value $||\widehat{\alpha},\mathfrak{F}||(x)(\chi_\Lambda)$ is
  the probability of reaching a state in $\mathfrak{F}$  from $x$
  via an execution fragment whose trace is a member of $\Lambda$: %%%%%%%%%%%%%%%%%%%%%%%%%%
\begin{thm}\label{theorem:saturation_formula_probabilistic}
  We have:
  \begin{align*}
    & ||\widehat{\alpha},\mathfrak{F}||(x)(\chi_\Lambda)
        = \mathcal{Q}(Exec(x,\Lambda,\mathfrak{F})).
  \end{align*}
 \end{thm}

\subsubsection{Infinite behaviour}

Let us focus on the infinite behaviour of $(\widehat{\alpha},\mathfrak{F})$ introduced in the previous section
given by \[||\widehat{\alpha},\mathfrak{F}||_\omega = (\mathfrak{f_F}\cdot
\widehat{\alpha}^+)^\omega: X\to \mathfrak{D}(\Sigma^\omega).\]
The following theorem gives us insight into what (some of) the values of the infinite behaviour function are.
To be more precise, we show that for a state $x$ and a subset $\Lambda \subseteq \Sigma^\ast$,
the value of the infinite behaviour of $x$ calculated for the characteristic
function of the set of all infinite sequences from $\Sigma^\omega$ with prefixes in $\Lambda$ equals to the probability
of $\mathfrak{F}$-accepting executions starting at $x$ whose trace prefix belongs to $\Lambda$.

\begin{thm}\label{theorem:infinite_beh_probabilistic}
  For any $\Lambda \subseteq \Sigma^\ast$ we have:
 \begin{align*}
    & ||\widehat{\alpha},\mathfrak{F}||_\omega : X\to (\Sigma^\omega \to [0,1])\to_\omega [0,1], \\
    & ||\widehat{\alpha},\mathfrak{F}||_\omega(x)(\chi_{\Lambda \cdot \Sigma^\omega }) =
    \mathcal{Q}(Exec(x,\Lambda) \cap AccExec(x,\mathfrak{F})),
  \end{align*}
  where $\Lambda \cdot \Sigma^\omega \defeq \{wv\in \Sigma^\omega \mid  w \in \Lambda\text{ and } v \in \Sigma^\omega \}$.
\end{thm}
\begin{proof} The proof is divided into three parts.

  \noindent \textbf{Part 1.}
  We will first show the statement holds for $\Lambda = \{\varepsilon\}$, i.e.\ we prove that
     \[
     ||\widehat{\alpha},\mathfrak{F}||_\omega(x)(\chi_{\Sigma^\omega}) =
     \mathcal{Q}(AccExec(x,\mathfrak{F})).\]
     Indeed, since
     \[
       AccExec(x,\mathfrak{F}) = \bigcap_{n\geq 0}\bigcup_{k\geq n} \{\mathfrak{p}\in Exec(x) \mid \mathfrak{p}_k \in \mathfrak{F} \}
     \]
     we have:
     \[
     \mathcal{Q}(AccExec(x,\mathfrak{F}) )= \lim_{n\to \infty} \mathcal{Q}(A_n^x),
     \]
     where $A_n^x \defeq \bigcup_{k\geq n} \{\mathfrak{p}\in Exec(x) \mid \mathfrak{p}_k \in \mathfrak{F} \}=
     Exec(x,\{\sigma\in \Sigma^\ast \mid \text{ length of }\sigma \geq n
     \},\mathfrak{F})$
     is a descending chain of $\Sigma(x)$-measurable sets. Let us now consider a family of maps $\{G_n:X\to (\Sigma^\omega \to
[0,1])\to_{\omega} [0,1]\}_n$ defined inductively as follows:
\begin{align*}
  G_0(x)(d) = 1 \text{ and } G_{n+1} = G_{n} \cdot \mathfrak{f_F}\cdot \widehat{\alpha}^+.
\end{align*}
Note that the sequence $\{G_n(x)(d)\}_n$ is descending for any fixed $x\in X$
and $d$ and that
\begin{align}
||\widehat{\alpha},\mathfrak{F}||_\omega(x)(d) = \lim_{n\to \infty} G_n(x)(d).
\label{equation:limit_omega}
\end{align}
We will now show that $G_n(x)(\chi_{\Sigma^\omega}) = \mathcal{Q}(A_{n}^x)$ for $n\geq 1$. For $G_1$ we have:
\begin{align}
 &  G_{1}(x)(\chi_{\Sigma^\omega}) = (G_0\cdot \mathfrak{f_F}\cdot \widehat{\alpha}^+)(x)(\chi_{\Sigma^\omega})=  \\
 &  \widehat{\alpha}^+(x)((\sigma,y) \mapsto G_0\cdot \mathfrak{f_F}(y)(\chi_{\Sigma^\omega | \sigma})
  \text{ and }v\mapsto \chi_{\Sigma^\omega}(v)   ) = \\
 &  \widehat{\alpha}^+(x)((\sigma,y) \mapsto G_0\cdot \mathfrak{f_F}(y)(\chi_{\Sigma^\omega})
  \text{ and }v\mapsto \chi_{\Sigma^\omega}(v)   ).\label{formula_conditional_probability_0}
\end{align}
In the above
\begin{align*}
& G_0\cdot \mathfrak{f_F}(y)(\chi_{\Sigma^\omega}) =
\mathfrak{f_F}(y)((\tau,z)\mapsto G_0(z)(\chi_{\Sigma^\omega | \tau})
\text{ and } v\mapsto \chi_{\Sigma^\omega}(v) ) =\\
& \mathfrak{f_F}(y)((\tau,z)\mapsto 1 \text{ and } v\mapsto 1 ) =
\left \{ \begin{array}{cc} 1 & y\in \mathfrak{F} \\ 0 & \text{ otherwise.} \end{array} \right.
\end{align*}
Hence, if we continue with~\ref{formula_conditional_probability_0} we get:
\begin{align*}
\widehat{\alpha}^+(x)\left((\sigma,y) \mapsto \left \{ \begin{array}{cc} 1 & y\in \mathfrak{F} \\ 0 & \text{ otherwise} \end{array} \right.
  \text{ and }v\mapsto 1   \right) = \mathcal{Q}(A_1^x).
\end{align*}
If we now assume by induction that the statement holds for some $n>1$ then
\begin{align*}
G_n\cdot \mathfrak{f_F} (y)(\chi_{\Sigma^\omega}) = \mathfrak{f_F}(y)((\tau,z)\mapsto \mathcal{Q}(A_n^z) \text{ and } v\mapsto 1) =
\left \{
 \begin{array}{cc}
    \mathcal{Q}(A_n^y) & y\in \mathfrak{F} \\
    0 & \text{ otherwise}.
 \end{array}
\right.
\end{align*}
Hence, by following a similar reasoning to the one applied to $G_1$ we get:
\begin{align*}
G_{n+1}(x)(\chi_{\Sigma^\omega}) = \widehat{\alpha}^+
\left(
(\sigma,y)\mapsto
\left \{
 \begin{array}{cc}
    \mathcal{Q}(A_n^y) & y\in \mathfrak{F} \\
    0 & \text{ otherwise}
 \end{array}
\right.
\text{ and }
v\mapsto 1
\right)  = \mathcal{Q}(A_{n+1}^x).
\end{align*}

\noindent \textbf{Part 2.} We will now show that the following holds:
\[
  ||\widehat{\alpha},\mathfrak{F}||_\omega(x)(\chi_{\{a_0a_1\ldots a_n\}\cdot \Sigma^\omega}) =
  \mathcal{Q}(Exec(x,\{a_0\ldots a_n\}) \cap AccExec(x,\mathfrak{F})).
\]
Assume $\mathfrak{s}=x=x_0\stackrel{a_0}{\to}\ldots \stackrel{a_{n-1}}{\to} x_n  \stackrel{a_{n}}{\to} x_{n+1}=x'$.
%Then $\chi_{trace(\mathfrak{s}\uparrow)} = \chi_{\{a_0 a_1\ldots a_{n}\}\cdot \Sigma^\omega}$ and
Then for
$\sigma \in \Sigma^\ast$ of length less than or equal to $n$ we have
$\chi_{\{a_0 a_1\ldots a_{n}\}\cdot \Sigma^\omega | \sigma}$
is equal $\chi_{\{a_k\ldots a_{n}\}\cdot \Sigma^\omega}$
if $\sigma = a_0\ldots a_{k-1}$ and it is the constantly equal to zero function otherwise.
This observation together with the fact that
\begin{align*}
& \mathcal{Q}(\mathfrak{s}\uparrow \cap AccExec(x,\mathfrak{F}))=
    \mathcal{Q}(AccExec(x,\mathfrak{F})\mid \mathfrak{s}\uparrow )\cdot Q(\mathfrak{s}\uparrow)=\\
& \mathcal{Q}(AccExec(x',\mathfrak{F}))\cdot \mathcal{Q}(\mathfrak{s}\uparrow)
\end{align*}
and induction allows us to prove the assertion.

\noindent \textbf{Part 3.} The statement from {Part 2.} can be easily generalized to
\[
  ||\widehat{\alpha},\mathfrak{F}||_\omega(x)(
  \chi_{
    \{w_1\}\cdot \Sigma^\omega\cup \{w_2\}\cdot \Sigma^\omega
  }
  ) =
  \mathcal{Q}(Exec(x,\{w_1,w_2\}) \cap AccExec(x,\mathfrak{F})),
\]
for any pair of incomparable words $w_1,w_2\in \Sigma^\ast$ with respect to lexicographic
ordering on $\Sigma^\ast$. Indeed, in this case the sets $\{w_1\}\cdot \Sigma^\omega$
and $\{w_2\}\cdot \Sigma^\omega$ are disjoint. If
$\sigma$ is not comparable with $w_1$ and $w_2$ then
$\chi_{ \{w_1\}\cdot \Sigma^\omega\cup \{w_2\}\cdot \Sigma^\omega |\sigma} $ is
constantly equal to zero function. However, if $\sigma$ is comparable wit $w_i$ then
for all such $\sigma$ which are sufficiently long we have:
$\chi_{ \{w_1\}\cdot \Sigma^\omega\cup \{w_2\}\cdot \Sigma^\omega |\sigma}
= \chi_{\{w_i\}\cdot \Sigma^\omega |\sigma} $. Hence, by the same argument as before we prove
the desired assertion. Note that the equation generalizes to any finite
set of incomparable words $\{w_1,\ldots,w_n\}$.

Finally, the general statement holds since the function
\[
||\widehat{\alpha},\mathfrak{F}||_\omega(x): (\Sigma^\omega \to [0,1]) \to_\omega [0,1]
\]
preserves suprema of $\omega$-chains. Indeed, let $\Lambda \subseteq \Sigma^\ast$ and note
that $\Lambda \cdot \Sigma^\omega = \Lambda'\cdot \Sigma^\omega$ for a countable subset
$\Lambda'\subseteq \Lambda$ of incomparable words $\Lambda'=\{w_1,w_2,\ldots\}$. Then
\[
\chi_{\Lambda\cdot \Sigma^\ast} = \chi_{\Lambda'\cdot \Sigma^\omega} =
\bigvee_n \chi_{\{w_1,\ldots,w_n\}\cdot \Sigma^\omega}.
\]
Hence,
\begin{align*}
 & ||\widehat{\alpha},\mathfrak{F}||_\omega(x)(
  \chi_{
    \Lambda\cdot \Sigma^\omega
  }
  ) =
 ||\widehat{\alpha},\mathfrak{F}||_\omega(x)(
  \chi_{
    \Lambda'\cdot \Sigma^\omega
  }
  ) =
 ||\widehat{\alpha},\mathfrak{F}||_\omega(x)(
  \bigvee_n \chi_{
   \{w_1,\ldots,w_n\}\cdot \Sigma^\omega
  }
  ) = \\
 & \bigvee_n ||\widehat{\alpha},\mathfrak{F}||_\omega(x)(
 \chi_{
   \{w_1,\ldots,w_n\}\cdot \Sigma^\omega
  }
  ) =
  \bigvee_n  \mathcal{Q}(Exec(x,\{w_1,\ldots, w_n\}) \cap AccExec(x,\mathfrak{F})) = \\
 &  \mathcal{Q}(Exec(x,\Lambda') \cap AccExec(x,\mathfrak{F}))=
 \mathcal{Q}(Exec(x,\Lambda) \cap AccExec(x,\mathfrak{F})).
\end{align*}
\end{proof}

\begin{exa}\label{example:probabilistic_auto_v2}
  Let us consider the probabilistic automaton from
  Example~\ref{example:probabilistic_auto} and put it
  into the framework of the Lawvere theory for the monad
  $\mathfrak{D}(\Sigma^\ast\times \mathcal{I}d+\Sigma^\omega)$.
  Below, we calculate (some values of) finite and infinite
  behaviours of the automaton derived from Example~\ref{example:probabilistic_auto}
  in a direct manner (i.e.\ without applying Theorem~\ref{theorem:saturation_formula_probabilistic} or
  Theorem~\ref{theorem:infinite_beh_probabilistic}).  By following the guidelines of
  Remark~\ref{remark:probabilistic_automata_coalgebraically}
  we obtain $(\alpha,\mathfrak{F})$,
  where $\alpha : X\to \mathcal{D}(\Sigma\times X)$ given  by $\alpha(s_0) = P(s_0,-,-)$ and
  $\alpha(s_1) = P(s_1,-,-)$. Moreover, $\widehat{\alpha}: X\to
  \mathfrak{D}(\Sigma^\ast \times X+\Sigma^\omega)$ is defined by
  \begin{align*}
     & \widehat{\alpha}(s_0) : (\Sigma^\ast \times \{s_0,s_1\}
     +\Sigma^\omega \to [0,1]) \to_\omega [0,1];
      d\mapsto \frac12 \cdot d(0,s_0) + \frac12\cdot  d(1,s_1)
  \end{align*}
  and $\widehat{\alpha}(s_1) = \widehat{\alpha}(s_0)$.
  Next, observe that by Lemma~\ref{lemma:saturation_for_probabilistic} the
  morphisms  $\widehat{\alpha}^+$
  and $\widehat{\alpha}^\ast$ are the least solutions to:
  \begin{align*}
    \widehat{\alpha}^+(s_0)(d) &  =  \widehat{\alpha}^+(s_1)(d)  =
      \frac12 \cdot \widehat{\alpha}^\ast(s_0)(d_{|0})
      + \frac12 \cdot \widehat{\alpha}^\ast(s_1)(d_{|1}) \\
    \widehat{\alpha}^\ast (s_0)(d) & =
    \max \{
        d(\varepsilon,s_0),
        \frac12 \cdot \widehat{\alpha}^\ast(s_0)(d_{|0})
          + \frac12 \cdot \widehat{\alpha}^\ast(s_1)(d_{|1})
      \} \\
    \widehat{\alpha}^\ast(s_1)(d)  & =
      \max \{
        d(\varepsilon,s_1),
        \frac12 \cdot \widehat{\alpha}^\ast(s_0)(d_{|0})
          + \frac12 \cdot \widehat{\alpha}^\ast(s_1)(d_{|1})
      \}.
  \end{align*}
  Consider $w=a_0 a_1\ldots a_n\in \Sigma^\ast$, $\chi_{\{w0\}}$, $\chi_{\{w1\}}$
  as in~\ref{equation_chi} and
  $\chi_{\mathfrak{F},\{w0\}}$, $\chi_{\mathfrak{F},\{w1\}}$ as
  in~\ref{equation:chi_extended}. Then by~\ref{equation:behaviour}
  \begin{align*}
    ||(\widehat{\alpha},\mathfrak{F})||(s_0)(\chi_{\{w0\}}) =
    \frac12 \cdot \widehat{\alpha}^\ast (s_0)( \chi_{\mathfrak{F},\{w0\}_{/0}})+
    \frac12 \cdot \widehat{\alpha}^\ast (s_1)(\chi_{\mathfrak{F},\{w0\}_{/1}})
  \end{align*}
  By carefully analysing the formula for $\widehat{\alpha}^\ast$ we conclude that, in our case,
  \[
    ||\widehat{\alpha},\mathfrak{F}||(s_0)(\chi_{\{w0\}}) = 0.
  \]
  Similarily, by~\ref{equation:behaviour} we have
  \begin{align*}
    ||\widehat{\alpha},\mathfrak{F}||(s_0)(\chi_{\{w1\}}) =
    \frac12 \cdot \widehat{\alpha}^\ast (s_0)( \chi_{\mathfrak{F},\{w1\}_{/0}})+
    \frac12 \cdot \widehat{\alpha}^\ast (s_1)( \chi_{\mathfrak{F},\{w1\}_{/1}}).
  \end{align*}
  In this case, however, a thorough analysis of the formula for $\widehat{\alpha}^\ast$ leads
  us to the following conclusion:
  \[
    ||\widehat{\alpha},\mathfrak{F}||(s_0)(\chi_{\{w1\}})=\left (\frac12\right)^{n+1}.
  \]

  Now, in order to compute $||\widehat{\alpha},\mathfrak{F}||_\omega$ first consider
  $\mathfrak{f_F}\cdot \widehat{\alpha}^+$ which is given by:
  \[
    \mathfrak{f_F}\cdot \widehat{\alpha}^+(s_i)(d) =
    \widehat{\alpha}^+(s_i)\left( (\sigma,s_0)\mapsto 0 \text{ and }
    (\sigma,s_1)\mapsto d(\sigma,s_1) \text{ and } v\mapsto d(v) \right).
  \]
  The morphism
  $||\widehat{\alpha},\mathfrak{F}||_\omega=(\mathfrak{f_F}\cdot \widehat{\alpha}^+)^\omega:
  X\to ( \Sigma^\omega \to [0,1]) \to_\omega [0,1]$
  is the greatest map satisfying
  $(\mathfrak{f_F}\cdot \widehat{\alpha}^+)^\omega =
  (\mathfrak{f_F} \cdot \widehat{\alpha}^+)^\omega
  \cdot \mathfrak{f_F}\cdot \widehat{\alpha}^+$. In particular, this means that
  \begin{align*}
    & (\mathfrak{f_F}\cdot \widehat{\alpha}^+)^\omega (s_i) (d) =
      (\mathfrak{f_F}\cdot \widehat{\alpha}^+)^\omega\cdot
      \mathfrak{f_F}\cdot \widehat{\alpha}^+ (s_i) (d) = \\
    & (\mathfrak{f_F}\cdot \widehat{\alpha}^+)(s_i)( (\sigma,s_j)\mapsto
      (\mathfrak{f_F}\cdot \widehat{\alpha}^+)^\omega (s_j) (d_{|\sigma})\text{ and }v\mapsto d(v)  )
   \end{align*}
   For $d=\chi_{\Sigma^\omega}$ we have $d_{|\sigma} = \chi_{\Sigma^\omega}$ and hence
   $(\mathfrak{f_F}\cdot \widehat{\alpha}^+)^\omega$ solves to the following equation for $x$:
   \begin{align}
    & x (s_i) (\chi_{\Sigma^\omega}) =
      (\mathfrak{f_F}\cdot \widehat{\alpha}^+)(s_i)( (\sigma,s_j)\mapsto
      x (s_j)
      (\chi_{\Sigma^\omega})\text{ and }v\mapsto 1  ). \label{equation:fixpoint}
   \end{align}
   In order to compute
   $(\mathfrak{f_F}\cdot \widehat{\alpha}^+)^\omega (s_i) (\chi_{\Sigma^\omega})$
   take  $d':\Sigma^\ast \times X+\Sigma^\omega\to [0,1]$ given by
   $(\sigma,s_0)\mapsto 0 \text{ and } (\sigma,s_1)\mapsto 1 \text{ and } v\mapsto 1$
   and note that it satisfies $d'_{|\sigma} = d'$. Moreover, it is not hard to see that
   $\widehat{\alpha}^\ast(s_i)(d')=1$\footnote{Indeed, $\widehat{\alpha}^\ast(s_1)(d')=1$
     follows trivially from the fact that $d'(\varepsilon,s_1){=}1$.
     By the recursive formula describing $\widehat{\alpha}^\ast$ and by induction
   we show that
   $\widehat{\alpha}^\ast (s_0)(d') \geq \frac12 + \frac14+\cdots + \frac{1}{2^n}$
   for any $n\in \mathbb{N}$. This proves the assertion.}. Therefore, the following equation holds:
   \begin{align*}
     & 1= \frac12 \cdot 1 + \frac12 \cdot 1 = \\
     &
     \frac12 \cdot \widehat{\alpha}^\ast(s_0)(d') + \frac12 \cdot \widehat{\alpha}^\ast (s_1)(d')
     = \widehat{\alpha}^+(s_i) (d') = \\
    & (\mathfrak{f_F}\cdot \widehat{\alpha}^+)(s_i)( (\sigma,s_j)\mapsto
     1\text{ and }v\mapsto 1  ).
   \end{align*}
   This proves that if we put
   \[(\mathfrak{f_F}\cdot \widehat{\alpha}^+)^\omega (s_i) (\chi_{\Sigma^\omega})=1\]
   then it satisfies~\ref{equation:fixpoint}. Thus,
   \[||\widehat{\alpha},\mathfrak{F}||_\omega (s_i)(\chi_{\Sigma^\omega}) =1.\]
\end{exa}

\begin{rem}\label{remark:kleene_for_probabilistc}
By  Theorem~\ref{theorem:probabilistic_fit_the_framework} probabilistic automata
can be put into the framework of Section~\ref{section:automata}. Hence, Kleene theorems
hold for any suitable choice of $\mathcal{A}$. In particular, we may take
$\mathcal{A}$ to be the least set of maps containing all morphisms of the form
\[
n\stackrel{\alpha}{\to}\mathcal{D}(\Sigma\times n)
\stackrel{\nu_{\Sigma\times n}}{\to }
\mathfrak{D}(\Sigma\times n)
\hookrightarrow \mathfrak{D}(\Sigma^\ast\times n+\Sigma^\omega)
\]
and satisfying the properties listed in the beginning of
Subsection~\ref{subsection:kleene_omega_reg}.
\end{rem}

\subsection{Summary}
The purpose of this section was to put probabilistic automata into
a monadic framework from Section~\ref{section:automata} and reason
about their (in)finite behaviours. We achieved this by introducing the
continuous continuation monad $\mathfrak{D}$ and viewing probabilistic automata
transition maps as coalgebras
\[
X\to \mathfrak{D}(\Sigma^\ast \times X+\Sigma^\omega).
\]
The monad $\mathfrak{D}(\Sigma^\ast \times \mathcal{I}d+\Sigma^\omega)$ gives rise
to a Kleisli category which satisfies~\ref{condition:0} -~\ref{condition:3} from
 Section~\ref{section:automata} making
it possible to consider finite and infinite behaviours of automata taken
into consideration. We proved that the behaviour maps encode probabilities of certain
events from the execution space. These probabilities were attained without changing the
underlying category: the type monad and the automata taken into consideration
are $\mathsf{Set}$-based. Additionally, Theorem~\ref{theorem:infinite_beh_probabilistic}
suggests that our infinite behaviour with BAC for probabilistic automata is
similar to the one presented in~\cite{urabe_et_al:LIPIcs:2016:6186}. However,
in \emph{loc.\ cit.} the base category for probabilistic systems was the category
of measurable spaces and measurable functions. Hence,
by Remark~\ref{remark:kleene_for_probabilistc}, Kleene theorems can be instantiated in
our setting directly, but it is not possible to do so in the
setting from~\cite{urabe_et_al:LIPIcs:2016:6186}.
%%%%%%%%%%%%%%%%%%%%%%%%
%%%%%%%%%%%%%%%%%%%%%%%%%
% Summary
% %
\section{Summary}
The purpose of this paper was to develop a coalgebraic (categorical) framework
to reason about abstract automata and their finite and infinite behaviours
satisfying BAC\@. We achieved this goal by constructing a monad suitable for handling
the types of behaviours we were interested in and defining them in the right
setting. A natural and direct consequence of this treatment was Theorem~\ref{theorem:kleene_regular} and Theorem~\ref{theorem:kleene_omega_regular},
i.e.\ a (co)algebraic characterization of regular and $\omega$-regular behaviour
for systems whose type is a $\mathsf{Set}$-based monad satisfying some
additional properties. Our theory of finite and infinite behaviour for abstract automata has been
successfully instantiated on: non-deterministic automata, tree
automata and probabilistic automata.

\subsubsection*{Future work}
Given our natural characterization of coalgebraic ($\omega$-)regular languages we
ask if it is possible to characterize it in terms of
a preimage of a subset of a finite algebraic structure. Especially, considering
the fact that by Theorem~\ref{theorem:least_fixpoint} the pair of hom-sets
$(\mathbb{T}(n,n),\mathbb{T}(n,0))$ equipped with suitable operations resembles
a Wilke algebra used in the algebraic characterization of these languages (see
e.g.~\cite{pin:automata} for details).

Our definitions of the operators $(-)^\ast$ and $(-)^\omega$
via the least and greatest fixpoints
suggest a connection between our line of work and $\mu$-calculus~\cite{K83a,DBLP:books/cu/BlackburnRV01, LecturesOnModalMuCalculus}.
In particular,
it would be interesting to clarify how our coalgebraic framework
fits into the framework of coalgebraic modal
$\mu$-calculus and its semantics (e.g.~\cite{DBLP:journals/corr/abs-1105-2246,DBLP:conf/icalp/FontaineLV10})
with an emphasis laid on non-classical systems,
e.g.\ probabilistic systems from Section~\ref{section:probabilistic_systems}.

\subsubsection*{Related work}
The first coalgebraic take on $\omega$-languages was presented in~\cite{DBLP:conf/cmcs/CianciaV12}, where authors put deterministic Muller
automata with Muller acceptance condition into a coalgebraic framework.  Our work is
related to a more recent paper~\cite{urabe_et_al:LIPIcs:2016:6186}, where Urabe
\emph{et al.} give a coalgebraic framework for modelling behaviour with B\"uchi
acceptance condition for $(T,F)$-systems. The main ingredient of their work is a
solution to a system of equations which uses least and greatest fixpoints. This
is done akin to Park's~\cite{park1981:10.1007/BFb0017309} classical
characterization of $\omega$-languages via a system of equations. In our paper
we also use least and greatest fixpoints, however, the operators we consider are
the two natural types of  operators $(-)^\ast=\mu x. \mathsf{id} \vee x\cdot
(-)$ and $(-)^\omega=\nu x. x\cdot (-)$ which generalize the language operators
$(-)^\ast$ and $(-)^\omega$ known from the classical theory of regular and
$\omega$-regular languages. The definitions of behaviours
of an automaton are presented in terms of simple expressions involving Kleisli
composition and the above operators.
This allows us to state and prove generic Kleene theorems
for ($\omega$-)regular input which was not achieved in~\cite{urabe_et_al:LIPIcs:2016:6186} and (in our opinion) would be difficult to
obtain in that setting. To summarize, the major differences between our work and~\cite{urabe_et_al:LIPIcs:2016:6186} are the following:
\begin{itemize}
\item we use the setting of systems with internal moves (i.e.\ coalgebras over a
monad) to discuss infinite behaviour with BAC, which is given in terms of a simple
expression using $(-)^\ast$ and $(-)^\omega$ in the Kleisli category,
\item we provide the definition of (in)finite behaviours of a system
and build a bridge between regular and $\omega$-regular behaviours
by characterizing them on a categorical level in terms of the Kleene theorems.
\end{itemize}

\noindent
Abstract finite automata have already been considered in the computer science
literature in the context of Lawvere iteration theories with analogues of Kleene
theorems stated and proven (see e.g.~\cite{DBLP:journals/tcs/Esik97,esik2011,esik2013,esikhajgato2009:ai,bloomesik:93}). Some of these results
seem to be presented  using a slightly
different language than ours (see Theorem~\ref{theorem:kleene_regular}
and e.g.~\cite[Theorem 1.4]{bloomesik:93}). We decided to state
Theorem~\ref{theorem:kleene_regular} the way we did, in order to make a direct
generalization of the classical Kleene theorem for regular input and to give a
coalgebraic interpretation which is missing in~\cite{DBLP:journals/tcs/Esik97,esik2011,esik2013,esikhajgato2009:ai,bloomesik:93}.  We should also
mention that the infinite behaviour with BAC was defined in \emph{loc.\ cit.}
only for a very specific type of theories (i.e.\ the matricial theories over an
algebra with an infinite iteration operator), which do not encompass e.g.
non-deterministic B\"uchi tree automata and their infinite tree languages or
probabilistic automata and their infinite behaviour.

\bibliographystyle{alpha}
\bibliography{biblio}

\end{document}